\documentclass[manuscript]{emulateapj}
\usepackage{natbib}
\usepackage{times}
\usepackage{amsmath}
\usepackage{color}

\slugcomment{To appear in Astrophysical Journal Supplement Series}

\shorttitle{NGVS-IR, I. A New Near-UV/Optical/Near-IR Globular Cluster Selection Tool}
\shortauthors{R.~P.~Mu\~noz et al.}

\begin{document}

\title{The Next Generation Virgo Cluster Survey -- Infrared (NGVS-IR)\footnotemark[$\dagger$]: I. A new Near-UV/Optical/Near-IR Globular Cluster Selection Tool} 
\footnotetext[$\dagger$]{Based on observations obtained with WIRCam, a joint project of CFHT, Taiwan, Korea, Canada, France, at the Canada-France-Hawaii Telescope (CFHT) which is operated by the National Research Council (NRC) of Canada, the ``Institute National des Sciences de l'Univers of the Centre National de la Recherche Scientifique" of France, and the University of Hawaii.}

\author{Roberto~P.~Mu\~noz$^{1,2}$, Thomas H.~Puzia$^{1}$, Ariane~Lan\c{c}on$^{2}$, Eric W. Peng$^{3,4}$, Patrick C\^{o}t\'e$^{5}$, Laura Ferrarese$^{5}$, John P. Blakeslee$^{5}$, Simona Mei$^{6,7,8}$, Jean-Charles Cuillandre$^{9}$, Patrick Hudelot$^{10}$, St\'ephane Courteau$^{11}$, Pierre-Alain Duc$^{12}$, Michael L. Balogh$^{13}$, Alessandro Boselli$^{14}$, Fr\'ed\'eric Bournaud$^{12}$, Raymond G. Carlberg$^{15}$, Scott C. Chapman$^{16}$,  Patrick Durrell$^{17}$, Paul Eigenthaler$^{1}$, Eric Emsellem$^{18,19}$, Giuseppe Gavazzi$^{20}$, Stephen Gwyn$^{5}$, Marc Huertas-Company$^{6,7}$, Olivier Ilbert$^{14}$, Andr\'es Jord\'an$^{1}$, Ronald L\"asker$^{21}$, Rossella Licitra$^{6}$, Chengze Liu$^{22,23}$, Lauren MacArthur$^{5}$, Alan McConnachie$^{5}$, Henry Joy McCracken$^{10}$, Yannick Mellier$^{10}$, Chien Y. Peng$^{24}$, Anand Raichoor$^{7}$, Matthew A. Taylor$^{1}$, John L. Tonry$^{25}$, R. Brent Tully$^{25}$, Hongxin Zhang$^{4}$}
\affil{
$^{1}$Instituto de Astrof\'isica, Facultad de F\'isica, Pontificia Universidad Cat\'olica de Chile, Av.~Vicu\~na Mackenna 4860, 7820436 Macul, Santiago, Chile\\
$^{2}$Observatoire astronomique de Strasbourg, Universit\'e de Strasbourg, CNRS, UMR 7550, 11 rue de l'Universite, F-67000 Strasbourg, France\\
$^{3}$Department of Astronomy, Peking University, Beijing 100871, China\\
$^{4}$Kavli Institute for Astronomy and Astrophysics, Peking University, Beijing 100871, China\\
$^{5}$Herzberg Institute of Astrophysics, National Research Council of Canada, Victoria, BC V9E 2E7, Canada\\
$^{6}$GEPI, Observatoire de Paris, CNRS, Universit\'e Paris Diderot, 5 Place J. Janssen, F-92190 Meudon Cedex, France\\
$^{7}$Universit\'e Paris Denis Diderot, 75205 Paris Cedex 13, France\\
$^{8}$California Institute of Technology, Pasadena, CA 91125, USA\\
$^{9}$Canada-France-Hawaii Telescope Corporation, Kamuela, HI 96743, USA\\
$^{10}$Institut d'Astrophysique de Paris, UMR 7095 CNRS \& UPMC, 98 bis Boulevard Arago, F-75014 Paris, France\\
$^{11}$Department of Physics, Engineering Physics and Astronomy, Queen's University, Kingston, ON, Canada\\
$^{12}$Laboratoire AIM Paris-Saclay, CNRS/INSU, Universit\'e Paris Diderot, CEA/IRFU/SAp, F-91191 Gif-sur-Yvette Cedex, France\\
$^{13}$Department of Physics and Astronomy, University of Waterloo, Waterloo, ON N2L 3G1, Canada\\
$^{14}$Aix Marseille Universit\'e, CNRS, LAM (Laboratoire d'Astrophysique de Marseille) UMR 7326, F-13388 Marseille, France\\
$^{15}$Department of Astronomy and Astrophysics, University of Toronto, Toronto, ON M5S 3H4, Canada\\
$^{16}$Institute of Astronomy, University of Cambridge, Madingley Road, Cambridge CB3 0HA, UK\\
$^{17}$Department of Physics and Astronomy, Youngstown State University, One University Plaza, Youngstown, OH 44555, USA\\
$^{18}$Universit\'e de Lyon 1, CRAL, Observatoire de Lyon, 9 av. Charles Andr\'e, F-69230 Saint-Genis Laval; CNRS, UMR 5574; ENS de Lyon, France\\
$^{19}$European Southern Observatory, Karl-Schwarzchild-Str. 2, D-85748 Garching, Germany\\
$^{20}$Universit\`a degli Studi di Milano-Bicocca, Piazza della Scienza 3, 20126, Milano, Italy\\
$^{21}$Max Planck Institute for Astronomy, K\"onigstuhl 17, 69117 Heidelberg, Germany\\
$^{22}$Center for Astronomy and Astrophysics, Department of Physics and Astronomy, Shanghai Jiao Tong University, 800 Dongchuan Road, Shangai 200240, China\\
$^{23}$INPAC, Department of Physics and Astronomy and Shanghai Key Lab for Particle Physics and Cosmology, Shanghai Jiao Tong University, Shanghai 200240, China\\
$^{24}$Giant Magellan Telescope Organization, 251 South Lake Avenue, Suite 300, Pasadena, CA 91101, USA\\
$^{25}$Institute for Astronomy, University of Hawaii, 2680 Woodlawn Drive, Honolulu, HI 96822, USA}
\email{rmunoz@astro.puc.cl}

\begin{abstract}
The NGVS-IR project (Next Generation Virgo Survey -- Infrared) is a contiguous near-infrared imaging survey of the Virgo cluster of galaxies. It complements the optical wide-field survey of Virgo (NGVS). The current state of NGVS-IR consists of $K_s$-band imaging of 4 $deg^2$ centered on M87, and $J$ and $K_s$-band imaging of $\sim\!16$ $deg^2$ covering the region between M49 and M87.~In this paper, we present the observations of the central 4 $deg^2$ centered on Virgo's core region.~The data were acquired with WIRCam on the Canada-France-Hawaii Telescope and the total integration time was 41 hours distributed in 34 contiguous tiles.~A survey-specific strategy was designed to account for extended galaxies while still measuring accurate sky brightness within the survey area.~The average $5\sigma$ limiting magnitude is $K_s\!=\!24.4$\,AB mag and the 50\% completeness limit is $K_s\!=\!23.75$\,AB mag for point source detections, when using only images with better than 0.7\arcsec\ seeing (median seeing $0.54\arcsec$). Star clusters are marginally resolved in these image stacks, and Virgo galaxies with $\mu_{K_s}\!\simeq\!24.4$\,AB mag arcsec$^{-2}$ are detected.~Combining the $K_s$ data with optical and ultraviolet data, we build the $uiK_s$ color-color diagram which allows a very clean color-based selection of globular clusters in Virgo.~This diagnostic plot will provide reliable globular cluster candidates for spectroscopic follow-up campaigns needed to continue the exploration of Virgo's photometric and kinematic sub-structures, and will help the design of future searches for globular clusters in extragalactic systems.~We show that the new $uiK_s$ diagram displays significantly clearer substructure in the distribution of stars, globular clusters, and galaxies than the $gzK_s$ diagram -- the NGVS+NGVS-IR-equivalent of the $BzK$ diagram, which is widely used in cosmological surveys. Equipped with this powerful new tool, future NGVS-IR investigations based on the $uiK_s$ diagram will address the mapping and analysis of extended structures and compact stellar systems in and around Virgo galaxies.
 \end{abstract}


\section{Introduction}

The proximity of the Virgo galaxy cluster, located at a distance of 16.5$\pm$0.2 Mpc \citep{mei07,blakeslee09}, makes it one of the closest and most thoroughly studied laboratories of star and galaxy formation in the nearby universe.~With about 2000 known galaxy members \citep{binggeli85, gavazzi03}, more than $10^4$ globular clusters (GCs) in M\,87 and M\,49 alone \citep{tamura06a, tamura06b, peng08} and a total GC population of $(6.48\pm1.44)\!\times\!10^4$ for the entire NGVS region (Durrell et al.~2013, {\it in prep.}), it offers vast sample statistics to investigate the evolution of stellar populations and the formation and assembly histories of their host galaxies in a relatively dense cluster environment.

The first detailed census of the Virgo galaxy cluster was published by \cite{ames30} and consisted of a catalog of 2278 galaxies brighter than 18th magnitude. This survey led to the identification of several background clusters and to the discovery of a Southern extension of the Virgo cluster, which are impressive results given the technical limitations existing at that time.~A quarter of a century later, \cite{reaves56} published the first comprehensive catalog of dwarf galaxies in the Virgo cluster based on the study of photographic plates taken with the Lick 20-inch astrograph. This work showed the presence of a large number of dwarf elliptical galaxies beyond the local group and started a new phase of the Virgo cluster research.~The most important contribution to assessing Virgo cluster membership was published by \cite{binggeli85} in the form of the famous Virgo Cluster Catalog (VCC).~This catalog consists of 2096 galaxies, covers an area of $140\;deg^2$ and is complete down to $B_T\leq 18$ mag.

\subsection{The Next Generation Virgo Cluster Survey}

The VCC has been the reference standard of Virgo for more than a quarter of a century and it is about to be superseded by the Next Generation Virgo Cluster Survey \citep[NGVS,][]{ferrarese12}. The NGVS is a multi-passband optical survey conducted with MegaCam at the Canada-France-Hawaii Telescope (CFHT) on Mauna Kea, Hawaii, a 3.6-meter telescope that benefits from superb seeing and transparency conditions.

The NGVS survey area covers 104 contiguous square degrees out to the virial radius around M\,87 and M\,49 to unprecedented photometric depths in the five optical filters\footnote{Several designations are used to denote MegaCam filters. For simplicity, we adopt $u^*,g,r,i$ and $z$, since of all filters only $u^*$ is significantly different from SDSS u.} $u^*,g,r,i$ and $z$.~With substantially sub-arcsecond seeing, a surface brightness sensitivity better than $\mu_g\!\simeq\!29.0$\,mag arcsec$^{-2}$ and a point-source detection limit of $g\!\simeq\!25.9$\,mag ($10\,\sigma$), this survey will extend our optical view of the galaxy luminosity function and the scaling relations of galaxies to luminosities more than five magnitudes below the faintest VCC galaxies\footnote{Unless stated otherwise, magnitudes quoted in this paper are given in the AB magnitude system. See Appendix~\ref{app:ABvsVega} for the numerical conversions between AB and Vega magnitudes for NGVS and NGVS-IR filters.}.

The five optical passbands of NGVS combined with the superior spatial resolution of the images ($i$-band seeing is better than $0.6\arcsec\!\simeq\!48$\,pc at Virgo distance) provide a first classification of the sources, but leave important areas of ambiguity.~In particular, the identification and study of compact stellar systems and their stellar population properties is challenged by the age-metallicity-extinction degeneracy of optical colors \citep{worthey94, arimoto96}.~The difficulties begin already at the level of sample definition:~The distribution of old metal-poor GCs merges into the parameter space of dwarf stars in optical color-color diagrams, while metal-rich GCs overlap with the locus of remote late-type galaxies. Around selected Virgo galaxies, the pointed observations of the HST/ACS Virgo Cluster Survey (ACSVCS) have allowed a less ambiguous determination of the GC population due to the improved spatial resolution of HST \citep{cote04, jordan09}.~Spectroscopic surveys are progressively extending the collection of well-classified objects to vaster areas, though with inevitable sparseness and depth limitations \citep[e.g.][]{park12, romanowsky12, schuberth12, pota13}.

\subsection{The Near-Infrared Complement: NGVS-IR}
The NGVS near-infrared project (NGVS-IR) is designed to complement the NGVS with deep, spatially well-resolved images at wavelengths longer than the $z$ band to provide wider sampling of the spectral energy distribution (SED). The first part of this project, and the subject of this paper, is a deep $K_s$-band survey conducted with the {\it Wide-field InfraRed Camera} \citep[WIRCam;][]{pug04} at CFHT that covers the central $2\times2$ square degrees around the Virgo cluster core, the location of the cD galaxy M\,87 and its vicinity. This area was the first fully surveyed in $ugriz$ as part of the NGVS pilot survey, and we will refer to this field as the pilot field hereafter. The NGVS-IR observations are currently being expanded beyond the pilot field with the 4-meter class {\it Visible and Infrared Survey Telescope for Astronomy} \citep[VISTA;][]{dalton06} at the European Southern Observatory (ESO) on Cerro Paranal in Chile.~These observations slightly overlap with the pilot field and cover regions around the brightest elliptical galaxy in Virgo, M\,49 in the filters $J$ and $K_s$, as well as a strip of $J$ and $K_s$-band images that connects the pilot field and the M\,49 field, and augments the coverage of the M\,86 group.~Taken altogether, the current NGVS-IR survey area provides a total contiguous spatial coverage of the central regions in Virgo of more than $\sim\!20$ deg$^2$.

\subsection{The Need for NGVS-IR}

The intermediate goals of the NGVS-IR survey are the development of a photometric selection method to identify globular clusters (GCs) in the Virgo core region, the study of the age-metallicity-mass relations of GCs in M87 and the intracluster medium, the characterization of the stellar populations of Ultra-Compact Dwarf galaxies (UCDs) by combining optical/near-IR colors, the measurement of accurate distances to bright Virgo galaxies by using the Surface Brightness Fluctuation (SBF) method \citep{tonry90}, the study of the star formation histories and mass profiles of several Virgo galaxies, and lastly, the determination of the galaxy mass function of the Virgo cluster.

Although the Virgo cluster has already been observed at near-IR wavelengths, none of the available near-IR surveys is deep enough to accomplish the NGVS science goals. The Two Micro All Sky Survey \citep[2MASS,][]{skrutskie06} provides uniform, precise photometry and astrometry over the entire celestial sphere in $J,H$ and $K_s$ bandpasses with $10\sigma$ limiting magnitudes of 15.8, 15.1 and 14.3 mag, respectively.~The 2MASS survey offers the largest coverage of the Virgo cluster in near-IR wavelengths, but its shallow photometry is not well suited for deep extragalactic studies.~Deeper photometry is provided by the UKIRT Infrared Deep Sky Survey \citep[UKIDSS,][]{lawrence07} Large Area Survey (LAS), which has a much smaller area than 2MASS but reaches $5\sigma$ limiting magnitudes of 20.5, 20.0, 18.8 and 18.4 mag in $Y,J,H$ and $Ks$ bandpasses, respectively.~The deepest near-IR photometry of Virgo galaxies corresponds to the Spectroscopic and H-band Imaging Survey of Virgo cluster galaxies \citep[SHiVir,][]{mcdonald11}, that consists of pointed $H$-band observations with the UH2.2 meter, UKIRT, and CFHT telescopes in good seeing conditions ($<\!0.6\arcsec$). While SHiVir provides deep $H$-band imaging for 286 VCC galaxies, it does not offer the one-to-one contiguous mapping with the optical NGVS data like NGVS-IR does.

One of the most immediate aims of NGVS-IR is to improve source classification in order to obtain the cleanest possible samples of objects of any given type, and of GCs in particular.~For this purpose, it was decided to produce a first set of mosaic images and catalogs with a focus on compact sources with small angular sizes.~As will be shown below, adding near-IR photometry to the optical data provides an impressive separation of Virgo GCs from foreground stars and background galaxies in near-UV/optical/near-IR color space. The $uiK_s$ diagram\footnote{For simplicity, we will indifferently use the letters $u$ and $u^*$ to refer to the MegaCam filter.}, in particular, proves to be an exceptional tool for future searches of GCs and compact stellar systems, such as Ultra-Compact Dwarfs \citep[UCDs, see][and references therein]{hilker11} around galaxies in a large variety of environments.~With more than 20\,deg$^2$ of near-IR data soon available for Virgo, we will improve the decontamination of the optically-selected samples currently analyzed within the NGVS collaboration and provide extensive studies of the GC distributions in luminosity, color, and space. Those distributions contain a record of GC formation mechanisms and GC system assembly histories. They are also a key test for stellar population models, which still struggle to exploit optical and near-IR photometry jointly \citep[e.g.][]{pessev08, taylor11, riffel11}. Finally, the construction of clean GC samples facilitates very efficient spectroscopic follow-up campaigns with little contamination by other sources. This, again, turns GCs into probes of galaxy cluster dynamics out to large scales \citep[e.g.][]{schuberth12, romanowsky12, pota13}

\begin{figure}
   \centering
   \includegraphics[width=8.5cm]{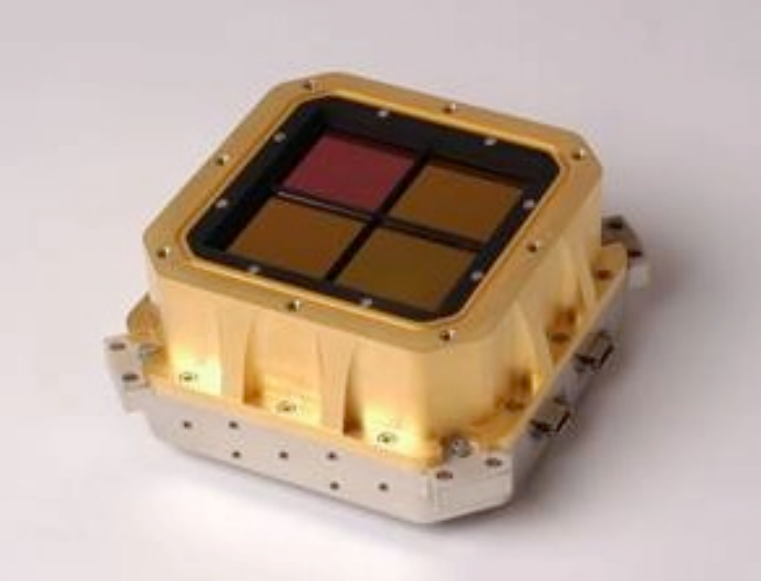} 
   \caption{Illustration of the $2\!\times\!2$ mosaic CFHT/WIRCam detector array. Each chip features a HAWAII2-RG near-infrared HgCdTe detector with 18\,$\mu$m pixels. The patrol field has $2040\!\times\!2040$ active pixels with a 4-pixel wide border of reference pixels per chip. The field of view of the full detector array is about $21\arcmin\!\times\!21\arcmin$ with 45\arcsec\ wide inter-chip gaps. {\it Image courtesy of Academia Sinica Institute of Astronomy and Astrophysics (ASIAA).}}
   \label{fig:cfhtwircam}
\end{figure}

A further motivation for NGVS-IR is the study of UCDs and the dwarf galaxy -- globular cluster transition. In the past decade, studies of the central regions of the Virgo and Fornax galaxy clusters have revealed dozens of massive compact objects in the magnitude range $-14 \la M_V \la -11$\,mag with half-light radii between 10 and 30\,pc \citep{hasegan05,jones06,misgeld11}.~Currently, two popular formation scenarios are that UCDs are the remnant nuclei of dwarf galaxies that were stripped by interactions \citep[e.g.][]{bekki03}, or alternatively that they are agglomerates of multiple clusters that merged early in their lifetimes during violent star-formation episodes such as those triggered by galaxy mergers \citep[e.g.][]{fellhauer02}. These scenarios can be tested through studies of the mass-metallicity relation \citep[e.g.][]{harris06}. NGVS will provide us with a larger sample of massive compact stellar systems within Virgo than ever before, and with NGVS-IR we will be able to quantify the ages, metallicities and masses of these objects to higher accuracy than with optical colors alone \citep[see e.g][]{puzia02, hempel04, pessev08}. 

\medskip

The paper is organized as follows.~We describe the details of our observations in section~2 and the data reduction steps in section~3.~Sections~4 and 5 contain a discussion of the point source photometry and completeness estimates, while section~6 includes the presentation of our first results. We summarize our work in section 7.

%%%%%%%%%%%%%%%%%%%%%%%%%%%%%%%%%%%%%%%%%%%%%%%%%%%%%%%%%

\section{Observations}
\subsection{Instrumental Setup}

\begin{table*}
\begin{center}
\caption{Journal of all CFHT/WIRCam observations, grouped according to CFHT observing programs.}
\label{tab:obs_full}
\begin{tabular}{@{}clr@{:}c@{:}lr@{:}c@{:}lllll}
  \hline\hline
  \multicolumn{1}{c}{Program\tablenotemark{a}} & \multicolumn{1}{c}{Tile} & \multicolumn{3}{c}{R.A.} & \multicolumn{3}{c}{Dec.} & \multicolumn{2}{c}{Observing date} & \multicolumn{1}{c}{$\rm{N}_{\rm{all}}$\tablenotemark{c}} & \multicolumn{1}{c}{$\rm{N}_{0.7}$\tablenotemark{d}} \\
   \multicolumn{1}{c}{($T_{\rm exp}$)}& & \multicolumn{3}{c}{(J2000)} & \multicolumn{3}{c}{(J2000)} & \multicolumn{1}{c}{Month, Year} & \multicolumn{1}{l}{MJD (55XXX)\tablenotemark{b}} & & \\
  \hline
  09BC26 & t1 & 12 & 33 & 21 & 11 & 40 & 09 & December, 2009 & 187,188 & 108 & 107 \\
  (7.6 h)   &  t2 & 12 & 32 & 04 & 11 & 40 & 13 & December, 2009 & 188,189 & 72 & 67 \\
	&	t3 & 12 & 30 & 48 & 11 & 40 & 14 & December, 2009 & 194,195 & 109 & 81 \\
	&	t7 & 12 & 33 & 21 & 11 & 59 & 07 & December, 2009 & 190,191 & 106 & 104 \\
	&	t8 & 12 & 32 & 05 & 11 & 59 & 08 & December, 2009 & 187,188 & 109 & 103 \\
	&	t9 & 12 & 30 & 48 & 11 & 59 & 13 & December, 2009 & 188,189 & 72 & 71 \\
	&	t10 & 12 & 29 & 33 & 11 & 59 & 09 & December, 2009 & 192,193 & 108 & 69 \\
	&	t11 & 12 & 28 & 17 & 11 & 59 & 14 & December, 2009 & 194,195 & 100 & 69 \\
	&	t14 & 12 & 32 & 05 & 12 & 18 & 09 & December, 2009 & 190,191 & 108 & 108 \\
	&	t16 & 12 & 29 & 33 & 12 & 18 & 15 & December, 2009 & 194,195 & 101 & 65 \\
	&	t22 & 12 & 29 & 33 & 12 & 37 & 09 & December, 2009 & 192,193 & 108 & 106 \\ 	
  \hline
  09BF22 & t4 & 12 & 29 & 32 & 11 & 40 & 13 & December, 2009 & 191,193 & 72 & 71 \\
  (1.5 h)	&  t17 & 12 & 28 & 16 & 12 & 18 & 13 & December, 2009 & 191,193 & 72 & 72 \\
	&	t20 & 12 & 32 & 05 & 12 & 37 & 12 & December, 2009 & 189 & 36 & 36 \\
	&	t31 & 12 & 33 & 21 & 13 & 15 & 12 & December, 2009 & 189 & 36 & 32 \\
  \hline
  10AC10 & t1 & 12 & 33 & 21 & 11 & 40 & 03 & March, 2010 & 279,280 & 134 & 70 \\
  (18.6 h) & t2 & 12 & 32 & 05 & 11 & 40 & 12 & March-April, 2010 & 280,281,315 & 112 & 39 \\
	&	t3 & 12 & 30 & 49 & 11 & 40 & 12 & April, 2010 & 288,289,290,291,308,310 & 112 & 96 \\
	&	t7 & 12 & 33 & 21 & 11 & 59 & 08 & March, 2010 & 283 & 108 & 108 \\
	&	t8 & 12 & 32 & 04 & 11 & 59 & 13 & March, 2010 & 279,280 & 140 & 90 \\
	&	t9 & 12 & 30 & 49 & 11 & 59 & 09 & March-April, 2010 & 280,281,315 & 109 & 32 \\
	&	t10 & 12 & 29 & 33 & 11 & 59 & 06 & March-April, 2010 & 283,287 & 116 & 71 \\
	&	t11 & 12 & 28 & 17 & 11 & 59 & 09 & April, 2010 & 288,289,290,291,308,310 & 108 & 99 \\
	&	t13 & 12 & 33 & 21 & 12 & 18 & 10 & April-July,  2010 & 310,315,342,374,377,378,379,380 & 292 & 240 \\
	&	t14 & 12 & 32 & 05 & 12 & 18 & 08 & March, 2010 & 283 & 108 & 108 \\
	&	t15 & 12 & 30 & 49 & 12 & 18 & 10 & April-July, 2010 & 310,315,342,374,377,378,379,380 & 301 & 196 \\
	&	t16 & 12 & 29 & 33 & 12 & 18 & 08 & April, 2010 & 288,289,290,291,308,310 & 108 & 86 \\
	&	t19 & 12 & 33 & 21 & 12 & 37 & 10 & April-July, 2010 & 310,315,342,374,377,378,379,380 & 296 & 247 \\
	&	t21 & 12 & 30 & 49 & 12 & 37 & 12 & April-May, 2010 & 311,313,315,343 & 175 & 86 \\
	&	t22 & 12 & 29 & 32 & 12 & 37 & 10 & March-April, 2010 & 283,287 & 112 & 99 \\
	&	t25 & 12 & 33 & 21 & 12 & 56 & 12 & April-May, 2010 & 311,313,315,343 & 169 & 131 \\
	&	t26 & 12 & 32 & 05 & 12 & 56 & 12 & April-May, 2010 & 311,313,315,343 & 172 & 121 \\
  \hline
  10AF03 & t4 & 12 & 29 & 33 & 11 & 40 & 08 & April, 2010 & 287,288,289 & 113 & 108 \\
  (13.4 h) & t12 & 12 & 27 & 00 & 11 & 59 & 15 & April, 2010 & 289,290 & 158 & 88 \\
	&	t17 & 12 & 28 & 17 & 12 & 18 & 08 & April, 2010 & 287,288,289 & 112 & 108 \\
	&	t18 & 12 & 27 & 01 & 12 & 18 & 10 & May, 2010 & 318,320,322,323 & 117 & 87 \\
	&	t20 & 12 & 32 & 04 & 12 & 37 & 11 & April-May, 2010 & 315,316,317 & 130 & 97 \\
	&	t23 & 12 & 28 & 17 & 12 & 37 & 11 & May-July, 2010 & 323,383 & 106 & 101 \\
	&	t24 & 12 & 27 & 01 & 12 & 37 & 15 & April, 2010 & 289,290 & 153 & 117 \\
	&	t27 & 12 & 30 & 49 & 12 & 56 & 06 & April, 2010 & 290,291 & 112 & 91 \\
	&	t28 & 12 & 29 & 33 & 12 & 56 & 08 & April, 2010 & 290,291 & 108 & 95 \\
	&	t29 & 12 & 28 & 17 & 12 & 56 & 12 & May-July, 2010 & 323,383 & 109 & 91 \\
	&	t30 & 12 & 27 & 01 & 12 & 56 & 07 & May, 2010 & 318,320,322,323 & 122 & 77 \\
	&	t31 & 12 & 33 & 21 & 13 & 15 & 07 & April-May, 2010 & 315,316,317 & 131 & 79 \\
	&	t32 & 12 & 32 & 05 & 13 & 15 & 17 & May-July, 2010 & 341,343,382,383 & 81 & 60 \\
	&	t33 & 12 & 30 & 49 & 13 & 15 & 07 & May-July, 2010 & 341,343,382,383 & 78 & 60 \\
	&	t34 & 12 & 29 & 33 & 13 & 15 & 10 & May-July, 2010 & 341,343,382,383 & 78 & 57 \\
	&	t35 & 12 & 28 & 17 & 13 & 15 & 08 & May-July, 2010 & 323,383 & 109 & 99 \\
	&	t36 & 12 & 27 & 01 & 13 & 15 & 05 & May, 2010 & 318,320,322,323 & 116 & 78 \\
\hline\hline
\end{tabular}
\tablenotetext{a}{The total exposure time for the corresponding semester in hours is given below the program identifier.}
\tablenotetext{b}{Modified Julian date.}
\tablenotetext{c}{Number of observed frames.~Note that the given number of frames refers to all science observations. In particular, there are no separate ``sky" observations with our pipeline reduction technique (see Section~\ref{txt:sky_subtractiontech}).}
\tablenotetext{d}{Number of frames with seeing lower than $0.7\arcsec$.}
\end{center}
\end{table*}

\begin{figure*}
   \centering
   \includegraphics[width=17.8cm]{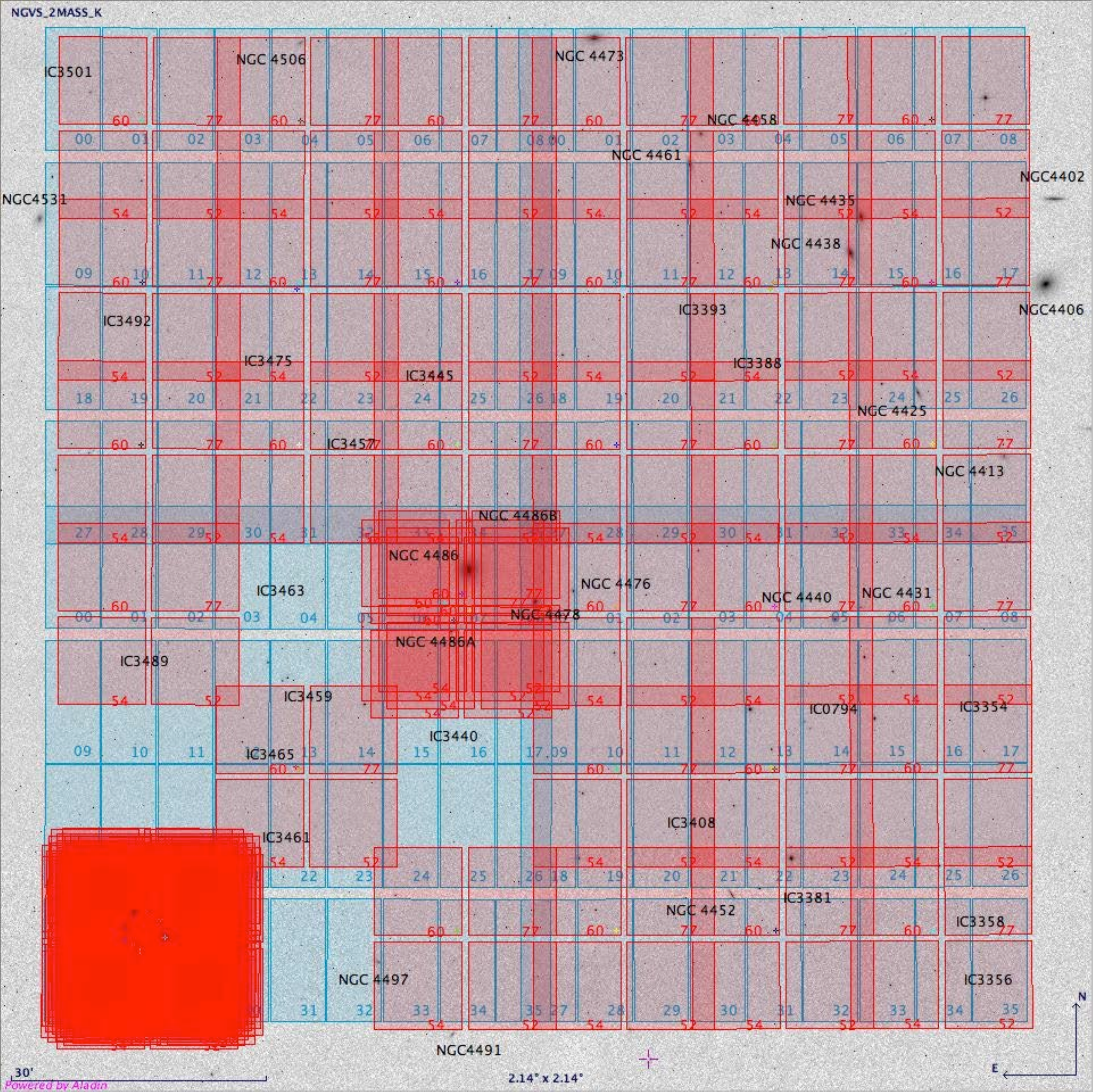}
   \caption{Illustration of the central $2.14^{\rm o}\!\times\!2.14^{\rm o}$ core region of the Virgo Cluster that was imaged by the NGVS-IR pilot project with CFHT/WIRCam in a $6\!\times\!6$ pointing mosaic (red pointings) and with CFHT/MegaCam in a $2\!\times\!2$ pointing mosaic (blue pointings).~The bottom-left tile of the WIRCam mosaic corresponds to tile 1 in Figure~\ref{fig:wircam_tiles} and shows the coverage achieved with 27 visits (see Section~\ref{txt:pointing_dither} for details), while tile 15, overlapping with the central giant elliptical M\,87 (NGC\,4486), shows the coverage of an individual 4-exposure dither (see Section~\ref{txt:exposure_dither}).~Note that four of the (red) WIRCam pointings in the lower left quadrant of the image that correspond to tiles 2, 7, 9, and 14 in Figure~\ref{fig:wircam_tiles} have been deliberately left out for illustration purposes of the tile-to-tile overlap regions produced by the two interlocking dither types.~All other (red) WIRCam tiles show the sky coverage of one individual exposure.~The most prominent galaxies are labeled with their corresponding NGC or IC numbers. The underlying image is a 2MASS K-band rendering of the field.}
   \label{fig:NGVSpilot}
\end{figure*}

The NGVS-IR observations for the pilot field were carried out at the Canada-France-Hawaii Telescope (CFHT), using the Wide-field InfraRed Camera \citep[WIRCam;][]{pug04}.~WIRCam is mounted at the prime focus and consists of four cryogenically cooled $2048\times2048$ HAWAII2-RG near-infrared (near-IR: $0.9\!-\!2.4\,\mu m$) HgCdTe detectors arranged in a $2\!\times\!2$ mosaic, with typical interchip gaps of 45\arcsec\ and a plate scale of 0.3\arcsec\,per pixel (see Figure~\ref{fig:cfhtwircam}).~The array is operated at $\sim\!80$ Kelvin with negligible dark current ($\sim\!0.05\,e^-$/sec) and $30\ e^-$ readout noise. The field of view (FOV) of the full mosaic is about $21\arcmin\!\times\!21\arcmin$ on the sky.~The observations took place over a series of observing programs spanning the time period from December 2009 to July 2010 split over the French and Canadian TAC time allocation (see Table~\ref{tab:obs_full}).~They are arranged in a mosaic of $6\!\times\!6$ WIRCam pointing or tiles, that cover $2\!\times\!2$ degrees of the Virgo cluster around the central giant elliptical galaxy M\,87 (see Figure~\ref{fig:NGVSpilot}), and match four pilot project CFHT/MegaCam pointings of the NGVS program \citep{ferrarese12}. All observations were carried out in Queue Service Observing mode (QSO) at an airmass less than 1.2 under clear observing conditions, when the estimated seeing was better than 0.8\arcsec.

\begin{figure*}
\centering
   \includegraphics[width=17.8cm]{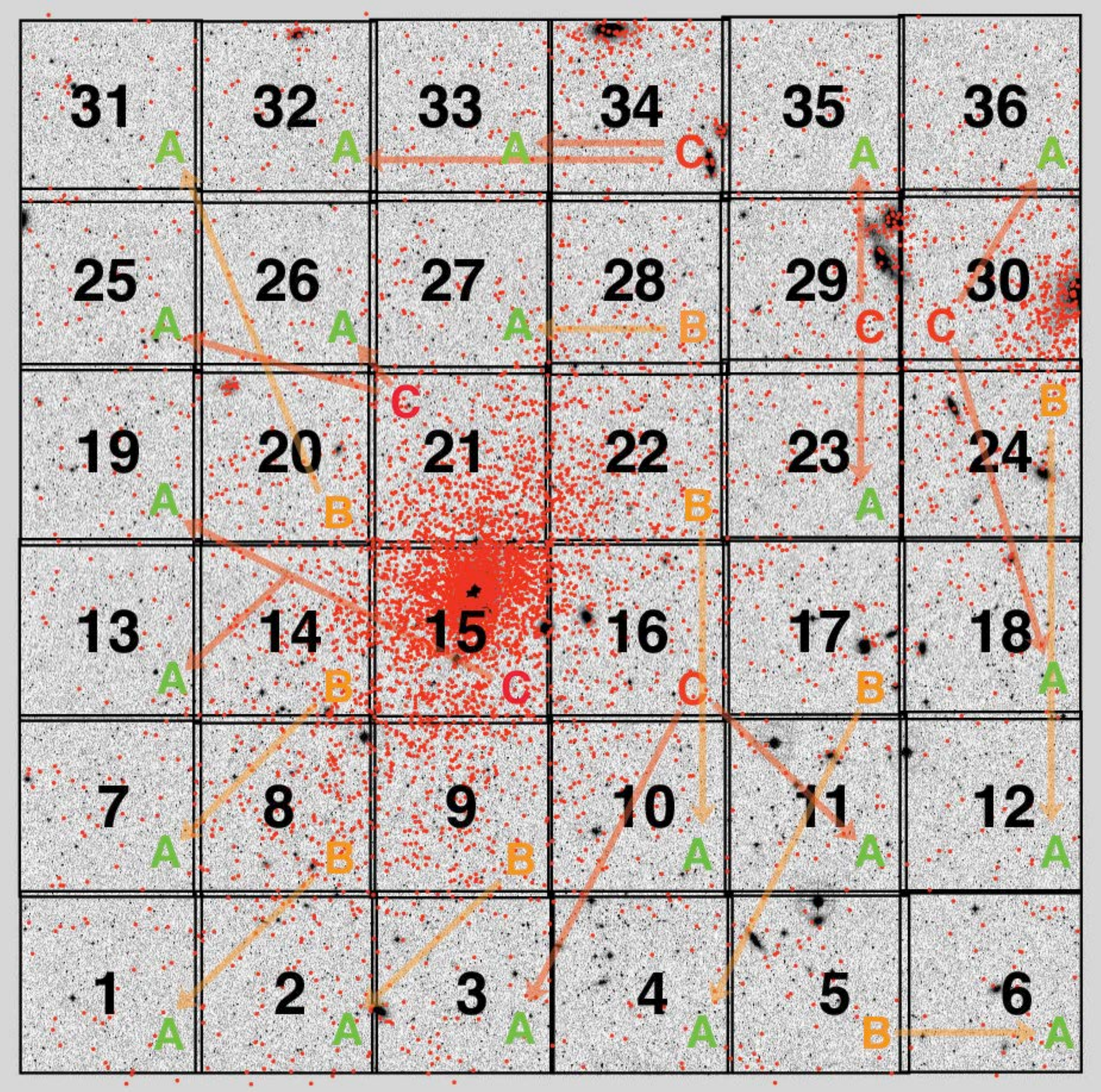} 
\caption{The CFHT/WIRCam pointing mosaic of the NGVS-IR pilot field that corresponds to Figure~\ref{fig:NGVSpilot}. The mosaic covers the inner $2\!\times\!2$ deg$^2$ around M87 with photometrically selected globular clusters (E. W. Peng, in preparation) shown as red dots on top of a 2MASS image. The arrows indicate the (A$_1$-C-A$_2$) and (A-B) tile-to-tile sequencing of the observations.}
\label{fig:wircam_tiles}
\end{figure*} 

\subsection{The Dithering Pattern}
\label{txt:dithering}
\subsubsection{General Considerations}
Pointed near-IR observations are usually executed with a dither pattern that sequentially observes the science target followed by a blank-sky region to account for the strongly varying near-IR sky surface brightness.~This strategy involves significant overhead and generates many sky observations that have generally no scientific use.~In large surveys of non-crowded fields one may use the area of interest itself to estimate the sky, by implementing dithered observations.~The dithering pattern also serves to fill the gaps between individual detector chips and defect areas of the detectors (of which WIRCam has quite a few).~The NGVS-IR field, however, targets an area containing very extended galaxies.~Surface brightness profile analyses of M87 found a $D_{25}\!\approx\!10\arcmin$ \citep{king78, RC3, liu05}, which is similar to the FOV of a single WIRCam chip.~We developed a dedicated dithering strategy to obtain sky estimates within the survey area despite the presence of large galaxies.~We use an individual integration time of 25 seconds to avoid saturation on the sky (the near-IR sky surface brightness on Mauna Kea\footnote{For a detailed description see the CFHT/WIRCam web pages at http://www.cfht.hawaii.edu.} varies around $\mu_{K_s}\!\approx\!16$ AB mag/arcsec$^{-2}$). A combination of small and large offsets, the latter of which are defined based on archival images of the field, are used to sample the sky on a timescale shorter than the typical timescales of variations in its brightness.

\begin{figure*}
   \centering
   \includegraphics[width=8.5cm]{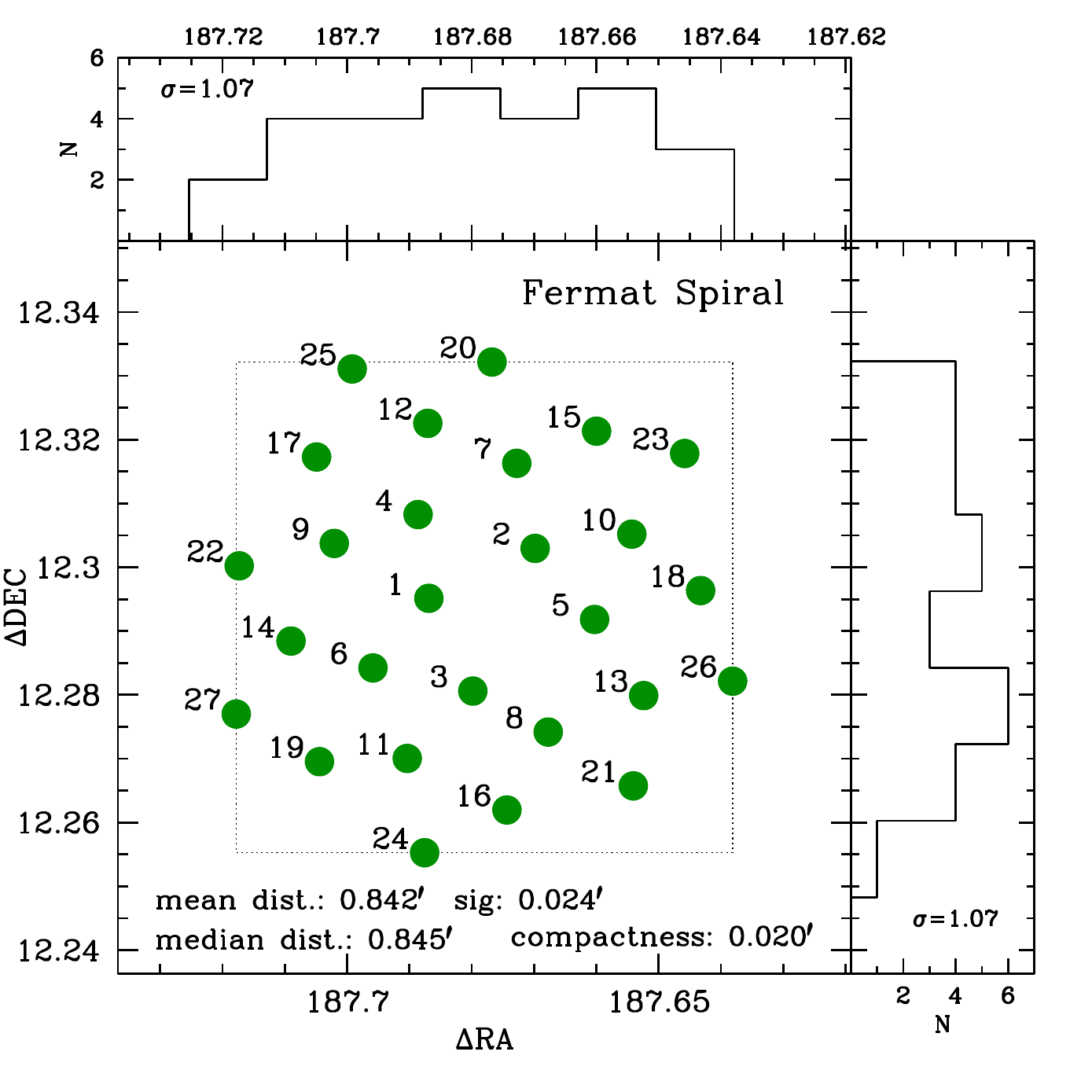} 
   \includegraphics[width=8.5cm]{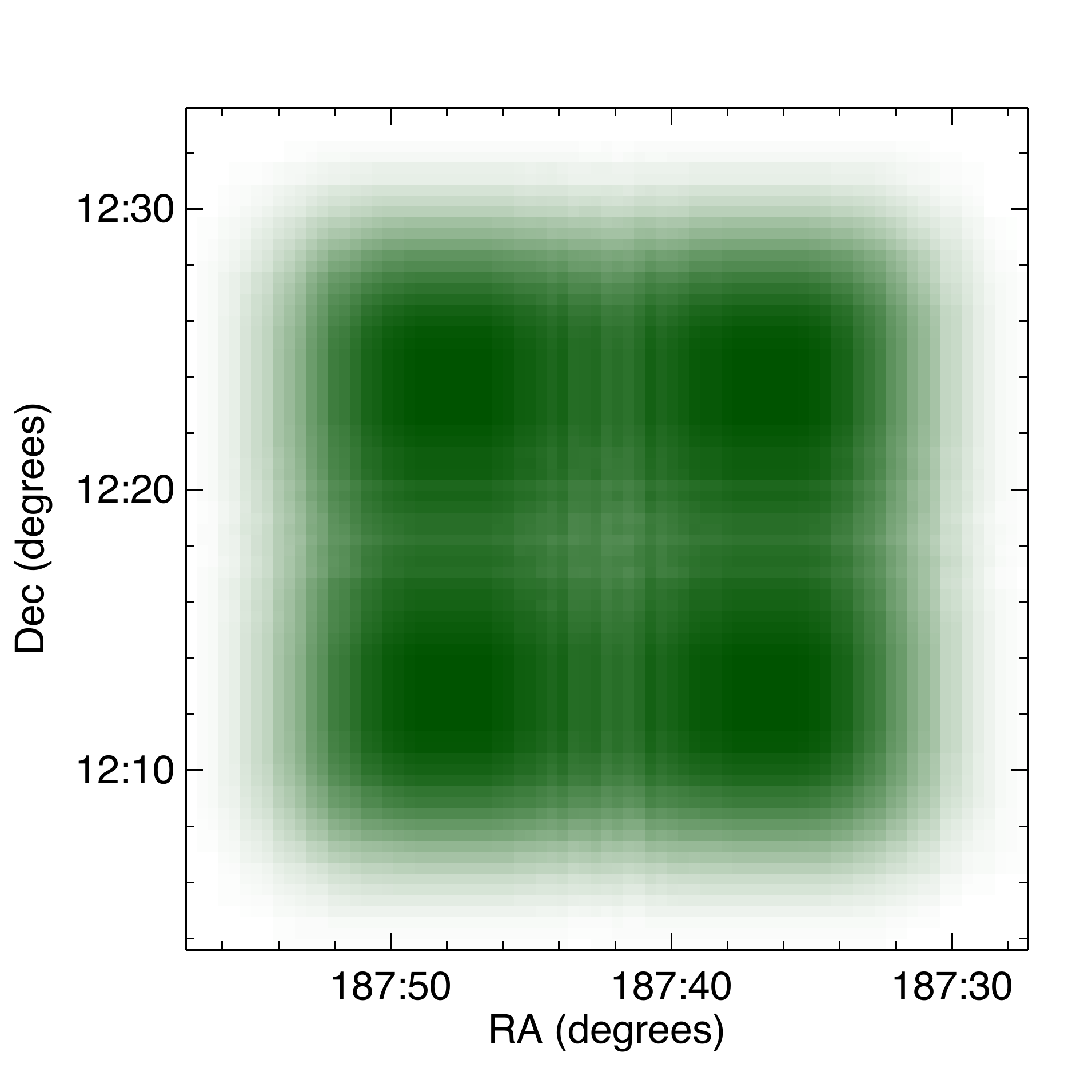}
   \caption{{\it (Left panel)}: Footprint of the {\sc Fermat}-spiral pointing dither pattern. The dimensions of the dotted box are ($\Delta$RA, $\Delta$DEC)$=(4.779\arcmin , 4.614\arcmin)$. The mean (median) distance between individual pointings is 0.842\arcmin\ (0.845\arcmin), while the standard deviation along the RA and DEC axis is $\sigma\!=\!1.07\arcmin$ in both cases. The compactness of the pattern is mean distance $\times\,\sigma\!=\!0.02$\arcmin. {\it (Right panel)}: Sky coverage map of the full F2D27 exposure sequence, including the exposure and pointing dither pattern. Dark green regions have the maximum number of $27\!\times\!4\!=\!108$ overlapping single exposures, while the number of sub-integrations decreases towards the lighter-green areas. The maximum variation in SNR in the critical inter-chip region is only 14\%.}
   \label{fig:fermatspiral}
\end{figure*}

\subsubsection{The Exposure Dither}
\label{txt:exposure_dither}
Each visit to one of the 36 pointings, or tiles, of the survey area is implemented as a sequence of 4 dithered 25~second exposures. This exposure dither is illustrated in Figure~\ref{fig:NGVSpilot} by the WIRCam tile footprint that overlaps NGC\,4486 (M87). The exposure dither pattern has a square geometry but this square is tilted with respect to edges of the detector. This facilitates the process of neutralizing smaller chip defects and covering the 45\arcsec\ inter-chip gap, while keeping each sub-integration safely in the linear regime of the detector. It also keeps individual visits to a given pointing reasonably short compared to usual timescales of sky surface brightness variations ($\sim 10\%$ in 10 minutes), so a given 4-exposure visit to one tile may be used to estimate the sky in tiles visited just before or just after.

\subsubsection{Tile-to-Tile Sequencing}
\label{txt:tile2tile}
After one $4\!\times\!25$ seconds visit is completed we always move the telescope to a new pointing in the 36-tile mosaic of the WIRcam survey. How these large offsets are chosen is based on the relative crowding of the tiles. Figure~\ref{fig:wircam_tiles} shows the mosaic overplotted on a 2MASS archive image, together with a sample of photometrically selected GCs indicated as red dots (Peng et al.~{\it in preparation}). Each tile is assigned a category A, B or C based on the crowding and spatial extent of objects that it covers.~We use two types of sequences: the (A$_1$-C-A$_2$) observing sequence combines the most crowded tiles (C-type) with two non-crowded tiles (A-type), while the shorter (A-B) sequence combines one of the moderately-crowded tiles (B-type) with one non-crowded tile (A-type).~The arrows in Figure~\ref{fig:wircam_tiles} illustrate which crowded (B) and (C) tiles are observed together in sequence with (A) tiles which are relatively devoid of crowded regions and extended objects.~This technique minimizes the required telescope slewing time and will allow accurate modeling of galaxy surface brightness profiles.

\subsubsection{The Pointing Dither}
\label{txt:pointing_dither}
Each (A-B) and (A$_1$-C-A$_2$) sequence is executed 27 times to reach the nominal exposure time of 45~minutes per pointing. We optimize the telescope nodding to simulate an on-tile dithering pattern.  In other words, subsequent repeats of a given (A-B) or (A$_1$-C-A$_2$) are offset.  
In contrast to the exposure dither (see above), we call this pattern the pointing dither. It guarantees that each time all pixels cover a different section of the sky while the extent of the pointing dither allows us to combat the chip gaps and larger chip defects.~The outer regions of the pointing dither pattern are covered by overlap regions between individual tiles of the entire $6\times6$ WIRCam mosaic.

An optimal dither pattern needs to be compact, be scalable, and prevent redundant pointings. Such a pattern is realized in nature by the arrangement of flower leaves (phyllotaxis) and seeds on flower heads, a prominent example, for instance, being the sunflower head. Mathematically these patterns are described by a {\sc Fermat} spiral \citep{vogel79}. This is a special type of the Archimedean spiral and can be analytically expressed as $r\!=\!c\, n^{1/\gamma}$ where $\gamma\!=\!2$ and $n\!=\!\theta/137.508^{\rm o}$, while $n$ is the index of the individual pointing.~The angle $137.508^{\rm o}$ is the {\it Golden Angle}.~Figure~\ref{fig:fermatspiral} shows the arrangement of the 27 pointings we use for moving between repeated (A$_1$-C-A$_2$) or (A-B) sequences.~We tested several Archimedean spirals with different scalings and found that the {\sc Fermat} spiral yields the least pixel-to-pixel variance in terms of x-y coordinate coverage (see the histograms in the left panel of Figure~\ref{fig:fermatspiral}).~In fact, the {\sc Fermat} spiral pattern is the most compact and homogeneous coverage of a 2-D surface without x-y pointing redundancy. We scale it to the dimensions ($\Delta{\rm RA}, \Delta{\rm DEC})=(7.605\arcmin , 7.851\arcmin)$ so that the central WIRCam inter-chip gap introduces the least variance on the final sky coverage map, which is illustrated in the right panel of Figure~\ref{fig:fermatspiral}.

The pointing dither and the exposure dither ($4\!\times\!25$ seconds) work together to homogeneously fill the entire survey area and we refer, in the following, to the full dither sequence as F2D27.~The individual components of F2D27 are illustrated in Figure~\ref{fig:NGVSpilot} where the exposure dither is shown for tile \#15 (center of the field), while the pointing dither is depicted for tile \#1 (lower left corner of the mosaic).~Each tile in the WIRCam mosaic is offset by 19\arcmin\ from its neighbour in $\Delta{\rm RA}$ and $\Delta{\rm DEC}$ to assure enough overlap between the individual mosaic tiles, so that the sampling of the WIRCam inter-chip gap remains the most critical for homogeneous sky modeling and similarly for the sampling of extended surface brightness structures, such as those of large galaxies.~A coverage analysis of this WIRCam inter-chip gap region shows that our F2D27 pattern covers all sky locations more than 64 out of $4\!\times\!27\!=\!108$ times, while the vast majority of the inter-chip gap is covered at least 80 times. This translates into a SNR fluctuation of only 14\% and is fully propagated by our pipeline to the variance maps.

In summary, the observing strategy includes the following exposure time outline for the two types of sequences:
\begin{eqnarray*}
\centering
({\rm A_1\!-\!C\!-\!A_2}) & \rightarrow & 3\cdot 27\times(4\times25) \sec  = 8100 \sec \\
({\rm A-B})                     & \rightarrow & 2\cdot 27\times(4\times25) \sec  = 5400 \sec
\end{eqnarray*}
and contains a total of six (A$_1$-C-A$_2$) sequences and nine (A-B) sequences that cover the entire field. Due to observing scheduling constraints, sequence (A6-B5) could not be observed and thus will not be discussed in this work.~On the other hand, some of the observed sequences benefited from more than the requested 27 visits.

%%%%%%%%%%%%%%%%%%%%%%%%%%%%%%%%%%%%%%%%%%%%%%%%%%%%%%%%%
\section{Data Reduction}

The data reduction process was divided into three major parts: the pre-processing stage, the main image processing, and the image post-processing. 

\subsection{Pre-Processing}

The processing of the raw data was done using the \`{}$\Gamma$iwi pipeline v2.0\footnote{http://www.cfht.hawaii.edu/Instruments/Imaging/WIRCam/\\IiwiVersion2Doc.html}.~The pipeline involves the following steps: flagging saturated pixels, non-linearity correction, dark subtraction, flat fielding and bad-pixel masking. 

The saturated pixels were identified in the raw images using a threshold value of 36000 ADU. A record of these pixels was kept and they were flagged with the value 65535 in the final pre-processed image. The non-linearity of the WIRCAM detectors is about 5\% at 30000 ADU and it was corrected at the beginning of the pre-processing. Dark frames were obtained at the beginning and the end of each observing night, and a master dark frame was computed by taking the median of 15 dark frames. The dark subtraction has a negligible impact on the pre-processed images, since the dark current of WIRCam detectors is significantly less than $1\,e^-$/sec. 

Twilight flat fielding was applied by the \`{}$\Gamma$iwi pipeline: Twilight flats are obtained at the beginning of each observing night and a daily twilight flat is built by taking the median of 15 consecutive frames.~The pipeline divides the semester into several sets of observing nights and then computes a master twilight flat on each set.~All science images were corrected by dividing by the corresponding normalized master flat.

The master bad pixel masks were built by analyzing the normalized master flats previously computed.~Bad pixels were identified by applying a sigma threshold algorithm. The bad pixel masks were constructed to have a value of 1 for good pixels and 0 for bad pixels.

We refer to the pre-processed images in the following as science images.

\subsection{Main Image Processing}
\label{txt:processing}

The processing consisted of masking cosmic-rays and satellite-trails in the science images, removing the residuals left by saturated stars, subtracting the sky from the images, computing the astrometric solution, applying the photometric calibration and producing the final stacked images.

\subsubsection{Cosmic-Ray and Satellite-Trail Removal}
The cosmic rays were identified by running the Laplacian cosmic ray identification algorithm \citep[\textsc{LACosmic};][]{dokkum01} on the science images. This algorithm is based on a variation of the Laplacian edge detection method and cosmic rays are identified by the sharpness of their edges. The code convolves the image with a laplacian kernel, then identifies cosmic rays as sharp borders in the image and creates a cosmic-ray mask for each science image. The best \textsc{LACosmic} parameters were obtained by performing a visual inspection of the masks and verifying that the peaks of non-saturated bright stars were not miss-identified as cosmic rays.

The detection of satellite tracks is not possible in the pre-processed images, since the sky brightness dominates the image while a typical satellite track is hundred times fainter than the sky. We run a quick sky removal on each image (target) that consisted of identifying the 10 closest images in observing time (sky), then taking the median of the sky images pixel by pixel and subtracting it from the target image.~The sky subtracted images were smoothed with a boxcar average of width 20 pixels and the coordinates of pixels with signal-to-noise ratio greater than 1.2 were registered.~Satellite tracks were identified as straight lines in the XY detector plane of the registered coordinates.

\subsubsection{Saturation Correction}
Bright and saturated stars leave residuals in the science images that last for about six subsequent images. First, the saturated bright stars were identified as those regions of at least four connected pixels with counts higher than 65535 and not masked as bad pixels in the master bad pixel mask. For chips 1 and 3, it was necessary to mask the subsequent three images, while for chips 2 and 4 it was necessary to mask the subsequent six images due to the different chip and detector characteristics. The science images were sorted according to their observing time, the pixels belonging to a saturated star were identified and the six (or three) consecutive images after saturation were masked at the corresponding coordinates.

\subsubsection{Sky Subtraction}
\label{txt:sky_subtractiontech}
One motivation for the (A$_1$-C-A$_2$) or (A-B) observational strategy presented in Section~\ref{txt:dithering} was to design a high-quality sky subtraction method. The observations in each (A$_1$-C-A$_2$) or (A-B) sequence were declared as target or sky tiles, and for each target image there was a set of sky images defined for computing and subtracting the sky in the corresponding target image. The sky images from A-type tiles are generally less crowded regions than target images, but do still contain a significant number of point-like and extended sources that need to be masked before computing the final sky image. The sky subtraction method is a two-step iteration process: the first iteration consists of computing a median sky image for each target image, then subtracting it from the target image and finally building a stacked target image from all such treated frames in a sequence; the second iteration consists of identifying the sources in the stacked target image, masking these sources in the each individual sky image, computing a new median sky and finally subtracting it from the target image.

The selection of the set of sky images to model and subtract the sky contribution in each target image depends on the amplitude and variance of the sky surface brightness and is based on the following three criteria: 
\begin{description}
\item[] $a)$ the category of the (A$_1$-C-A$_2$) or (A-B) observing sequence that the target image is part of,
\item[] $b)$ the time difference between sky images and the target image time stamp,
\item[] $c)$ the need for a sufficient number of sky images to compute a reliable median sky image. 
\end{description}

The first criterion is related to the sky-target dithering sequence explained in Section~\ref{txt:tile2tile}, and consists of the following set of rules: 1) if the science image belongs to type-A tile, then sky images are selected from type-A tiles of the same sequence; 2) if the science image belongs to a type-B tile, then sky images can be selected from type-A and type-B tiles; 3) if the science image belongs to a type-C tile, then sky images can be selected only from type-A tiles. 

The second criterion consists of choosing sky images close enough in time to the target image. We used only those sky images taken within equal or less than 15 minutes before and after the corresponding target image observation. In particular, at the beginning and the end of each observing run there were sets of sky images containing just 8 images, and as it will be explained by the third criterion, that number of images is below the recommend number of frames to model the sky.

The third criterion consists of defining the sufficient number of sky images that is necessary to compute a reliable median sky image that robustly captures the sky variations. We ran several quality tests on the sky images and the conclusion was that the minimum number of frames for this survey is ten sky images given the variability of the near-IR sky throughout our observing campaign.~This criterion was applied as follows: we sorted the sky images according to their time stamp with respect to the target image, then we computed the maximum between 10 and the number of sky frames that fulfilled the first and second criterion, and the final set of sky images was defined by applying that maximum number.

Once the set of sky images is defined for each target image, we compute the pixel-to-pixel median sky image and subtract it from the corresponding target image. 

\subsection{Post-Processing}
\label{txt:postprocessing}
We applied post-processing routines to the images because some of the processed images were still showing visible systematics. These post-processing steps consist of removing the slope from each amplifier detector and subtracting large-scale variations of the sky background.

\subsubsection{Residual Amplifier Differences}
Each WIRCam detector has 32 amplifiers with different gains that result in background level offsets that differ by $\sim\!10\%$ in the twilight flat-field images. Most of the signal of the amplifiers disappears after flat-field correction and removing the sky during the pre-processing pass, but there is still a non-negligible contribution that appears as horizontal bands through the image. To model and subtract this residual, we start with sky-subtracted images obtained from the processing pass and select the pixels belonging to each amplifier, taking the median along the x-axis of the detector. Then we fit the variation with a linear relation as a function of the y-axis coordinate and finally subtract this model from each horizontal band in each of the sky-subtracted images.

\begin{table}
\begin{center}
\caption{Summary of the average zero points of photometric nights.}
\begin{tabular}{@{}lrrr@{\,$\pm$\,}l@{}}
\hline\hline
Program & MJD & $\rm{N}_{\rm phot}^a$ &  \multicolumn{2}{c}{$\overline{zp}_K$} \\
    \hline
    09BC26 & 55187 & 71 & 23.16 & 0.01 \\
    	  & 55190 & 138 & 23.14 & 0.02 \\
	  & 55191 & 71 & 23.14 & 0.02 \\
	  & 55192 & 55 & 23.12 & 0.01 \\
	  & 55194 & 123 & 23.13 & 0.02 \\
	  & 55195 & 159 & 23.14 & 0.01 \\
    \hline
    09BF22 & 55191 & 71 & 23.12 & 0.01 \\
    \hline
    10AC10 & 55280 & 128 & 23.14 & 0.02 \\
     & 55283 & 44 & 23.11 & 0.02 \\
     & 55287 & 129 & 23.14 & 0.01 \\
     & 55288 & 60 & 23.10 & 0.02 \\
     & 55290 & 3 & 23.13 & 0.01 \\
     & 55291 & 33 & 23.13 & 0.02 \\
    	 & 55308 & 59 & 23.12 & 0.02 \\
	 & 55310 & 256 & 23.15 & 0.03 \\
	 & 55311 &  78 & 23.15 & 0.02 \\
	 & 55342 &  84 & 23.14 & 0.02 \\
	 & 55343 & 111 & 23.15 & 0.01 \\
	 & 55374 &  56 & 23.11 & 0.03 \\
	 & 55379 &  37 & 23.13 & 0.03 \\
	 & 55380 &  70 & 23.11 & 0.02 \\
    \hline
    10AF03 & 55287 &   4 & 23.15 & 0.01 \\
      & 55288 & 144 & 23.11 &  0.01 \\
      & 55290 &  38 & 23.14 &  0.02 \\
      & 55291 & 132 & 23.11 &  0.02 \\
      & 55316 & 143 & 23.12 &  0.02 \\
      & 55317 &  32 & 23.13 &  0.02 \\
      & 55320 &  36 & 23.12 &  0.02 \\
      & 55322 & 114 & 23.11 &  0.02 \\
      & 55341 &  54 & 23.15 &  0.01 \\
      & 55343 &  46 & 23.15 &  0.02 \\
      & 55382 &  57 & 23.12 &  0.02 \\
      & 55383 &  41 & 23.14 &  0.02 \\
    \hline
\end{tabular}
\label{tab:zp}
\tablecomments{$^{\rm a}$Number of photometric frames}
\end{center}
\end{table}

\subsubsection{Large-Scale Variations}
After close examination of the gain corrected and sky subtracted images, we found large-scale variations of the image background still present after the sky removal, with an amplitude of $\sim$0.3\% of the typical sky. These feature are likely to be produced by a variation in the thermal stability of the detector and/or internal reflections within the instrument due to the presence of large galaxies. They have a non-negligible effect on the stacked the images. 

In order to improve the background flatness, we  mask the sources and model the large-scale background structure in each frame. For this purpose, we first use {\sc Swarp} \citep{bertin02} to build a stacked image using all the sky-subtracted frames computed in the processing stage. We run {\sc SExtractor} on the stacked images \citep[v2.5.0,][]{bertin96} and obtain the segmentation maps with pixels associated with the sources.~To remove the outer low-surface brightness parts of galaxies, we increase the sizes of the masked regions by 50 to 800 pixels, depending on the size of the galaxy, and project the source mask back into the individual pre-stack sky-subtracted images.~The large-scale variations are computed on each of the four WIRCam detectors independently. Each detector is divided into a grid of $16\!\times\!16$ cells. Those cells in which more than 50\% of the pixels are masked are discarded. In all other cells a median background value is computed. A large-scale variation image is then computed using the {\sc Kriging} linear interpolation algorithm as implemented in {\sc IDL} (v8).~This background model is then subtracted to produce the final science-grade images.~This method has some limitations when dealing with galaxies with  extremely extended surface brightness profiles.~However, these limitations have no impact on the analyses and results presented in this paper.

\subsection{Astrometric Calibration}
We calculate the relative astrometric solution with the {\sc Scamp} software package \citep[v1.7,][]{bertin06}, which reads the catalogs produced by {\sc SExtractor} and cross-matches the source positions against an astrometric reference catalog, such as 2MASS \citep{skrutskie06}.~We run {\sc Scamp} using a maximum search range of $\sim\!0.15$\arcmin\ ({\sc position\_maxerr}=0.15) and a minimum signal-to-noise ratio of 40 ({\sc sn\_thresholds}\,=\,40, 80) to match reference stars.

We compute the distortion maps for each of the WIRCam detector chips and find that the pixel scale varies in a radially symmetric way by at most 0.5\% between the center of the $2\!\times\!2$ detector mosaic and the outer radius of the WIRCam FOV. The astrometric calibration against the 2MASS reference frame is based on $\sim\!10^3$\,stars and the resulting astrometric world coordinate solution accuracy has an approximately gaussian distribution with a FWHM smaller than 0.02\arcsec.

\subsection{Photometric Calibration}
\begin{figure}
\centering
   \includegraphics[width=8.9cm, bb=80 20 750 530]{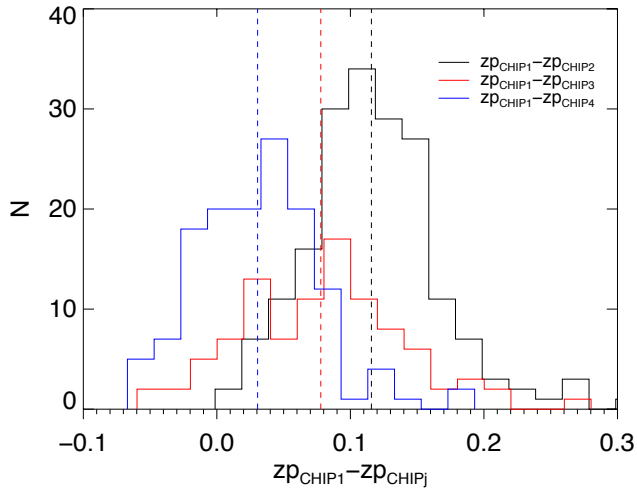} 
\caption{Distribution of the difference between the zero-points of detector 1 and detectors 2, 3 and 4.~CHIPS 2 and 4 show a peaked distribution and CHIP 3 a broader distribution. The average zero-point offsets of CHIP-$j$ with respect to CHIP 1 are  0.12, 0.08 and  0.03 for CHIPS 2, 3 and 4, respectively.}
\label{fig:zp_offset}
\end{figure} 

We use the 2MASS point source catalog \citep[2MASS-PSC;][]{cutri03} to perform the preliminary photometric calibration of the sky subtracted images.~We select 2MASS point-like sources with photometric quality flags equal to A and B (i.e.~reliable photometry) that correspond to sources with S/N ratio greater than 7 and magnitude error less than 0.16 mag.

The photometric calibration was done using {\sc Scamp} as for the astrometric calibration.~{\sc Scamp} distinguishes between two types of exposures: those observed under photometric conditions and the others. The information about the photometric nights was obtained from the QSO weather reports and the images observed during those nights were visually inspected to define a clean sample of exposures taken in photometric conditions.~The zero-point of the photometric calibration was computed using
\begin{eqnarray}
m_{\rm std} & = & m_{\rm inst} + zp - k X \label{eqn:mstd},
\end{eqnarray}
where $m_{\rm std}$ is the 2MASS magnitude, $m_{\rm inst} $ is the instrumental magnitude measured on the sky-subtracted images, $zp$ is the zero-point of the photometric calibration, $k$ is the airmass term, and $X$ is the airmass of the observation.

The four WIRCam detectors have different gains and in principle a $zp$ should be computed per exposure and detector, but the number of reliable 2MASS stars per detector can be as low as ten.~We opt for computing the $zp$ offsets between the WIRCam detectors using groups of exposures, and then keep these offsets fixed to assess $zp$ variations between exposures in the group.~The $zp$ offsets are computed by grouping the exposures every 15 minutes, then crossmatching the detected sources with reliable 2MASS point-like sources and computing the median value of $(m_\mathrm{std}\!-\!m_{\rm inst})$ for all the stars on each detector. Figure~\ref{fig:zp_offset} shows the distribution of the difference between the $zp$ values of chip 1 and and those of chips 2, 3 and 4 ($zp$ offset), where the distribution for chip 2 is found to be the most concentrated. The $zp$ offsets for detectors 2, 3 and 4 are 0.12, 0.08 and 0.03 mag, respectively.

For computing the airmass term ($k$) in Equation~\ref{eqn:mstd}, we first compute the mean value of $(m_\mathrm{std}\!-\!m_{\rm inst})$ on each exposure and then apply a linear regression analysis using only photometric nights.~The estimated value of $k$ for this survey is $k\!=\!0.041$.~We use Equation~\ref{eqn:mstd} to estimate the $zp$ for exposures observed in photometric nights and add the value as {\sc phot\_zp} to the header.

For exposures observed in non-photometric nights, {\sc Scamp} runs an internal crossmatch of point-like sources and then adjusts the $zp$ for those exposures by minimizing Equation~\ref{eqn:zp_scamp}. We adopt the same parameters as in the previous section, but this time we use a S/N ratio of 20 or higher in order to have a larger sample of stars for doing the photometric calibration ({\sc sn\_thresholds}=20). We compute
\begin{equation}
\chi^2 = \sum_s \sum_a \sum_ {b>a} w_{ab} (zp_a + m_{{\rm inst},a} - zp_b + m_{{\rm inst},b} )^2 \label{eqn:zp_scamp}
\end{equation}
where $a$ and $b$ refer to exposures that contain the point-like source $s$ in the overlapping region, and $w_{ab}$ is the non-zero weight for the pair of detections in exposures $a$ and $b$, which are computed via $w_{ab}=1/\sigma^2_{\rm phot}$ where $\sigma^2_{\rm phot}$ is the square sum of all contributing photometric variances.

\begin{figure}[!t]
\centering
   \includegraphics[width=8.4cm, bb=20 18 382 392]{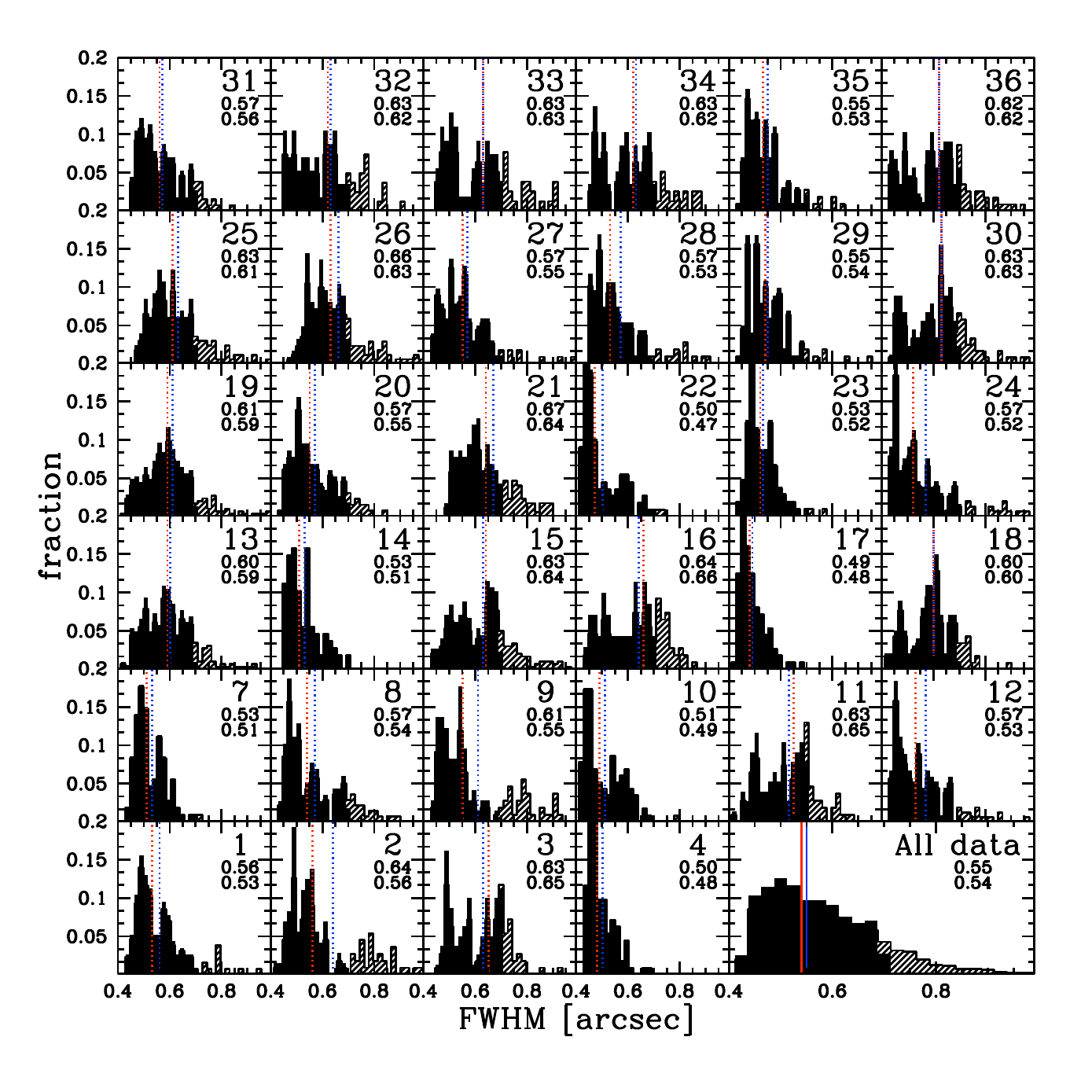} 
\caption{The seeing quality distributions of all images ({\it hatched histograms}) for each NGVS-IR pilot field tile that contributed to the final mosaic ({\it solid histograms}). The tile number is given in the upper right corner of each panel, as well as the mean (upper value) and median value of each distribution (lower value), considering all frames available for each tile. The mean and median values are marked by blue and red vertical dotted lines, respectively. Note that only images with a seeing better than 0.7\arcsec\ were used in this work, marked by the solid histograms. In the bottom right corner of the panel (we recall that no data were obtained for tiles 5 and 6), we illustrate the seeing distribution of all images of the NGVS-IR pilot field, and provide the corresponding mean and median values of only the images with a seeing better than 0.7\arcsec\, which are rendered as solid vertical lines.}
\label{fig:seeing}
\end{figure}

The instrumental magnitude, $m_{\rm inst}$, in Equation~\ref{eqn:mstd} is measured on the sky-subtracted images with a circular aperture photometry of radius $4\times \overline{\rm FWHM}$, where $\overline{\rm FWHM}$ is the mean value of the full-width-at-half-maximum of the bright stars in each exposure.~Figure~\ref{fig:seeing} shows the distribution of the FWHM of point-like sources of all the images of the survey.~The FWHM distribution is not homogeneous and covers the range between 0.4\arcsec\ and 1.0\arcsec.~Thus, the photometry aperture is large enough to include more than 95\% of the total light.~The photometric zero points adopted in this work are shown in Table \ref{tab:zp}.~The $\overline{zp}_K$ refers to the mean zero point of science frames observed in the respective night and the quoted uncertainty to its standard deviation.

\subsection{Stacking and Data Quality Assessment}
\label{txt:stacks}
The sky-subtracted images were stacked using the {\sc Swarp} software \citep[v2.19,][]{bertin02} after the astrometric and photometric calibrations were completed.~{\sc Swarp} uses the astrometric solution calculated by {\sc Scamp} and combines several frames to produce a final image of a given size and pixel scale. We use the $i$-band quadrants of the NGVS pilot field area as a reference for building the same size field-of-view $K_s$-band stacked images, since it helps the subsequent analysis. We note that the pixel scale of WIRCam images is 0.3\arcsec, while the pixel size of our final stacked images is 0.186\arcsec\ and that we build the stacked images using only images with FWHM smaller than 0.7\arcsec\ (see Figure~\ref{fig:seeing}).~The median seeing of all selected data is 0.54\arcsec.~The NGVS-IR pilot field covers a total area of 3.98 degrees$^2$, and consists of 34 slightly overlapping tiles with coverage of about 26.5\arcmin$\times$26.5\arcmin.~For data handling purposes, we sub-divide the pilot field area into four quadrants roughly corresponding to about 1 degrees$^2$ each.

The large majority of the sky subtracted images in the survey are super-critically sampled (above the Nyquist limit).~We use the {\sc Lanczos-2} resampling algorithm which offers a good compromise between improving the image resolution and minimizing the number of artifacts introduced by the resampling routine.~The images are combined using a sigma clipping algorithm that computes the dispersion on a pixel-by-pixel basis and rejects pixels beyond $\pm3\sigma$ of the mean value.

\begin{figure*}[!t]
\centering
\includegraphics[width=18cm]{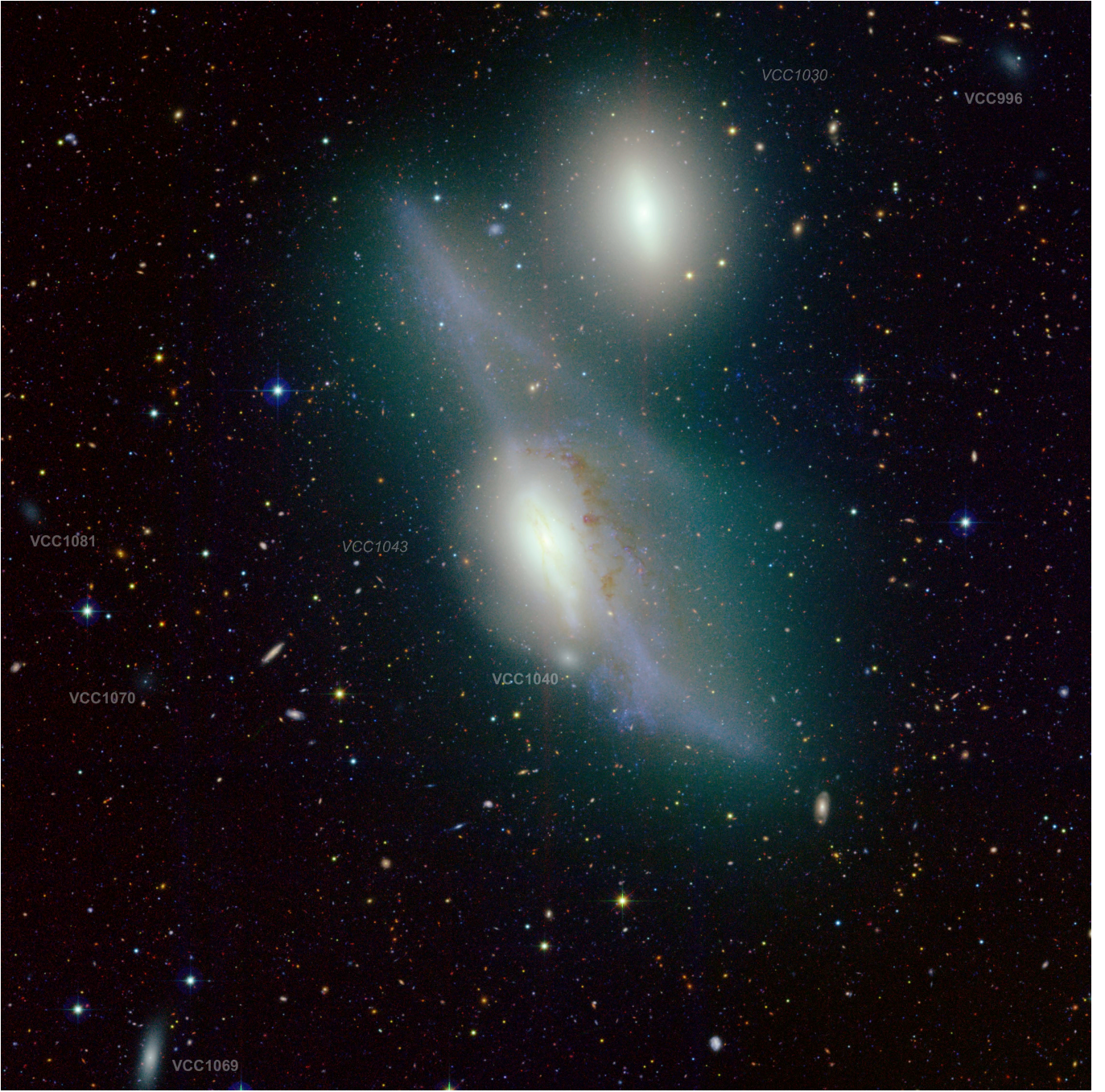}
\caption[vcc1043 uiK color composite]{Sample color composite from the combined NGVS+NGVS-IR dataset of a $11\arcmin\!\times\!11\arcmin$ ($53\!\times\!53$ kpc$^2$) sample field of view around the merger remnant NGC\,4438 (VCC\,1043, center of the image) and its bright companion NGC\,4435 (VCC\,1030, see also Figure~\ref{fig:NGVSpilot}). The image is an RGB  composite using the $u^*$ (blue), $i$ (green), and $K_s$-band (red) data and a logarithmic luminosity stretch. Besides the impressively detailed substructure in the merger remnant, numerous low-surface brightness dwarf galaxies are visible.~We labeled five VCC dwarf galaxies, which are among the faintest members of the VCC catalog and are all clearly detected in our $K_s$-band images. Note that the vast majority of sources with small angular sizes are red background galaxies. This extremely deep near-UV/optical/near-IR photometry is well suited for finding galaxy clusters in the redshift range $0.5\!\leq\!z\!\leq\!1$ that are in the process of formation.}
\label{fig:vcc1043}
\end{figure*}

The NGVS-IR image stacks, which this paper is based on, reach $K_s$-band surface brightnesses fainter than $\mu_{K_s}\!\approx\!24$ AB mag/arcsec$^{-2}$.~In some of our NGVS-IR stacks we can make out dwarf galaxies with $\mu_K\!\approx\!24.4$ AB mag/arcsec$^{-2}$.~The extended emission of 94\,\% of the known VCC galaxies can be visually seen in the final images.~However, we emphasize that the data products discussed here were not obtained with the purpose of optimizing surface brightness sensitivity and we, therefore, postpone any study of luminosity profiles to future work.~Specific data reduction procedures will also be required for the study of near-IR surface brightness fluctuations in Virgo galaxies.~In Figure~\ref{fig:vcc1043} we illustrate the data quality with a color stack of an $11\arcmin\!\times\!11\arcmin$ cutout around the interacting galaxy pair NGC\,4435+4438, which includes five of the faintest VCC dwarf galaxies and even fainter low-surface brightness dwarfs all of which are detected on the $K_s$ stacks.

Before performing detailed completeness test, we note that visual inspection of the images shows that our NGVS-IR WIRCam survey achieves a $K_s$ point-source sensitivity of $\sim\!23$\,mag, thus reaching about 2.9\,mag deeper in this passband than the UKIDSS Large Area Survey, the most extensive and deepest recent survey covering the NGVS pilot field \citep{lawrence07, lawrence12}. NGVS-IR compares in surface brightness sensitivity with the pointed $H$-band observations of the SHiVir survey\footnote{http://www.astro.queensu.ca/virgo/} of bright Virgo galaxies of \citep{mcdonald11}.~These, however, are not contiguous images and were taken with several instruments in varying observing conditions.

\begin{figure*}[!t]
\centering
\includegraphics[width=8.9cm]{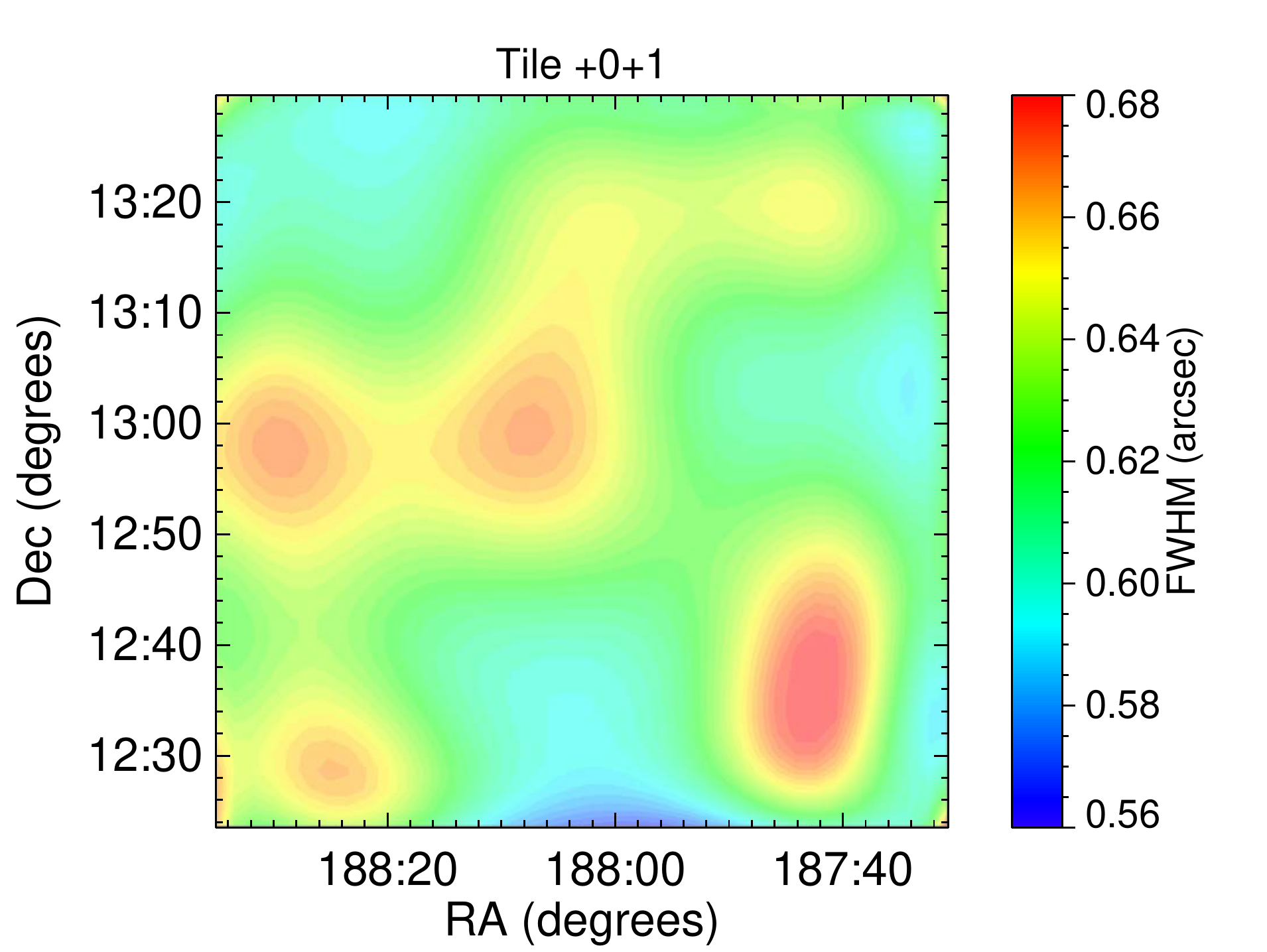}
\includegraphics[width=8.9cm]{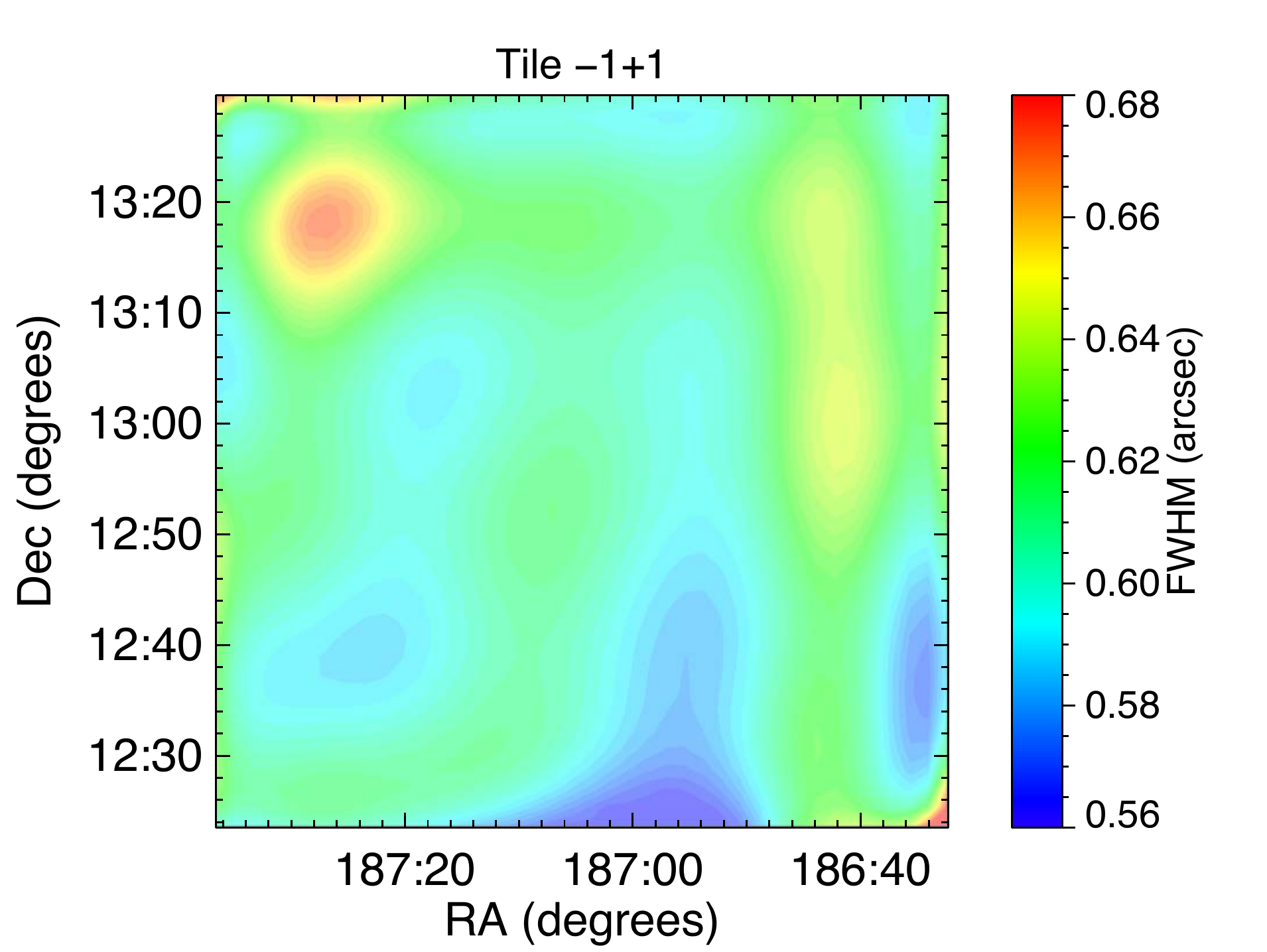}
\includegraphics[width=8.9cm]{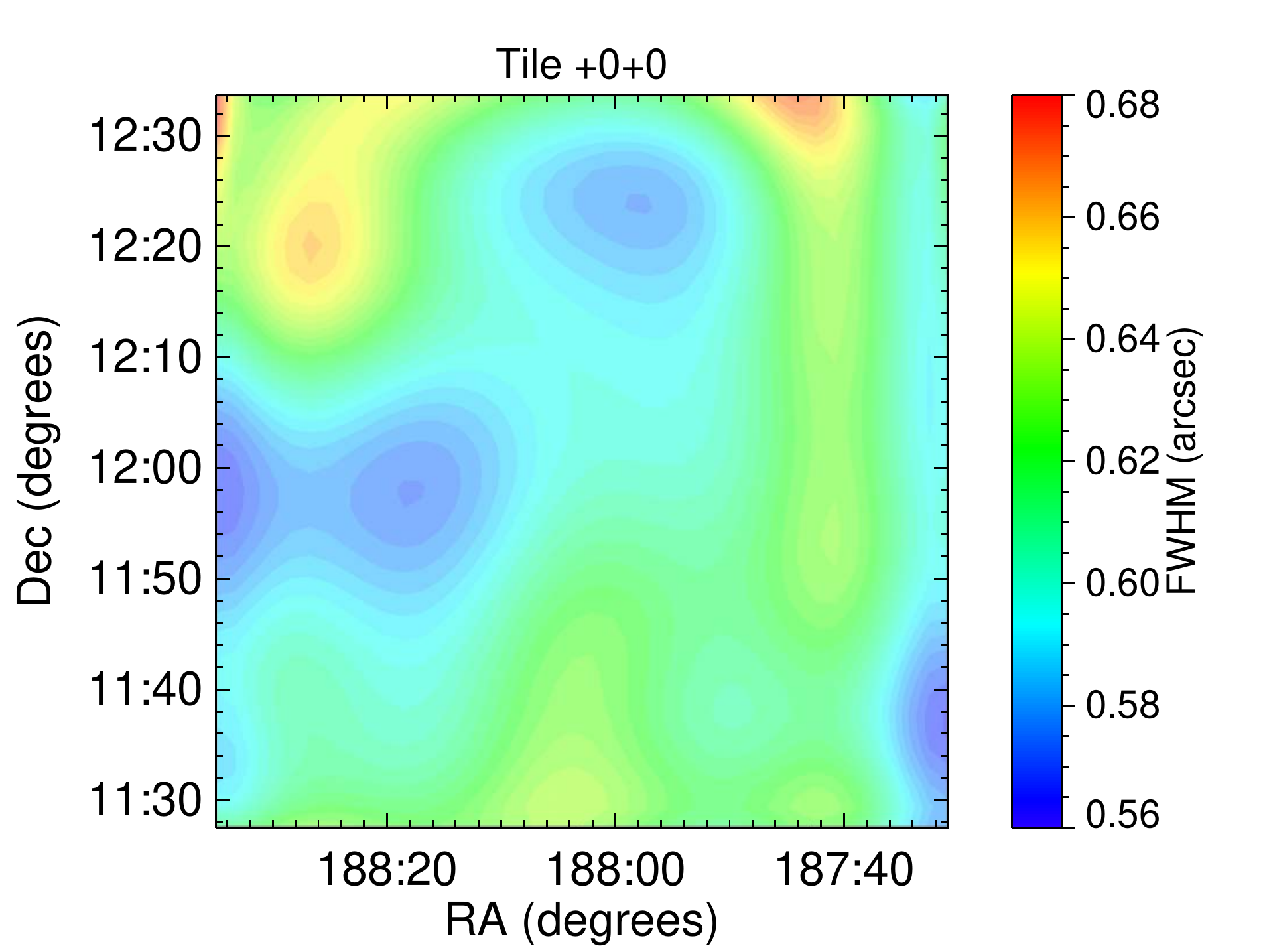}
\includegraphics[width=8.9cm]{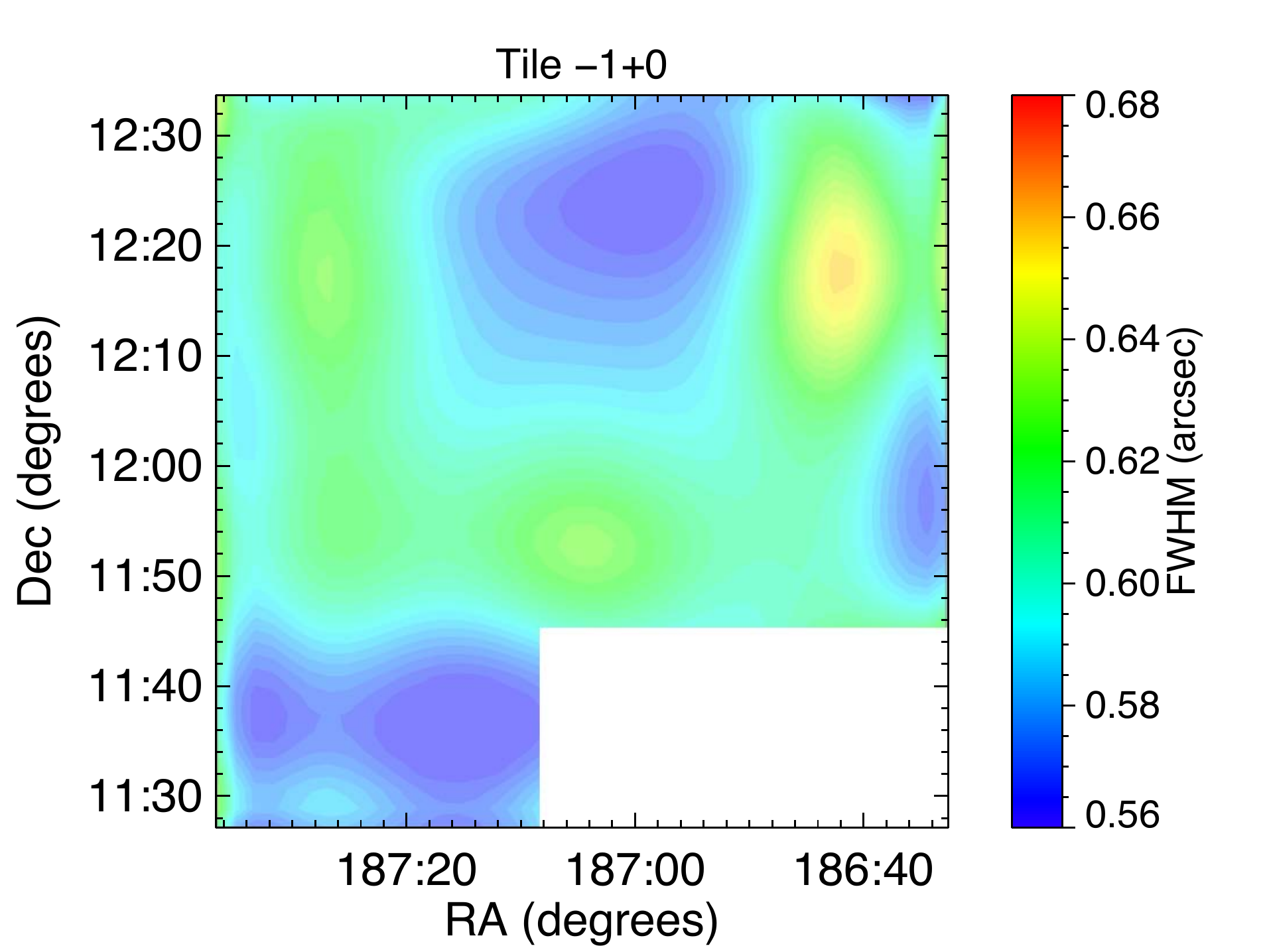}
\caption{Image quality maps for the NGVS-IR pilot field split into the four quadrants corresponding to the NGVS MegaCam tile $(0,0)$, $(+1,0)$, $(+1,+1)$, and $(0,+1)$ \citep[see][for details]{ferrarese12}. The color bar parameterizes the full-width at half maximum (FWHM) of point-like sources in units of arc seconds over the survey area, the maximum variation of which is $\sim\!20\%$ which is mainly due to the seeing distributions of observations in each tile of the NGVS-IR mosaic as illustrated in Figure~\ref{fig:seeing}.}
\label{fig:psf_fwhm}
\end{figure*} 

\vskip 1cm

%%%%%%%%%%%%%%%%%%%%%%%%%%%%%%%%%%%%%%%%%%%%%%%%%%%%%%%%%
\section{Photometry}
\label{txt:photometry}

Source detection in the $K_s$ stacks and photometric measurements were performed with {\sc SExtractor} \citep[v2.5.0][]{bertin96} and {\sc PSFex} \citep{bertin11}.~After a first {\sc SExtractor} run for source detection, {\sc PSFex} produces local models of the point spread function (PSF), that can then be used in a second run of {\sc SExtractor} to obtain integrated PSF magnitudes of stars.~As input for {\sc PSFex} we use a clean sample of stellar sources, obtained by combining a compactness and a color criterion (see Sect.\,\ref{sec:uiK_plane}), and by rejecting any star that saturated the detector in one or more individual images that entered the stack. The PSF was modeled separately across the four individual quadrants of the pilot field (see Figure~\ref{fig:psf_fwhm}).~Each of these quadrants consists of $3\!\times\!3$ WIRCam pointings, and PSF variations are dominated by pointing-to-pointing variations in the corresponding FWHM distribution functions (see Figure~\ref{fig:seeing}).~The spatial variations of the PSF are modeled with polynomials of degree seven to allow for changes on the typical scale resulting from this pattern.

\begin{figure*}[!t]
\centering
\includegraphics[width=8.9cm]{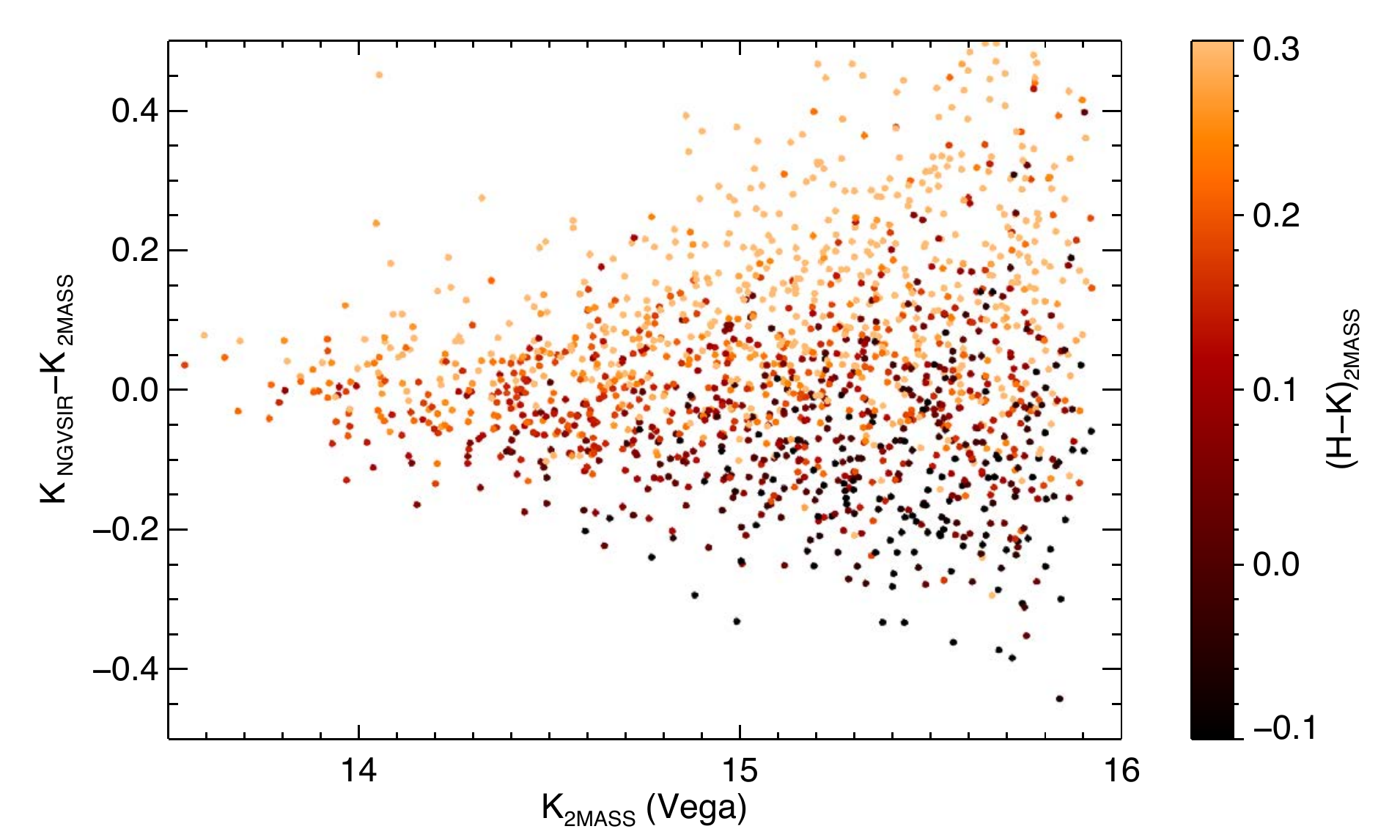}
\includegraphics[width=8.9cm]{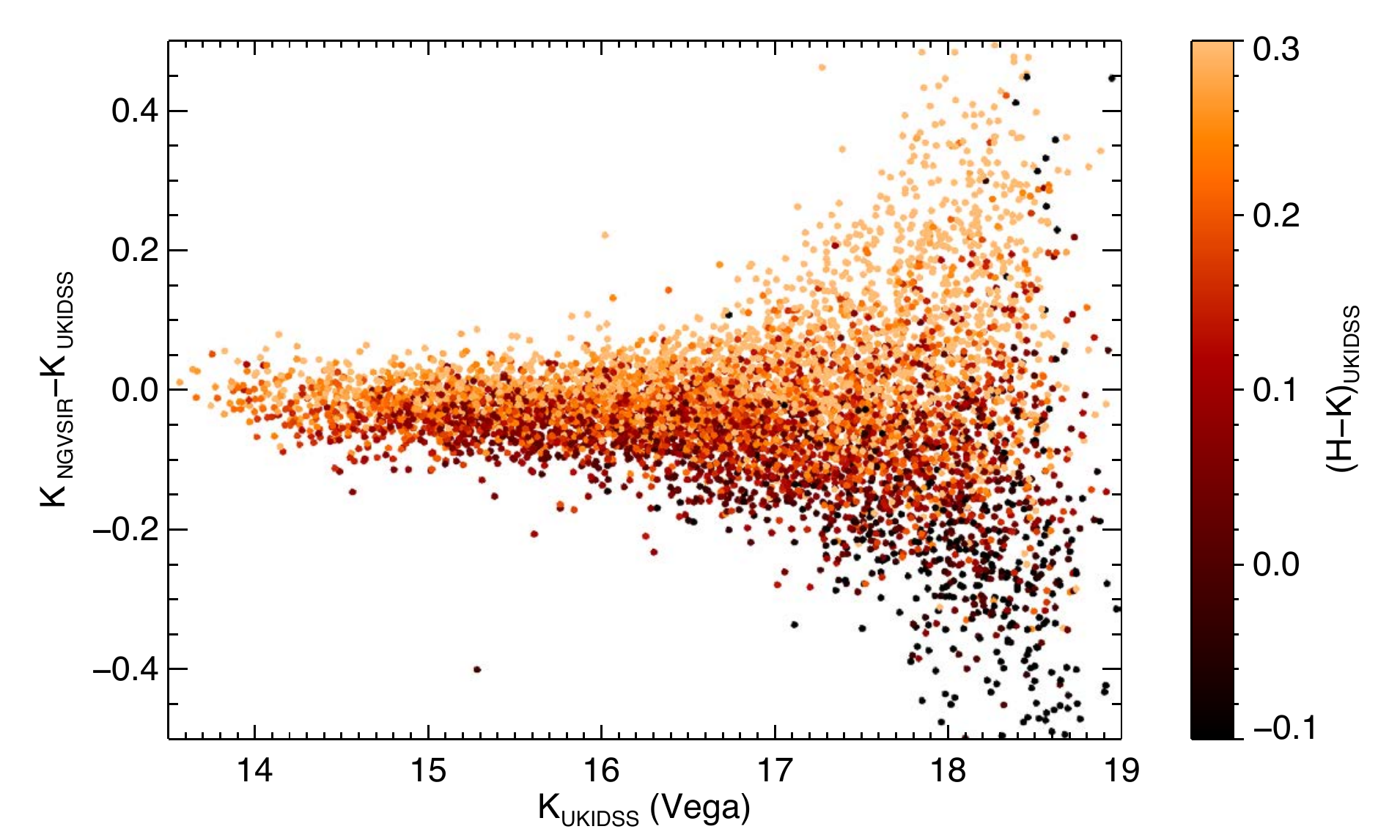}
\caption{The photometric calibration accuracy of the NGVS-IR pilot field data vs.~2MASS ({\it left}) and UKIDSS data ({\it right}). Objects in this plot are point sources according to their $K_s$-band morphology and NGVS/NGVS-IR colors and match coordinates in UKIDSS or 2MASS data to better than 1\arcsec. The right-hand side color bar encodes the $H\!-\!K$ color measured in the corresponding comparison system and illustrates significant residual color terms in the photometric calibration. To avoid conversion of UKIDSS and 2MASS magnitudes into the AB system, the plots show point source magnitudes in the Vega magnitude system. Note the significantly different photometric depth (i.e.~$\sim\!3$ mag) between the 2MASS and UKIDSS data. The data distributions in both panels are entirely dominated by the photometric uncertainties of the 2MASS and UKIDSS data.}
\label{fig:ngvsir_photcal}
\end{figure*}

\begin{figure*}[!t]
\centering
\includegraphics[width=8.9cm]{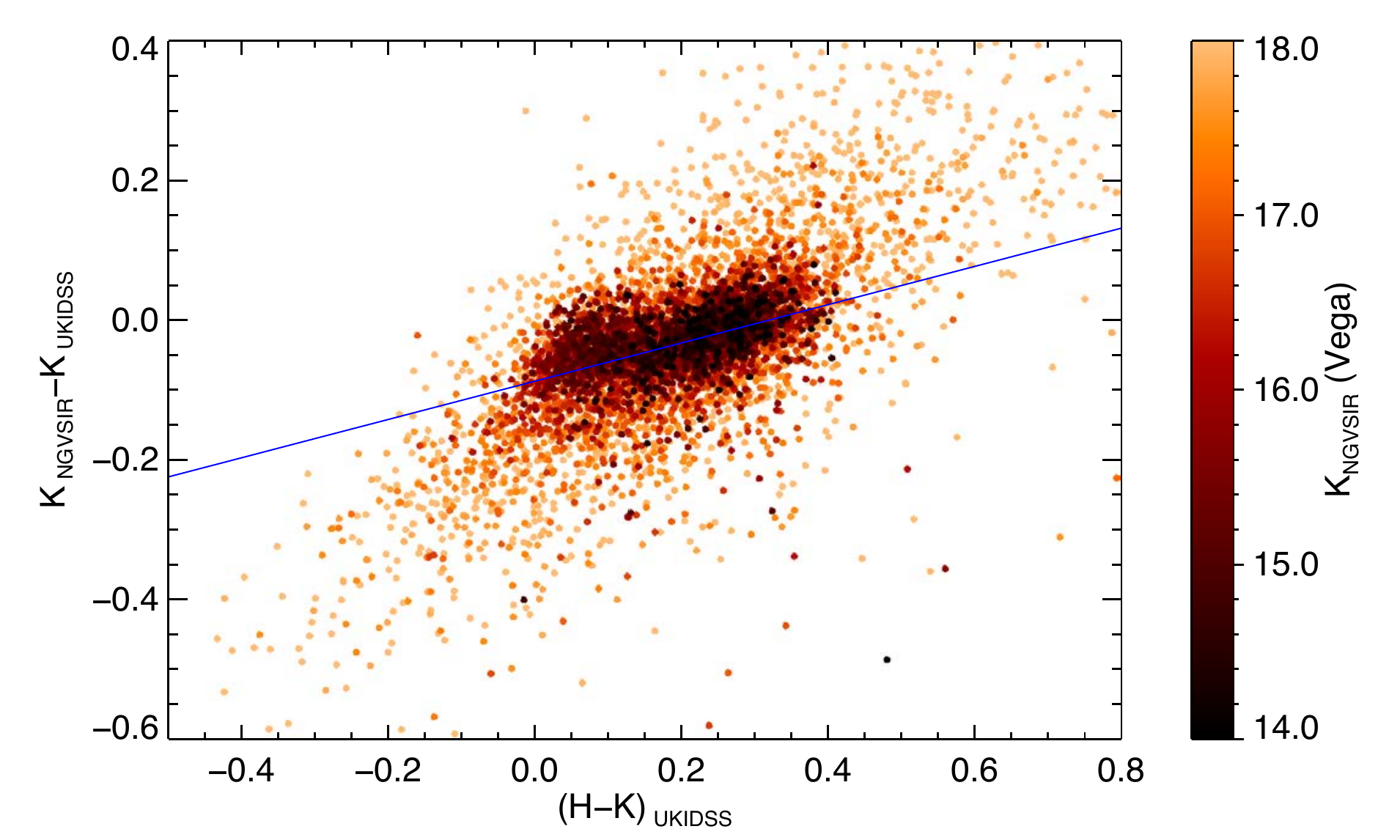}
\includegraphics[width=8.9cm]{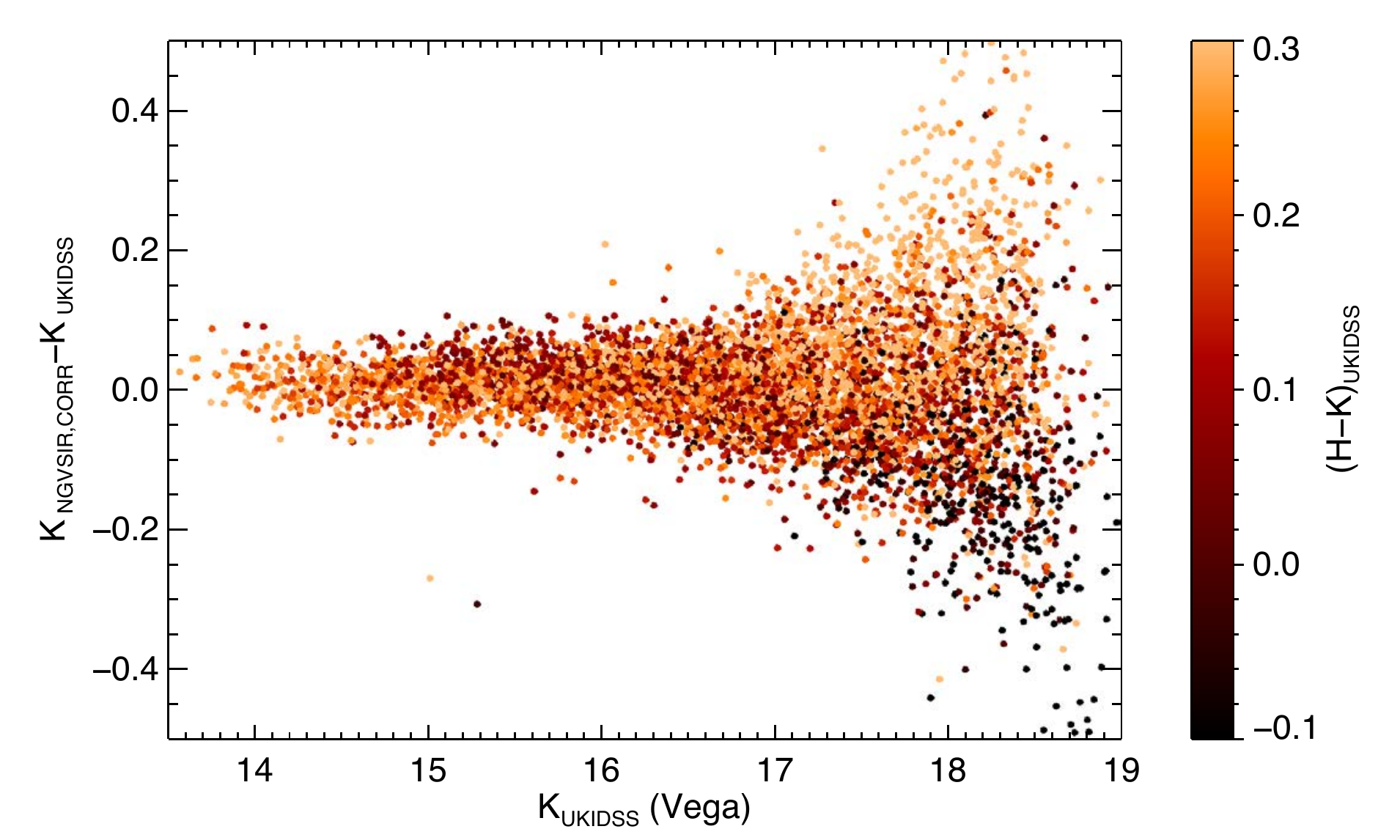}
\caption[]{({\it Left panel}): Color term resulting from the differences between the UKIDSS and WIRCam $K_s$ filters. The shown stellar sample is the same as shown in the right panel of Figure~\ref{fig:ngvsir_photcal}. WIRCam $K_s$ aperture photometry (in Vega mag) is used for color coding, and the best linear fit is indicated by the solid line. ({\it Right panel}): Same as right panel of Figure~\ref{fig:ngvsir_photcal}, but after applying the color-term correction to the $K_s$ NGVS-IR data. The symbols color is parametrized according to the $(H\!-\!K)_{\rm UKIDSS}$ color as indicated by the color scale next to the panel.}
\label{fig:ngvsir_photqual}
\end{figure*}

With the {\sc SExtractor} parameters {\sc detect\_thresh}=1.4, {\sc analysis\_thresh}=2.0,~{\sc detect\_minarea}=6 and {\sc deblend\_mincont}=0.000001, a total of $\sim$450,000 sources are detected and about 17,800 of these are identified as stars.~The accuracy of our measurements are best described by focusing on these point sources.~In Figure~\ref{fig:ngvsir_photcal} we compare the WIRCam $K_s$ photometry with the corresponding 2MASS and UKIDSS data.~The dispersion in both panels is predominantly due to the photometric errors in 2MASS and UKIDSS, respectively.~The UKIDSS data \citep[DR8,][]{lawrence07} provides a tighter calibration relation than 2MASS.~Both panels exhibit a significant color term.~The left panel of Figure~\ref{fig:ngvsir_photqual} shows that the color term relating WIRCam and UKIDSS magnitudes seems to depend on the $K_s$-band magnitude.~At the bright end, we see the actual effect of the different shapes of the $K_s$ filters of WIRCam and UKIDSS, while at the faint end the slope mostly reflects large uncertainties in the UKIDSS $K_s$ photometry.~For the UKIDSS native magnitudes (in the Vega photometric system, see Appendix~\ref{app:ABvsVega}) we have:
\[
\begin{array}{ccc}
(K - K_{\rm UKIDSS})_{\rm Vega} & = & -0.09 + 0.27\cdot (H-K)_{\rm UKIDSS} \\
     &   &                   \quad \mathrm{for}\, K\!<\!17.0\,{\rm mag}\\
%(K - K_{\rm UKIDSS})_{\rm Vega} & = & -0.16 + 0.63\cdot (H-K)_{\rm UKIDSS} \\     
%     &   &                   \quad \mathrm{for\ K\geq 17.0}\\
\end{array}
\]
The right panel of Figure~\ref{fig:ngvsir_photqual} illustrates the external photometric accuracy after this simple color-term correction. The photometric comparison between UKIDSS and WIRCam data is affected by a statistical uncertainty of 0.05 mag down to about $K_s\!=\!17$ Vega mag, which corresponds to $\sim18.82$ AB mag.

\section{Completeness Estimates}
The photometric completeness of point-sources in the final image stacks was computed by adding artificial stars with a range of magnitudes at randomly selected positions into the image. We use a set of {\sc IDL} and {\sc Python} scripts to generate a list with the positions and magnitudes of artificial stars, avoiding areas with the presence of saturated stars and very extended galaxies, e.g. the central regions of M87.

The first step consists of generating the positions of artificial stars.~We run {\sc SExtractor} on the final images and generate the segmentation maps.~We use these maps to avoid adding stars on top of a galaxy or other high density regions. Then, the image is divided into a 2D grid of 200-pixel wide bins and we randomly generate 10 positions in each bin. For each one of these positions we check that there is no collision with other sources (real or artificial) in a radius of 16 pixels, thereby avoiding artificial crowding effects. The final list consists of about 100,000 sources for each final quadrant image.

\begin{figure}[!t]
\includegraphics[width=9.1cm]{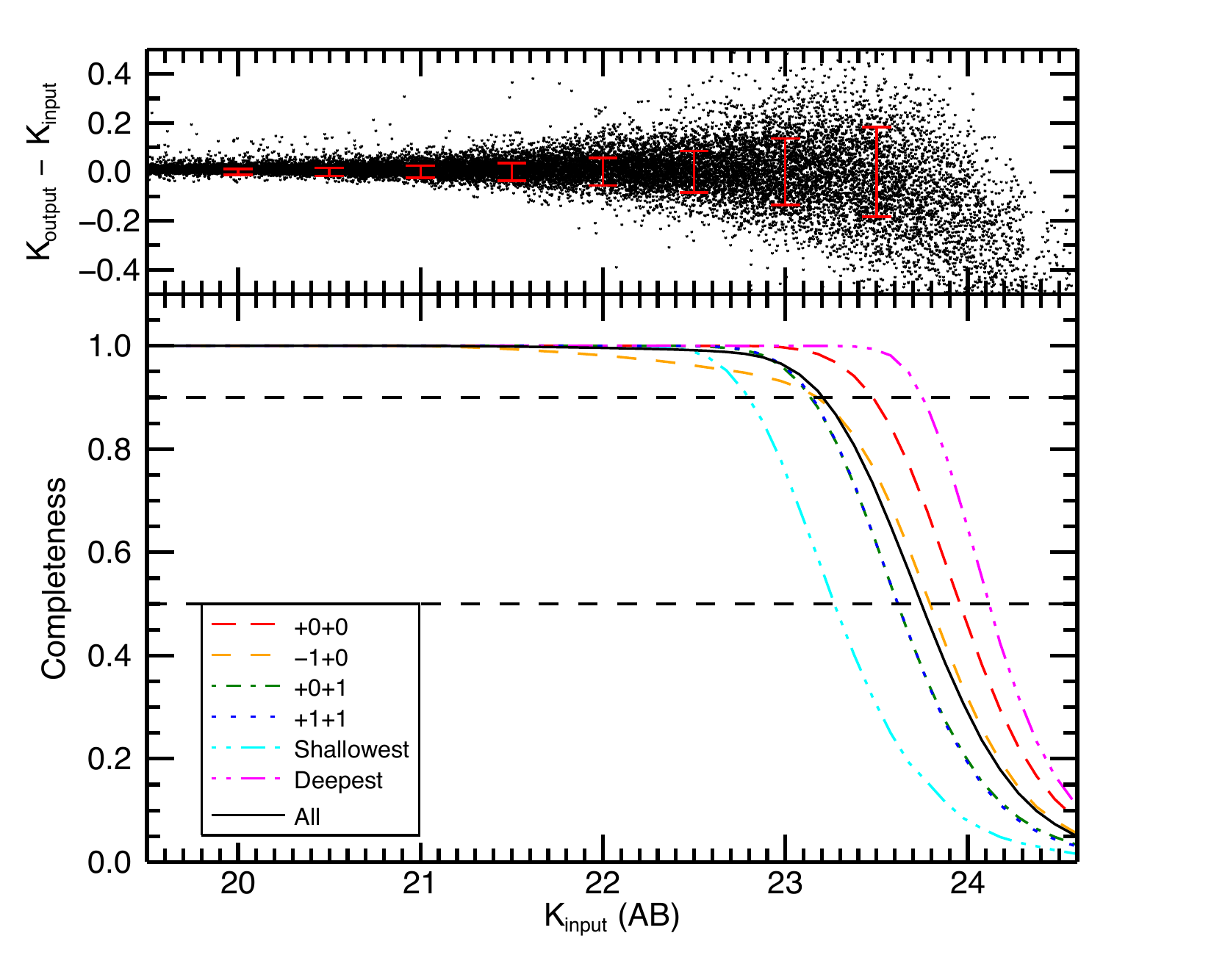}
\caption[Completeness mag curve]{({\it Top panel}): Internal photometric errors for tile +0+0 of the NGVS-IR. The errors were computed by adding artificial stars to the stacked image and then running {\sc SExtractor} to do the photometry. ({\it Bottom panel}): Photometric completeness curves for the four quadrants of the NGVS-IR Pilot program region, plus the deepest and shallowest regions in the survey. The horizontal dashed lines correspond to the 90\% and 50\% completeness.}
\label{fig:completeness_curve}
\end{figure} 

In the second step we generate the magnitudes of the artificial stars. First, we run {\sc PSFex} on the final images and compute a position-dependent PSF model. We use this PSF model to properly add a point-like source in different positions on the image. Then, we generate a homogeneous distribution of magnitudes in the range $18.0\!<\!K_{\rm Vega}\!<\!22.0$ mag and assign them to the stars generated in the previous step ($K_{input}$).

The third step adds the artificial stars to the image at the corresponding positions, with the generated magnitudes using the appropriate PSF model. We use a highly efficient multi-thread {\sc Python} script for reading the list of coordinates and building the point-like sources according to the PSF model computed with {\sc PSFex} using all available CPU cores.

We run 200 realizations of this procedure for each NGVS-IR quadrant ending up with $2\!\times\!10^7$ artificial objects per quadrant. We run {\sc SExtractor} with the same parameters used for the original stacks (see Section~\ref{txt:photometry}) and cross-match the positions of the detected sources with the artificial stars generated in step one, hereafter called recovered stars.~The top panel of Figure~\ref{fig:completeness_curve} shows the difference between the magnitude of artificial stars in tile +0+0 as measured with {\sc SExtractor} ($K_{output}$) and assigned in the mock catalog ($K_{input}$).~This figures illustrates the internal photometric errors of the NGVS-IR survey and demonstrate the excellent quality of the stacked images.~In order to estimate the completeness,  we adopt a magnitude bin of 0.1 mag and compute the ratio between the number of recovered stars and the number of artificial stars as a function of magnitude and position. The bottom panel of Figure~\ref{fig:completeness_curve} shows the completeness fraction as a function of the input AB magnitude for each of the four quadrants, the entire survey, and the deepest and shallowest regions in the pilot field.~The 90\% completeness magnitude for point-source detections and the corresponding $5\sigma$ limiting magnitude together with a conservative surface brightness limit estimate for the NGVS-IR quadrants are summarized in Table \ref{tab:completeness}. 

\begin{figure*}[!t]
\centering
\includegraphics[width=8.9cm]{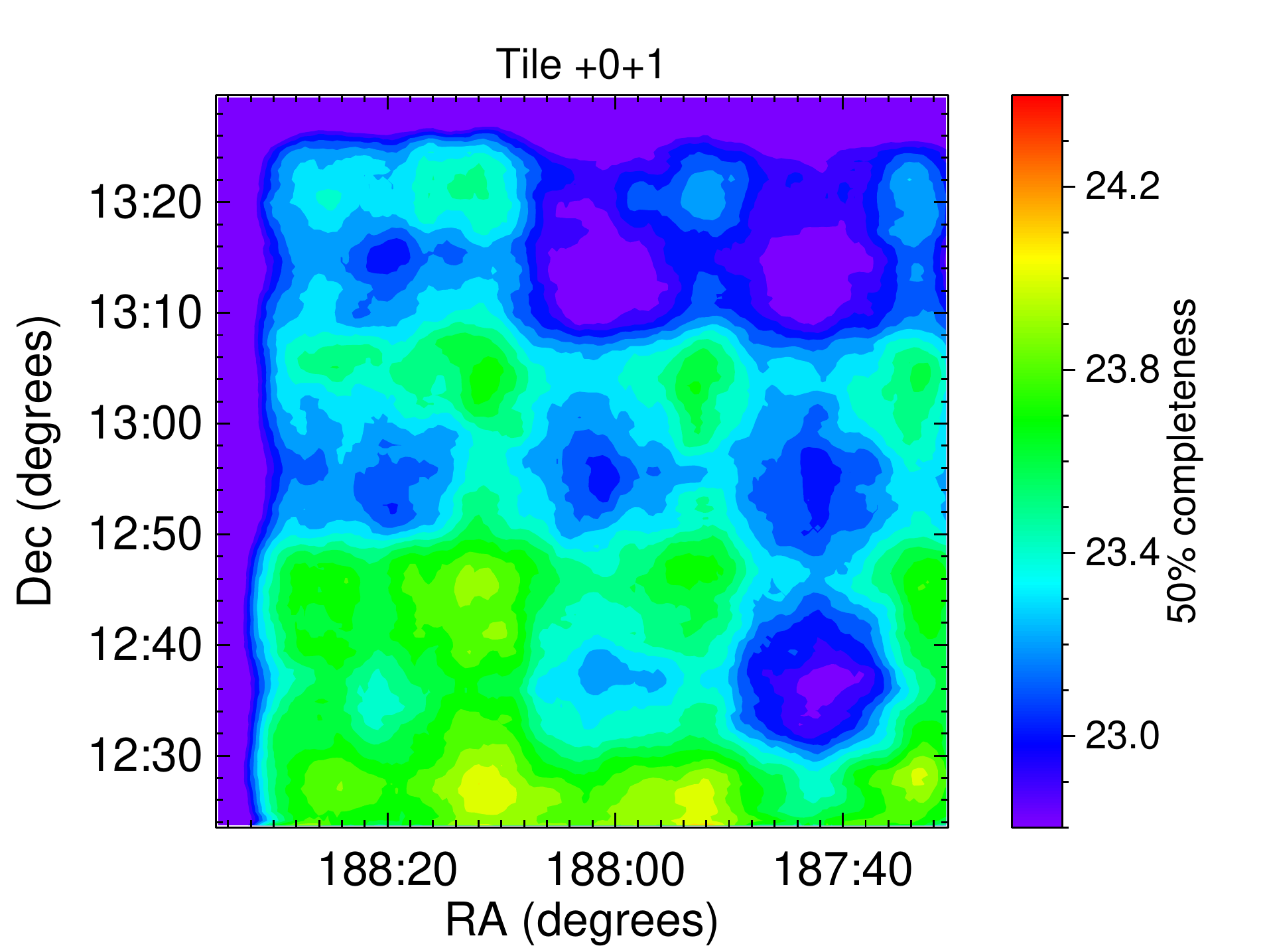}
\includegraphics[width=8.9cm]{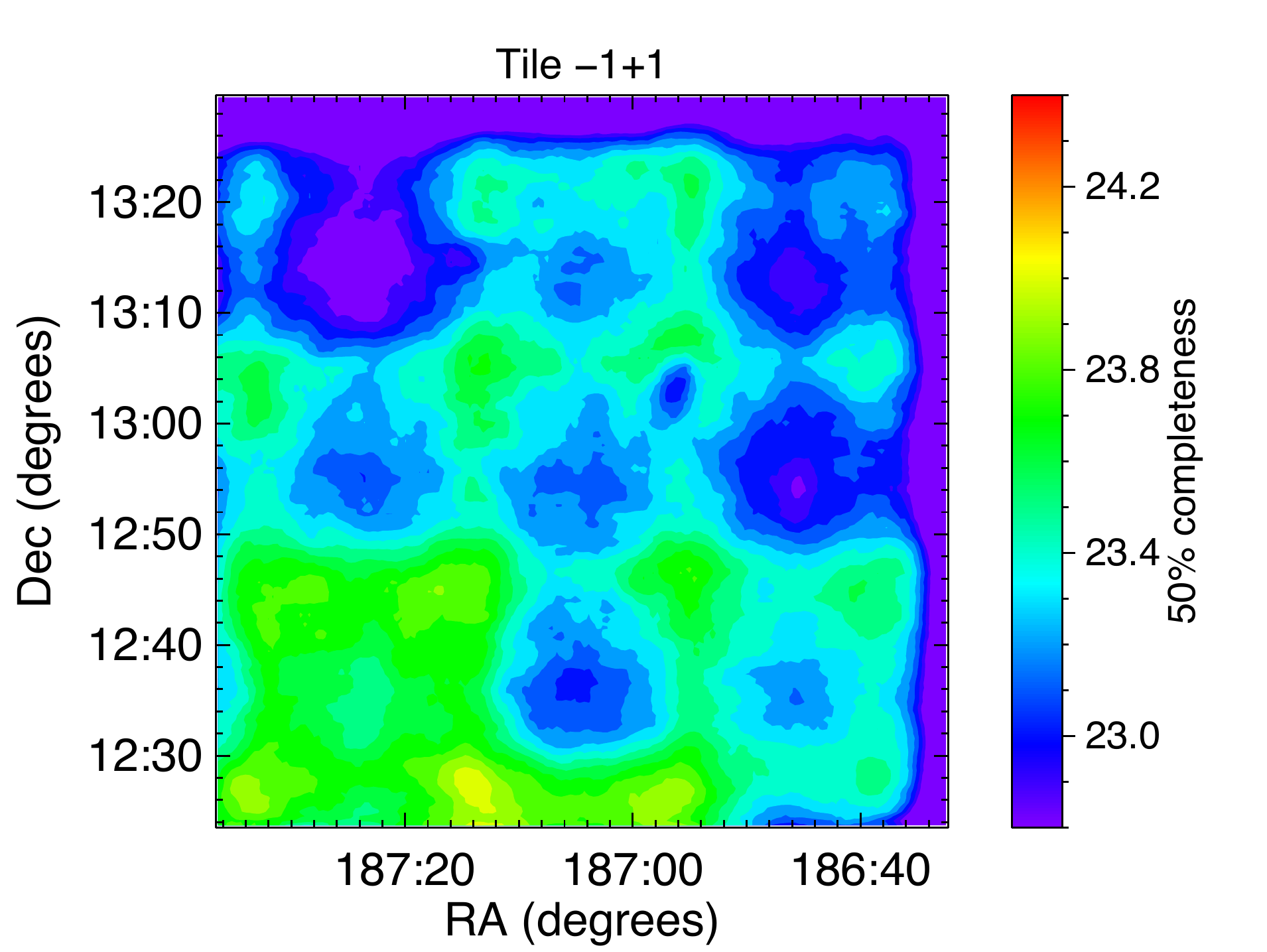}
\includegraphics[width=8.9cm]{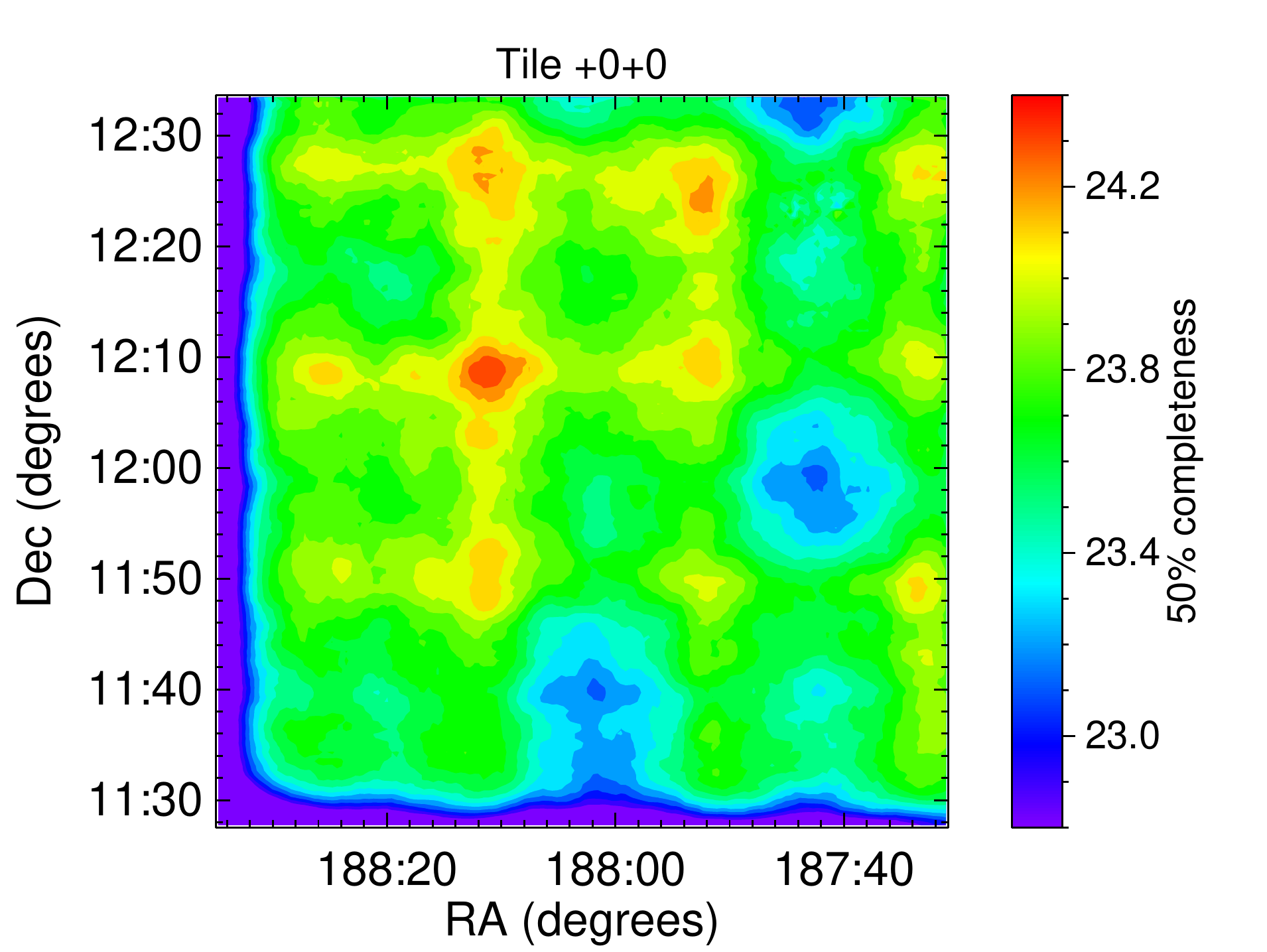}
\includegraphics[width=8.9cm]{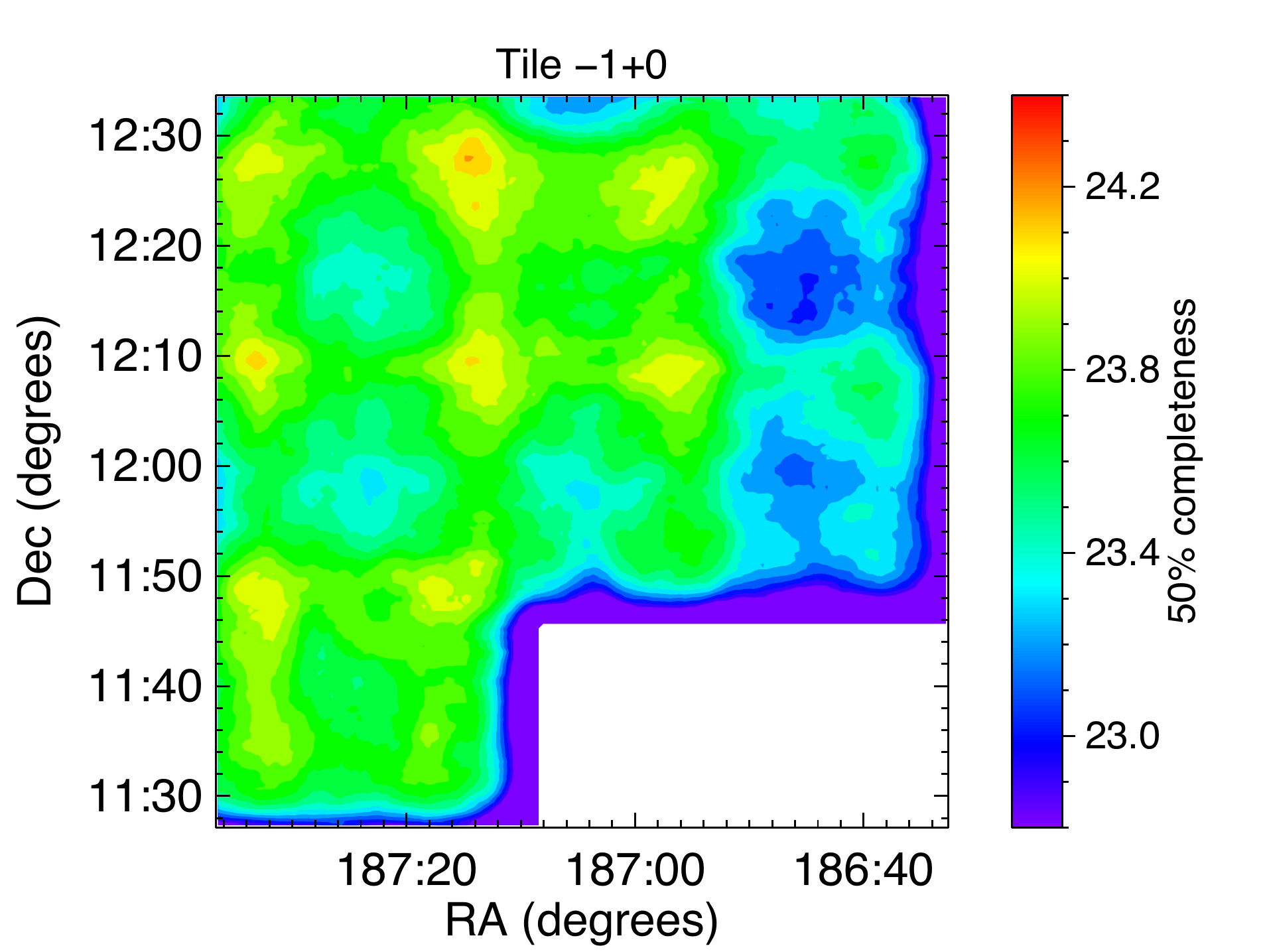}
\caption[Completeness mag map]{Illustration of the photometric depth variations for each of the NGVS-IR pilot field quadrants.~The color bar corresponds to the 50\% completeness magnitude of point-like source detections.~The structures in each field are mainly due to the seeing distribution of the corresponding tiles and their overlap regions.}
\label{fig:psf_fwhm_compl}
\end{figure*} 

Figure~\ref{fig:psf_fwhm_compl} illustrates the variations of all the completeness magnitudes across the four quadrants. The main reason for the fluctuations visible across the field stems from the photometric quality, i.e.~FWHM distribution (see Figure~\ref{fig:seeing}) in each individual WIRCam tile contributing to the final quadrant stack.

\begin{table}
\begin{center}
\caption{Completeness magnitude of the NGVS-IR stacks \label{tab:completeness}}
\begin{tabular}{cccc}
\hline\hline
Tile & $K_{AB, 90\%}$ & $K_{AB, 5\sigma}$ & $\mu_{K_s, {\rm AB, 5\sigma}}$\\ 
\hline
    $+0+0$ & 23.70 & 24.64  & 25.11\\
    $-1+0$ & 23.51  & 24.61  & 25.08 \\
    $+0+1$ & 23.32 & 24.16  & 24.63 \\
    $+1+1$ & 23.34 & 24.23  & 24.70 \\
\hline
\end{tabular}
\tablecomments{The $5\sigma$ limiting magnitude is measured inside a circular aperture with 0.7\arcsec\ radius and marks the limiting surface brightness estimate.}
\end{center}
\end{table}

%%%%%%%%%%%%%%%%%%%%%%%%%%%%%%%%%%%%%%%%%%%%%%%%%%%%%%%%%
\section{Discussion}

\subsection{Color-Color Diagrams}
\begin{figure*}[!t]
\centering
\includegraphics[width=14.5cm,bb=35 10 555 420]{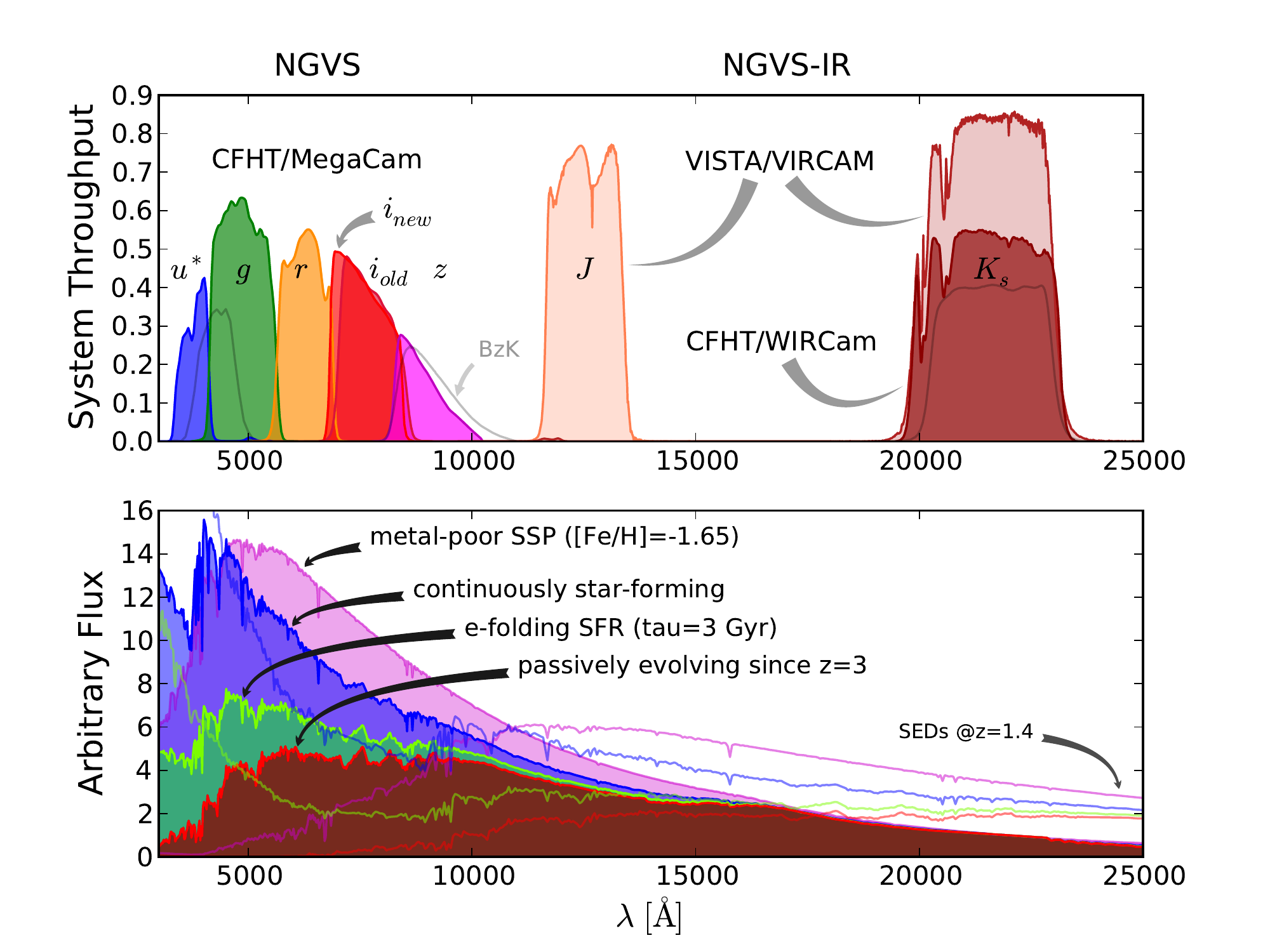}
\caption{{\it Top panel}: Direct comparison of system throughput curves, including detector efficiency, effects of telescope optics, and atmospheric absorption for all NGVS+NGVS-IR filters.~Note that the NGVS-IR $K_s$-band filter mainly differs in detector efficiency and optics between the two instruments, i.e.~CFHT/WIRCam vs.~VISTA/VIRCAM.~For comparison purposes, we plot the $BzK$ system throughput curves that were used in \cite{daddi04} in light grey shading.~Note that the atmospheric absorption effects impacting the $K_s$-band curve in the $BzK$ system are not included. {\it Bottom panel}: Comparison of four different SED types: 12 Gyr old metal-poor SSP with [Fe/H]~$\!=\!-1.65$ dex, constant SFR, e-folding SFR with $\tau\!=\!3$ Gyr, and passively evolving starburst since $z\!=\!3$, taken from {\sc P\'egase} \citep{fioc97}. Note that for SSPs with higher [Fe/H] the metal-poor SSP approximates the passively evolving starburst SED, that has been calculated for solar chemical composition. For easy comparison all SED curves at $z\!=\!0$ were normalized to the flux at $2.2\,\mu m$, roughly corresponding to the effective wavelength of the $K_s$ band.~The open SED curves correspond to the same four SED types as they would appear at $z\!=\!1.4$.}
\label{fig:sedplot}
\end{figure*}

Adding more filters and a successively wider SED coverage to the investigation of astronomical objects adds more diagnostic power to the analysis and delivers ultimately more robust and astrophysically meaningful results \citep[e.g.][]{park05, puzia07}.~Color-color planes are efficient tools for the classification of sources in large-scale imaging surveys \citep[e.g.][]{daddi04, faber07}. By combining near-UV and optical data from NGVS and near-IR photometry from the NGVS-IR, the advantage of spanning the entire spectral range of stellar emission from the atmospheric UV cutoff at $\sim\!3200$\,\AA\ to the near-infrared at $\sim\!2.5\,\mu$m becomes clear when one is confronted with the vastly improved system throughputs and SED coverage of the combined NGVS+NGVS-IR filter set as illustrated in Figure~\ref{fig:sedplot}.

To construct a first NGVS+NGVS-IR source catalog, we cross-matched the NGVS-IR catalog positions with those of the NGVS $u^*griz$ catalog, which contains approximately half a million objects. The distribution of angular separations between $K_s$ and $i$-band coordinates peaks at 0.09\arcsec, and 95\% of the detected sources (of small angular size) have a separation smaller than 0.3\arcsec.~This small bias is due to the marginally resolved nature of background galaxies which constitute a large fraction of the matched catalog and introduce a random  offset in the matching accuracy that appears as a systematic in one radial dimension. Future dedicated versions of those matched catalogs will be built with object-dependent procedures that allow for self-consistent optical and near-IR measurements, in particular for extended sources. However, independent measurements and subsequent matching are fully sufficient for the qualitative purposes of this paper. 

Historically, the optical/near-IR color-color plane that is most widely used in conjunction with large extragalactic surveys is the $BzK$ diagram which combines $(B\!-\!z)$ and $(z\!-\!K)$ color indices or equivalents thereof. The diagram was highlighted by \cite{daddi04} as a means of identifying both passively evolving and star-forming galaxies located at redshifts larger than 1.4 or so.~It was published later for a variety of deep fields, confirming the original segregation between low and high-redshift sources \citep[e.g.][]{lane07, mccracken10, mccracken12, bielby12}. The $BzK$ technique is now being frequently used for studies of the clustering properties, intrinsic morphologies, stellar populations, and the X-ray luminosities of $BzK$-selected galaxies \citep[e.g.][]{lin12, yuma11, yuma12, ly12, rangel13}. Ultra-deep follow-up spectroscopy campaigns confirm the efficient selection of this type of high-redshift, star-forming galaxies that suffers $\sim\!10\%$ contamination by other sources \citep{kurk13}.

\subsection{The NGVS+NGVS-IR analog of $BzK$: the $gzK_s$ color-color plane}
\label{sec:gzK_plane}
\begin{figure*}[!t]
\centering
\includegraphics[width=15cm]{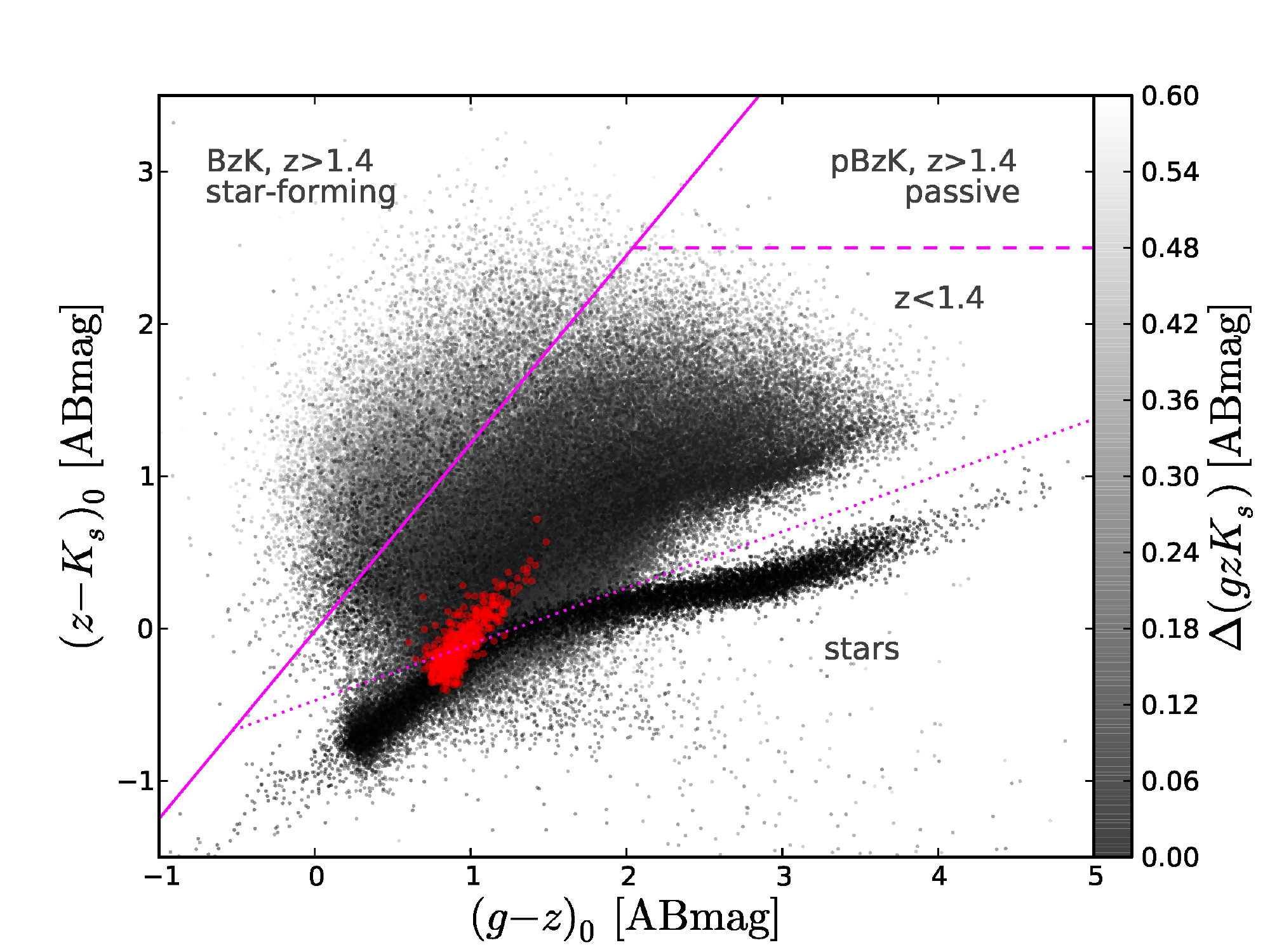}
\includegraphics[width=15cm]{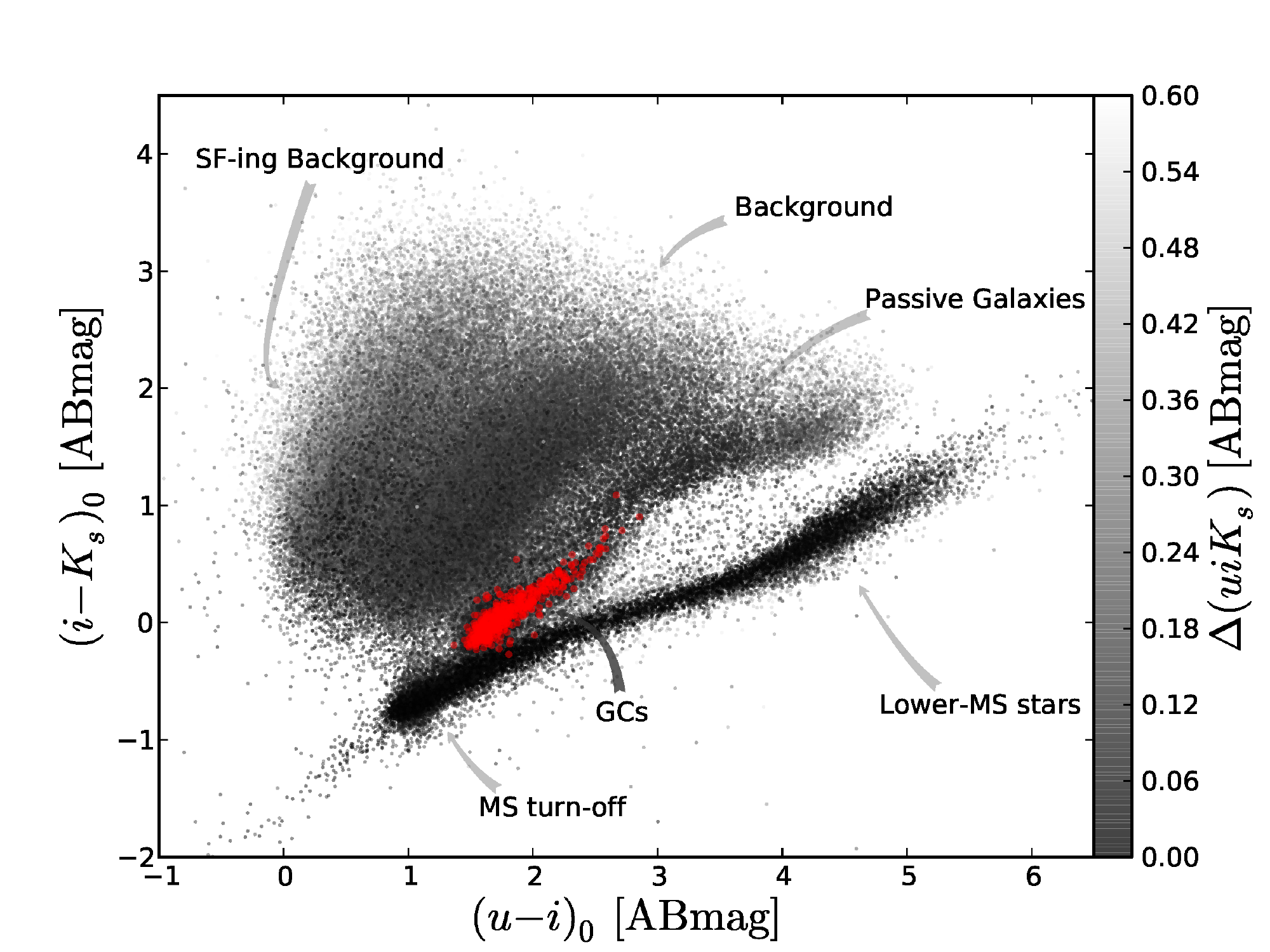}
\caption[uiK color-color diagram]{Illustration of the $gzK_s$ ({\it top panel}) and $uiK$ ({\it bottom panel}) color-color diagrams of all the objects in the NGVS pilot field ({\it grayscale dots}).~The symbol shading is parametrized by the total photometric error of each filter contributing to each color plane.~All magnitudes are in the AB system and include the reddening corrections from \cite{schlafly11}.~The red dots mark spectroscopically confirmed globular clusters with the systemic velocity of Virgo cluster galaxies ({\it see text for details}).~The magenta solid line shows the criteria defined by \citet{daddi04} to isolate $z\!>\!1.4$ star-forming galaxies, the dashed line to separate $z\!>\!1.4$ passively evolving galaxies from nearby systems, and the dotted line to separate foreground stars from galaxies.~The corresponding $BzK$ areas are labeled accordingly in the top panel. We indicate in the bottom panel characteristic object sequences (i.e.~stars, GCs, and passively evolving galaxies) and other overdensities (locations of normal and star-forming background galaxies) that are discussed in section~\ref{sec:uiK_plane}.}
\label{fig:giKuik_withGCs}
\end{figure*}

In terms of SED coverage of the involved filters, the closest analog to the $BzK$ diagram that can be constructed from our combined NGVS+NGVS-IR data is the $gzK_s$ diagram which is shown in the top panel of Figure~\ref{fig:giKuik_withGCs}, which illustrates our pilot field sample greyscale-coded by the Gaussian propagated total error of each contributing filter (see also Figure~\ref{fig:sedplot} for a comparison of the corresponding $gzK_s$ vs.~$BzK$ system throughput curves).~The mean foreground extinction towards M87 is $A(g)\!\simeq\!0.076$, $A(z)\!\simeq\!0.029$, and $A(K_s)\!\simeq\!0.007$ mag, and was taken from \cite{schlafly11}\footnote{See Appendix~\ref{app:extinction} for a discussion on the small color dependence of the extinction coefficients.}.~The following analysis includes the corresponding extinction correction terms.~We observe various characteristic object overdensities and sequences in the $gzK_s$ plane, such as the narrow sequence of foreground Galactic stars and other less constrained object distributions including actively star-forming and passively evolving galaxies at various redshifts that were discussed in previous works \citep[e.g.][]{bielby12, merson13}.~We use equations (1)--(8) provided in \cite{bielby12} to transform the $BzK$ object selections into our $gzK_s$ color-color plane.~In particular, we use the median color $(u\!-\!g)\!=\!-0.194$ mag of our entire sample and obtain for the selection of star-forming galaxies at $z\ga1.4$ the following relation (shown as solid magenta line in Figure~\ref{fig:giKuik_withGCs}): 
\begin{equation}
(z\!-\!K_s)_0 > 1.233\times(g\!-\!z)_0-0.017.
\end{equation}
To select passively evolving galaxies at $z\ga1.4$ we obtain (illustrated as dashed magenta line in Figure~\ref{fig:giKuik_withGCs}):
\begin{equation}
(z\!-\!K_s)_0 < 1.233\times(g\!-\!z)_0-0.017 \cap (z\!-\!K_s)_0 > 2.5.
\end{equation}
To select foreground stars and separate them from galaxies our transformations yield (dotted magenta line in Figure~\ref{fig:giKuik_withGCs}):
\begin{equation}
(z\!-\!K_s)_0 < 0.37\times(g\!-\!z)_0 + 0.474
\end{equation}

The corresponding relations shown in the top panel of Figure~\ref{fig:giKuik_withGCs} illustrate the quality of separating the classically defined $BzK$ galaxies at $z\!>\!1.4$ from the general locus of more nearby galaxies and the stellar sequence.~However, it is quite evident that even with the NGVS+NGVS-IR high-quality data, a clear separation of stellar foreground from the background galaxy population is not entirely possible based on the $gzK_s$ color-color diagram alone.

\begin{figure*}[!t]
\centering
\includegraphics[width=8.9cm, bb=10 200 600 600]{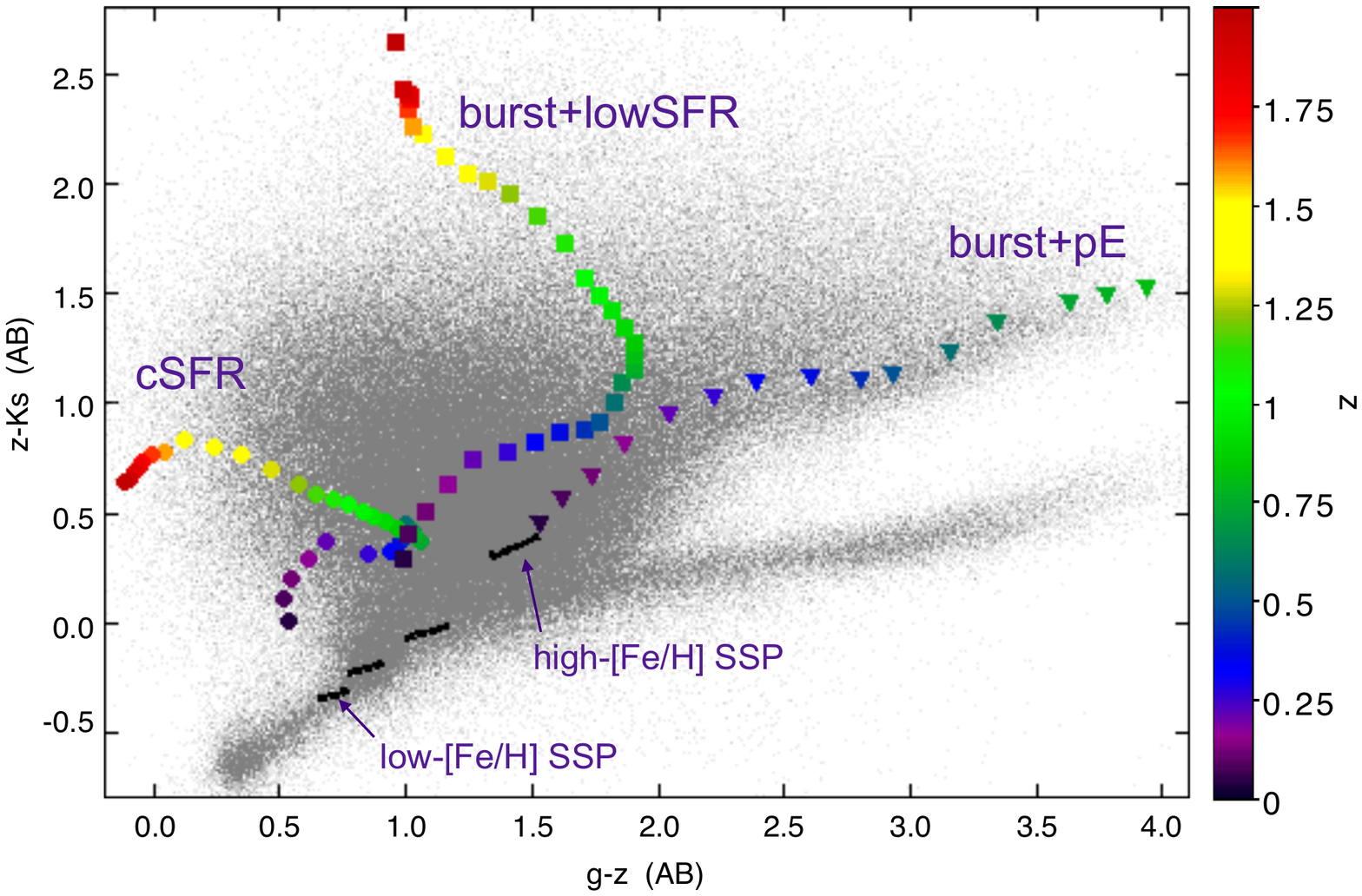}
\includegraphics[width=8.9cm, bb=10 200 600 600]{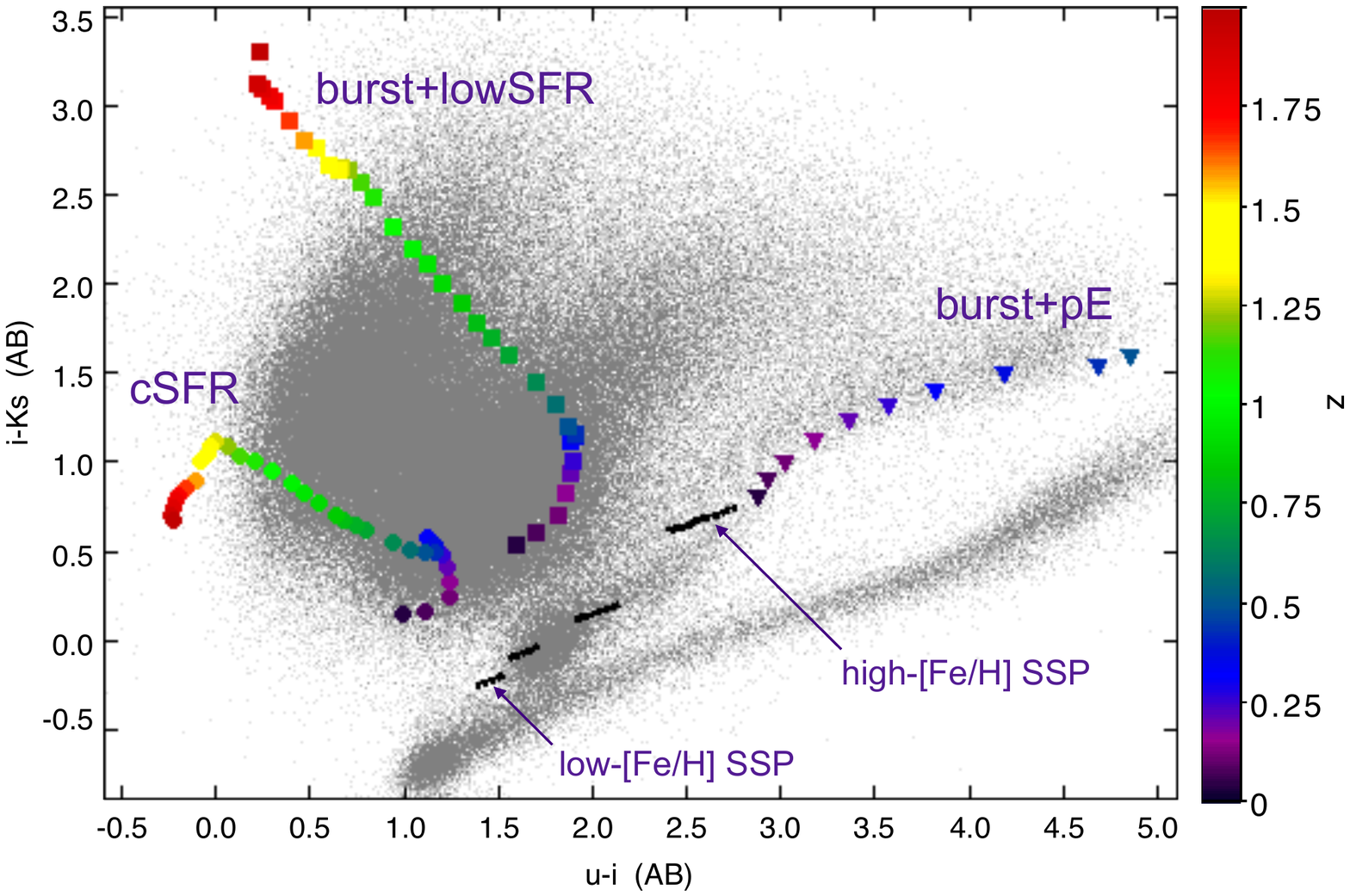}
\caption[giK color-color diagram with models]{({\it Left panel:}) Model locations on the $gzK_s$ diagram. The finely-dotted, black lines represent sequences of old single-age stellar populations (at redshift 0), for metallicities $Z=0.0004, 0.001, 0.004, 0.02$ (from bluer to redder colors) using the {\sc P\'egase} population synthesis models \citep{fioc97}. The large colored symbols represent the evolutionary paths of observed colors for galaxies formed at redshift 3 or above with three different star formation histories: (i) galaxy with a constant star-formation rate (cSFR), (ii) galaxy that formed the majority of its stars at high redshift, but star formation continues at a lower constant rate (burst+lowSFR), (iii) a ``red-and-dead" galaxy that formed all its stars at redshift 3 and evolved passively ever since (burst+pE). ({\it Right panel:}) Model locations in the $uiK_s$ diagram with symbols as in the left panel.}
\label{fig:giKuik_withTracks}
\end{figure*}

Overall the $gzK_s$ color-color plane is the tool of choice for the study of the redshifted universe beyond Virgo due to the superior photometric depth of the $g$ filter compared with the $u^*$ band observations. However, unlike any other deep optical/near-IR survey, NGVS+NGVS-IR also contains a nearby structure: Virgo itself. The most striking difference between our color-color diagrams and those of other surveys is not due to Virgo galaxies -- these occur in negligible numbers in any given part of the diagram --, but due to GCs that are concentrated around the massive elliptical galaxy NGC\,4486 (M87) and other giant ellipticals located in the central regions of Virgo (see Figure~\ref{fig:wircam_tiles}).~To demonstrate the locus of $gzK_s$ colors of Virgo GCs we highlight radial-velocity confirmed GCs in Figure~\ref{fig:giKuik_withGCs} as red dots (E. Peng et al., {\it in preparation}).~This GC locus overlaps with the colors of stars at the blue end, and is heavily contaminated by background galaxies towards redder colors, in the $gzK_s$ diagram.

\subsection{The $uiK_s$ Color-Color Plane}
\label{sec:uiK_plane}
For a photometry-based selection of GCs and other compact stellar systems, such as UCDs and dwarf galaxies, we show in the following that the most powerful color combination is given by the $uiK_s$ diagram, which combines high-quality near-UV, optical, and near-IR photometry from NGVS and NGVS-IR.~The bottom panel of Figure~\ref{fig:giKuik_withGCs} shows the $uiK_s$ color-color diagram, in which the data is deredden with the values extracted from the \cite{schlafly11} maps using $A(u)\!\simeq\!0.097$, $A(i)\!\simeq\!0.039$, and $A(K_s)\!\simeq\!0.007$ mag.~The typical structures seen in the $gzK_s$ diagram (top panel of Figure~\ref{fig:giKuik_withGCs}) that were classified in previous deep-field surveys appear much more prominent and better defined in the $uiK_s$ color-color plane. The stellar sequence seen in the $uiK_s$ diagram remains very clearly identified in this new plane and appears more separated from the main cloud of galaxies. This will be particularly useful for the analysis of stellar age and metallicity distribution functions in the Virgo overdensity described in \cite{ferrarese12}.~Several new narrowly defined features in the $uiK_s$ plane become visible at intermediate colors that we attempt to classify qualitatively in the following.~Most importantly, however, we note that the contamination of the GC locus is very significantly reduced with the use of the $uiK_s$ filter combination.

\subsubsection{An Efficient Tool for Star Cluster Selection}
In order to better understand the various features in the $uiK_s$ plane, we overplot in Figure~\ref{fig:giKuik_withTracks} predictions of simple stellar population (SSP) model calculations based on a customized version of the population synthesis code {\sc P\'egase} \citep{fioc97} that includes the exactly matched throughput functions for all NGVS+NGVS-IR filters.~While in the $gzK_s$ plot these SSP models coincide with the overlap region between stars and galaxies, in the $uiK_s$ plane they fall right on top of a sharply defined sequence which we identify as GCs.~We note that the metallicity and age coverage of the shown SSP model predictions, i.e.~$Z\!=\!0.0004, 0.001, 0.004, 0.02$ and $t\!=\!8$ to 13 Gyr (each running from bluer to redder colors), agrees fairly well with the GC candidate locus, in particular in the $uiK_s$ plane, and is consistent with what is expected from previous photometric and spectroscopic studies of stellar populations in typical Virgo GCs located in the vicinity of M87 \citep[e.g.][]{hanes86, cohen98, hanes01, kisslerpatig02, jordan02, tamura06b, peng09, yoon11, forte13}.~We verify the fidelity of the GC selection with objects that have the systemic radial velocity of the Virgo cluster using spectroscopy available in the literature \citep{hanes01,strader11} and recently obtained with the multi-object, moderate-dispersion spectrograph {\sc Hectospec} at MMT (E. Peng et al., in preparation). In Figure~\ref{fig:giKuik_withGCs} we overplot such radial-velocity confirmed Virgo GCs in the $gzK_s$ and $uiK_s$ color-color diagrams as red symbols.~At this point we defer a more quantitative analysis of the Virgo GC stellar populations to a future paper and continue with the investigation of the other sequences, keeping in mind the $uiK_s$-based photometric GC selection.

\begin{figure*}[!t]
\centering
\includegraphics[width=8.98cm]{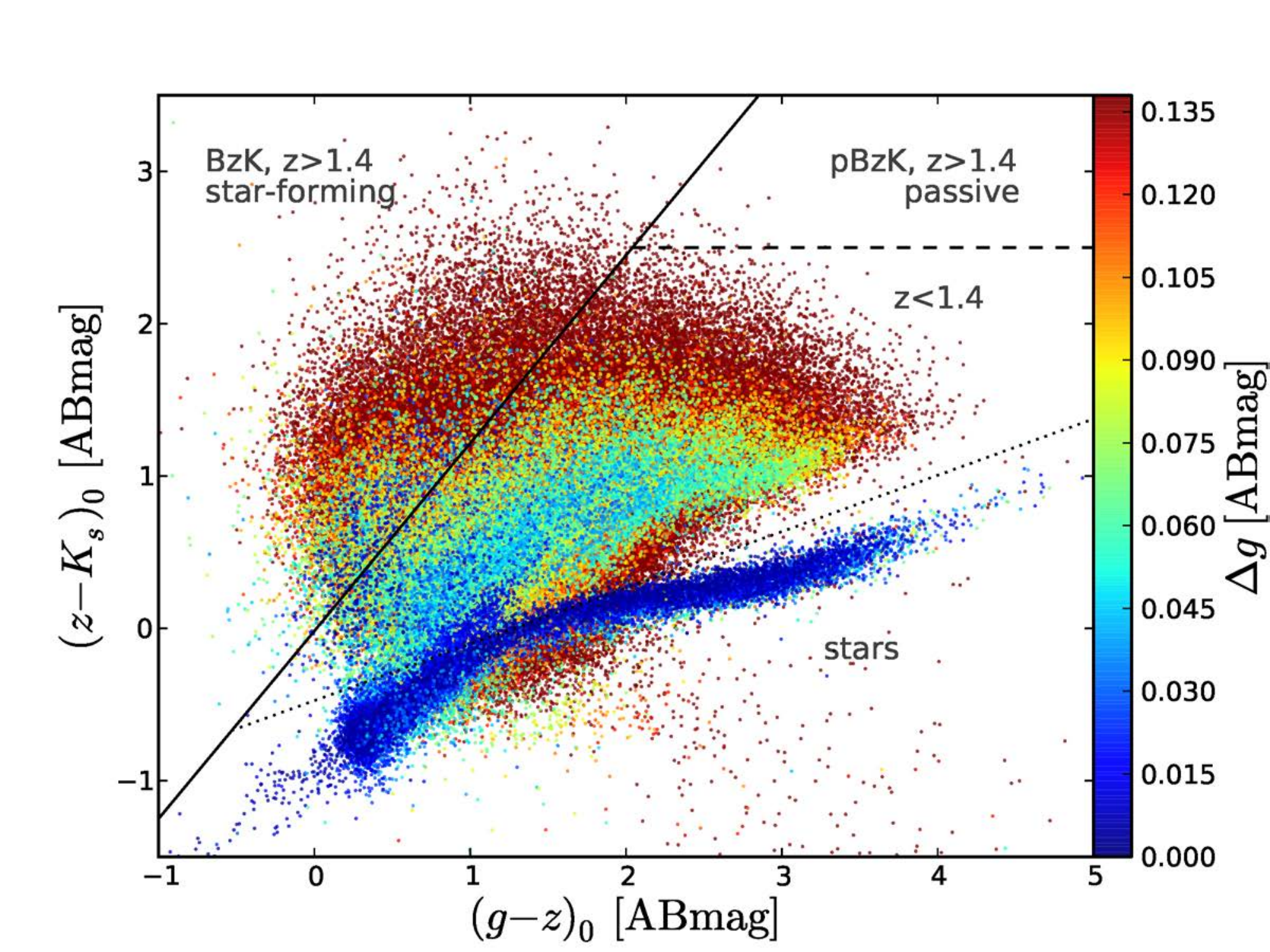}
\includegraphics[width=8.98cm]{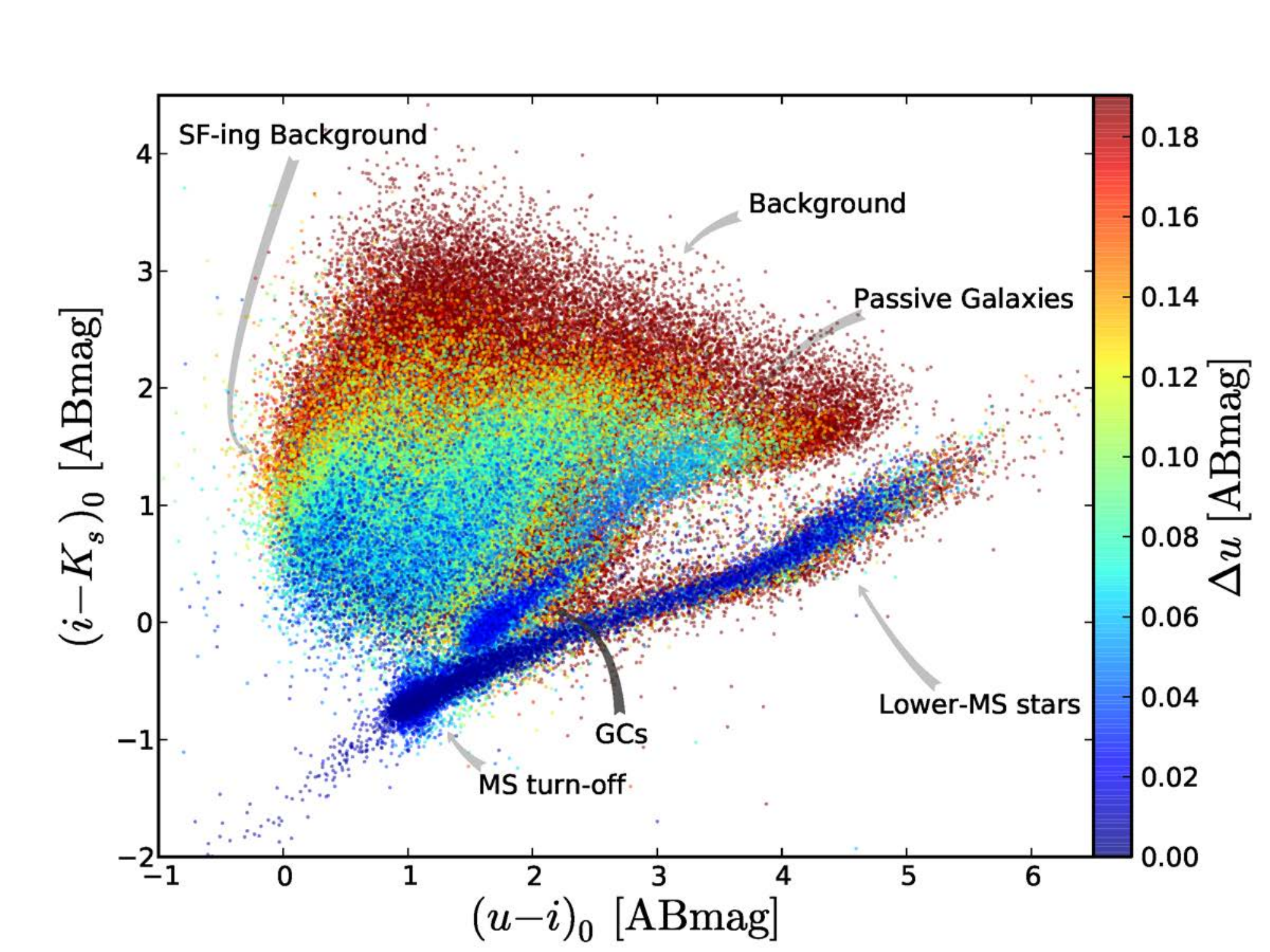}
\includegraphics[width=8.98cm]{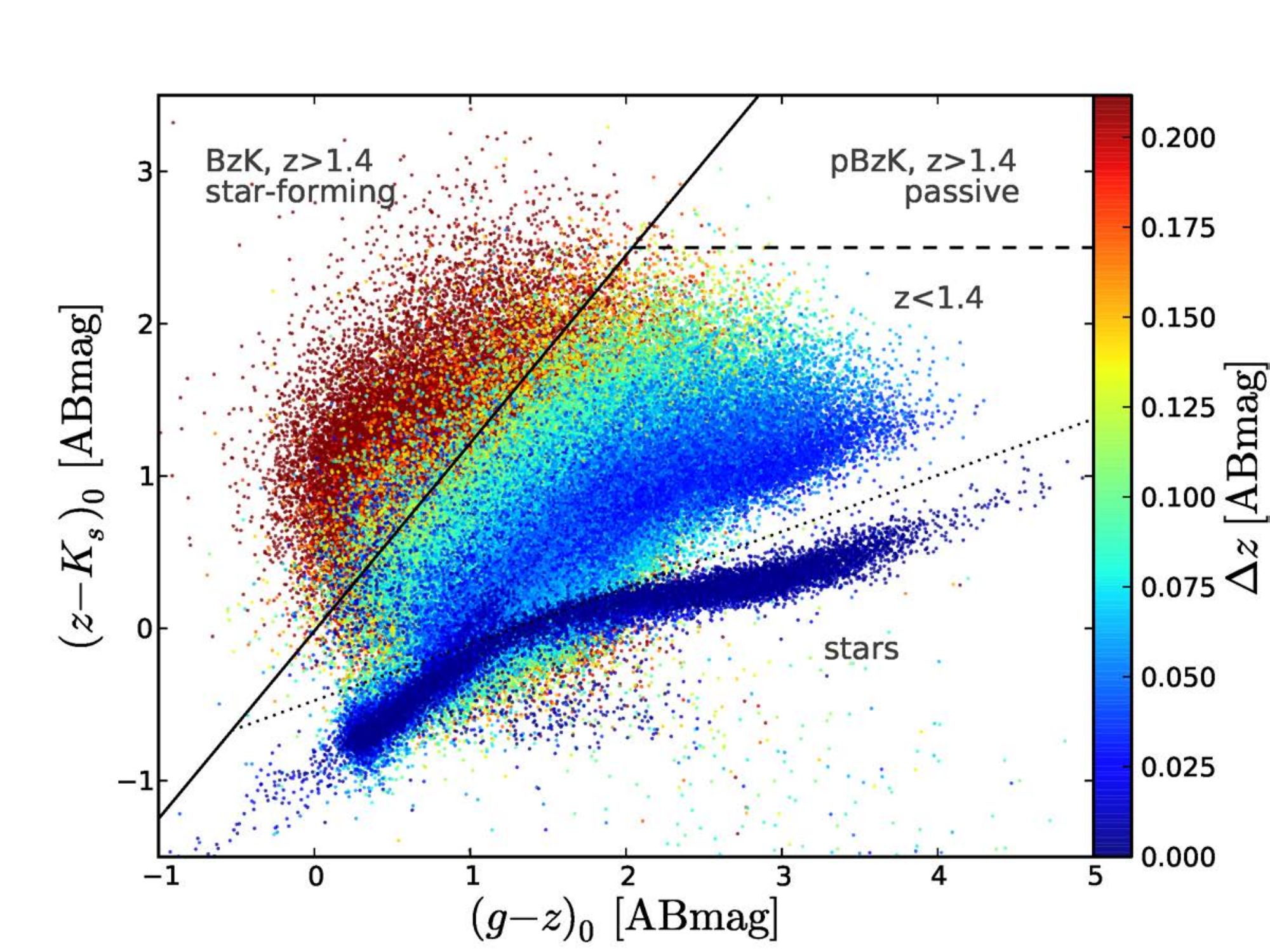}
\includegraphics[width=8.98cm]{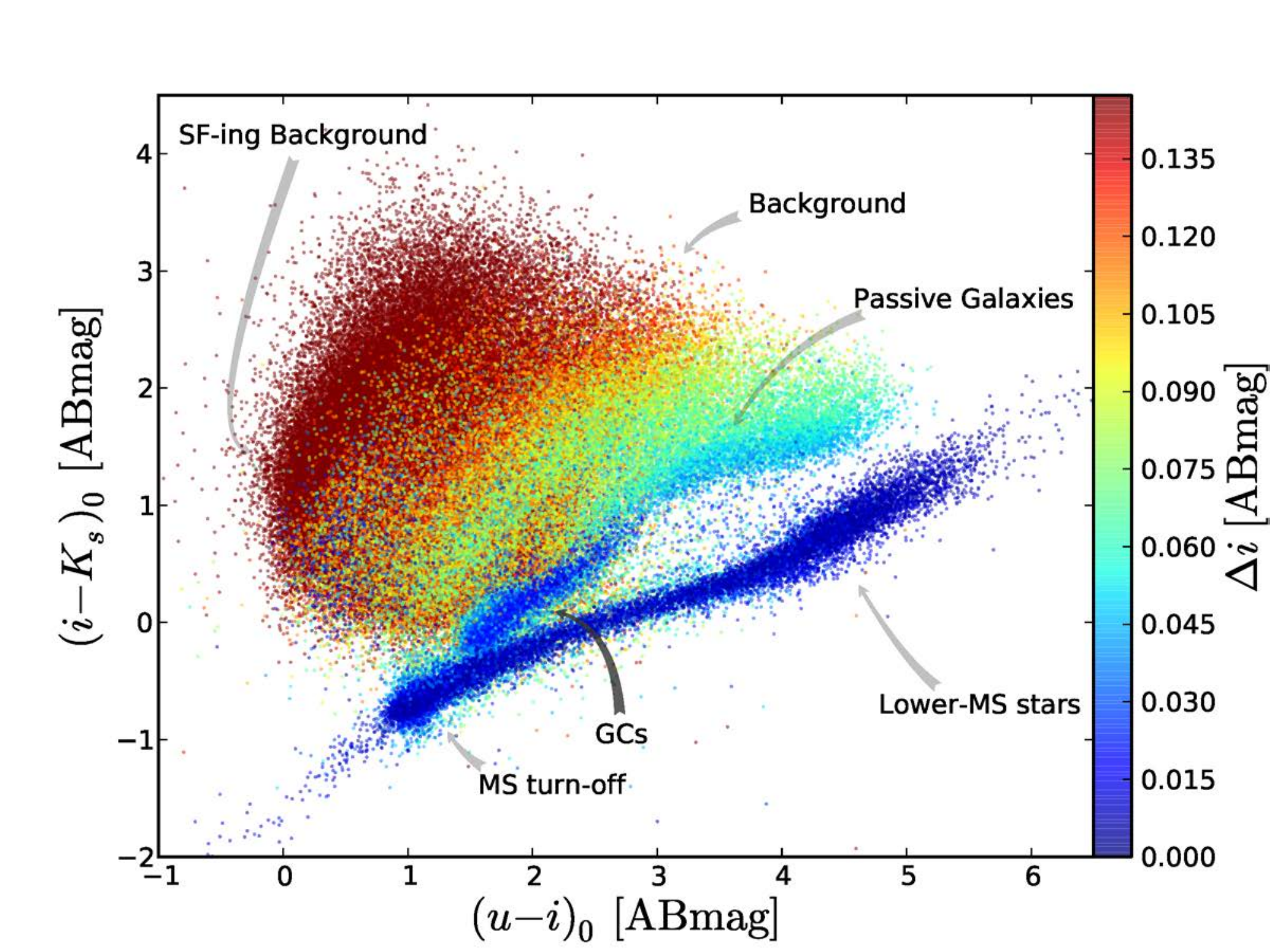}
\includegraphics[width=8.98cm]{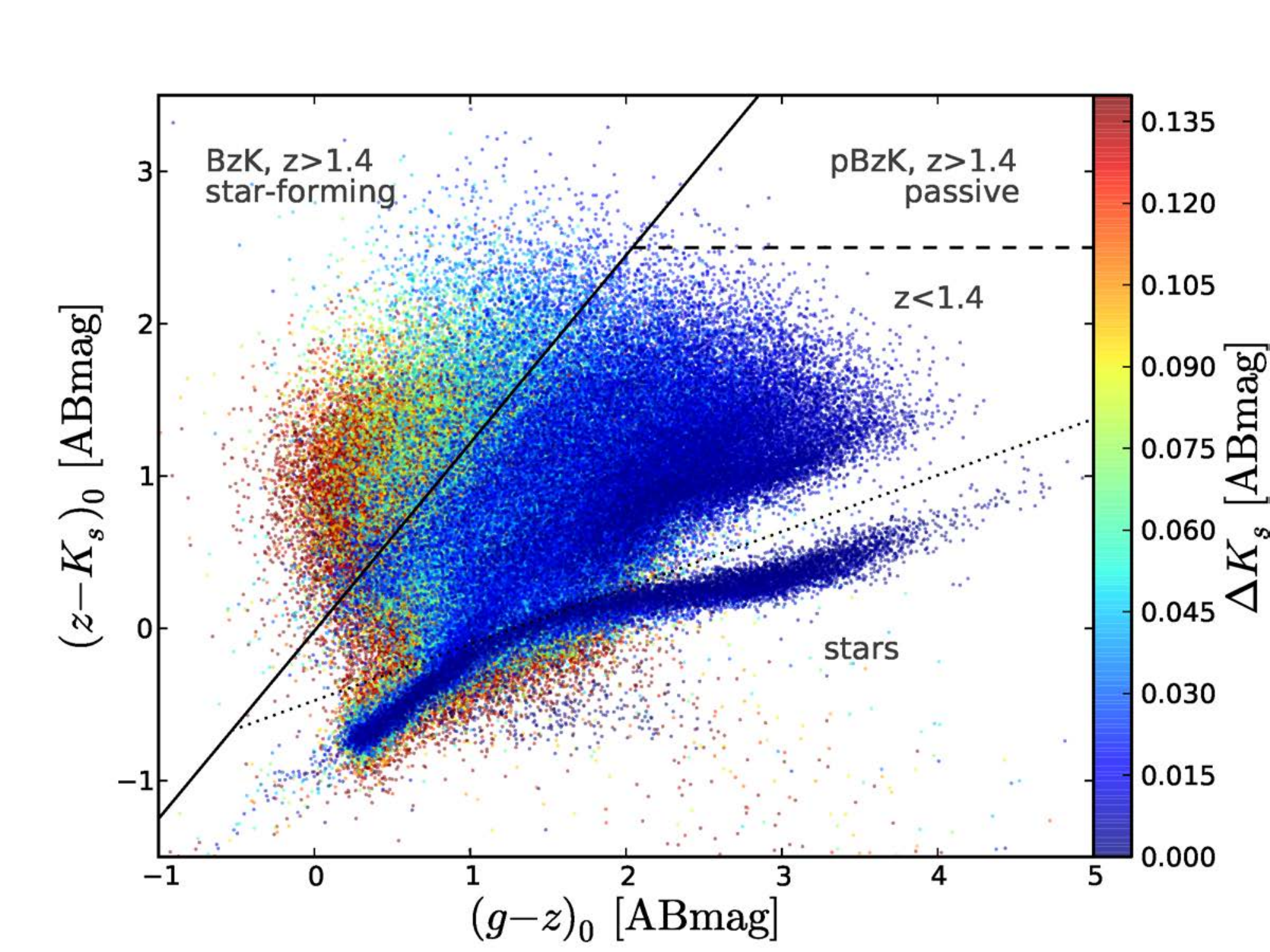}
\includegraphics[width=8.98cm]{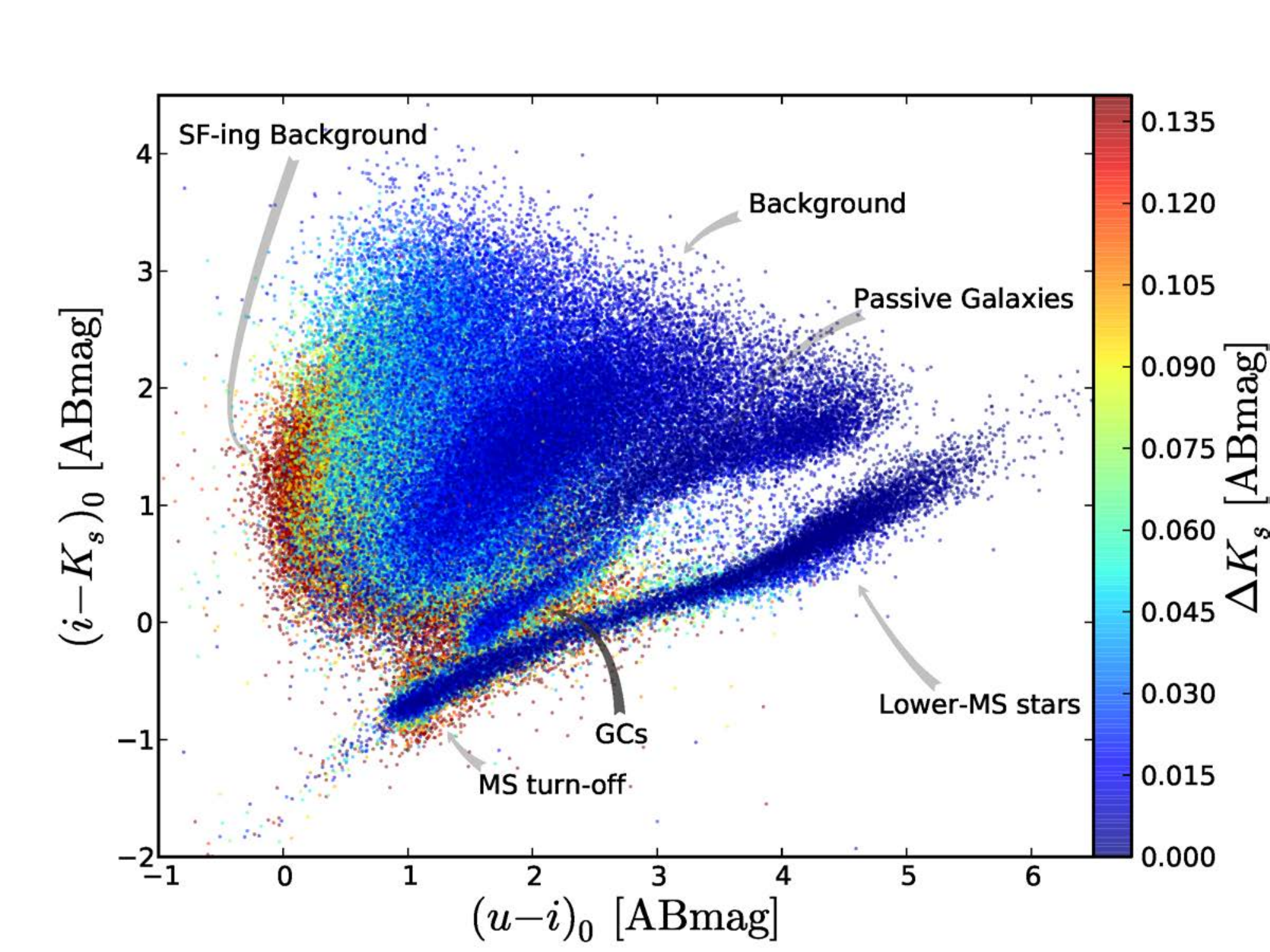}
\caption{({\it Left panels:}) Illustration of the $gzK_s$ color-color diagram, which is the best approximation to the classic $BzK$ diagram using the NGVS+NGVS-IR filter set.~The plots show the entire dataset in the Virgo pilot field. The lines mark the separations between star-forming galaxies at $z>1.4$, passively evolving galaxies at $z>1.4$, galaxies at $z<1.4$ and foreground stars as defined by \cite{daddi04}. Colorbars indicate the color parametrization of the symbols in each panel by one of the constituent $gzK_s$ filters, i.e.~in the $g$-band ({\it top}), $z$-band ({\it middle}) and the $K_s$ filter ({\it bottom}). ({\it Right panels}): The corresponding $uiK_s$ color-color diagram of objects in the NGVS pilot field region. The colorbars indicate the photometric error parametrization in the $u$-band ({\it top}), $i$-band ({\it middle}), and the $K_s$-band ({\it bottom}).}
\label{fig:giKuik_errdistrib}
\end{figure*}

\subsubsection{The Redshift Evolution of Galaxies in the $uiK_s$ Plane}
In addition to the SSP model color predictions, we plot in Figure~\ref{fig:giKuik_withTracks} the redshift evolution in the $gzK_s$ and $uiK_s$ color-color space of three prototypical SEDs of galaxies born at $z\!=\!3$ with different types of star-formation histories.~These tracks were computed using the NGVS+NGVS-IR instrument throughput curves (see Figure~\ref{fig:sedplot}) with a customized version of the population synthesis code {\sc P\'egase} \citep{fioc97},~but other population synthesis codes show very similar trends \citep[see also][]{dahlen13}.~In the order of numerically increasing colors, the tracks correspond to 1) a continuously star-forming galaxy, 2) a galaxy that formed the majority of its stellar content at high redshift and star formation continues at a lower constant rate, 3) a passively evolving galaxy that formed all its stellar content at $z\!=\!3$.

We notice that, in contrast to the $gzK_s$ plane, galaxies with any level of star formation are expected to be fairly distinct in their $uiK_s$ color properties from GCs and foreground stars from $z\!=\!0$ out to redshifts $z\!\simeq\!1.5$ and higher.~We find that the hook in the redshift evolution of star-bursting galaxies at $z\!\approx\!0.5\!-\!1$ in the $gzK_s$ color-color plane is responsible for much of the contamination at colors $(g\!-\!z)\!\simeq\!1.0$ and $(z\!-\!K_s)\!\simeq\!0\!-\!0.5$ mag, which is the mean locus of metal-poor and intermediate-metallicity GCs.~Such a contamination is entirely avoided in the $uiK_s$ plane which is due to the mostly blue-ward evolution of the $(u\!-\!i)$ color when redshifting the steeply increasing near-UV part of star-forming galaxy SEDs.~The redshift evolution of the corresponding passively evolving SEDs overlaps with the local galaxy background of Virgo and its diagnostic power is only limited by the depth of our $u^*$-band photometry.~The shallower magnitude limit of the $u^*$-band data is apparent when the blue edge of the galaxy distribution in each diagram of Figure~\ref{fig:giKuik_withTracks} is compared.~Although the redshift distribution of the background galaxies in the $uiK_s$ plane cuts off at lower values as compared to the $gzK_s$ plane, especially along the sequence of passively evolving galaxies (the cut-off is near $z\!\approx\!0.5$ instead of near $z\!\approx\!1$), the $uiK_s$ plane provides overall a clearly linear and powerful selection diagnostic of star-forming galaxies in the redshift range $z\!\simeq\!0.5\!-\!1.5$.~Because of this property and the photometric depth of the NGVS+NGVS-IR data we are in the position of searching for galaxy clusters at redshifts $z\!\ga\!1.0$, which will be described in detail in future papers.

\subsubsection{The Diagnostic Power of the $uiK_s$ Plane}
The reason for the superior GC-star and GC-galaxy separation power of the $uiK_s$ plane lies in the combination of the near-UV, optical, and near-IR color information, compared to the classic optical/near-IR $gzK_s$ plot (see Fig.~\ref{fig:giKuik_withGCs}).~While the optical/near-IR $(i\!-\!K_s)$ color probes the effective temperature of the red giant branch (RGB) stellar population, the near-UV/optical $(u\!-\!i)$ color probes the stellar fluxes blue- and redward of the $4000$\,\AA\ break and is, therefore, sensitive to hot stellar evolutionary phases.~This leads to a very efficient GC-galaxy separation due to the sensitivity of the $u$-band to hot stellar components in star-forming galaxies, combined with the redshift evolution of galaxy SEDs (see Figs.~\ref{fig:sedplot} and \ref{fig:giKuik_withTracks}).~The efficient GC-star separation in the $uiK_s$ diagram is due to the fact that at a given $(i\!-\!K_s)$ color (i.e.~effective RGB temperature), the $u$-band picks up the additional hot stellar component flux generated by horizontal branch stars, making GCs appear bluer in the $(u\!-\!i)$ color.

It is worth pointing out that attempts at photometrically separating extragalactic GCs from foreground stars and background galaxies exist in the recent literature, however, based on a combination of purely optical/near-IR \citep{puzia02} or purely near-UV/optical colors \citep{kim13}.~Neither of these color-color planes can separate foreground stars, GCs, and background galaxies.~The only earlier study that combined near-UV/optical/near-IR photometry of relatively small GC samples was attempting to study the age distribution function of GCs in giant elliptical galaxies \citep{hempel04}, and was based on much poorer photometric quality.~To our knowledge, the $uiK_s$ diagnostic plane presented in Figure~\ref{fig:giKuik_withGCs} is the first such plot based on high-quality near-UV/optical/near-IR photometry that allows for an efficient GC-star and GC-galaxy separation.

In order to assess the required photometric quality for robust GC-star and GC-galaxy separation, we plot in Figure~\ref{fig:giKuik_errdistrib} the $gzK_s$ and $uiK_s$ color-color plane and parametrize the symbol colors by the photometric uncertainty in each contributing filter.~It becomes immediately clear that the photometric depth, i.e.~photometric error distribution function in each filter, influences in a non-linear, and sometimes quite dramatic way the selection of $BzK$ galaxies as defined in \cite{daddi04}.~The corresponding relations are shown in the left panels (as in Figure~\ref{fig:giKuik_withGCs}) and illustrate the quality of selecting the classically defined galaxy samples as a function of photometric depth. It is also quite evident that even with the NGVS+NGVS-IR high-quality data, a clear separation of stellar foreground from the background galaxy population is not entirely possible based on the $gzK_s$ color-color diagram alone, in particular Virgo GCs cannot be robustly identified.

The right panels of Figure~\ref{fig:giKuik_errdistrib} demonstrate the high potential for studies of stellar populations in the nearby universe. We find that datasets with a photometric accuracy of $\Delta u\la0.05$, $\Delta i\la0.05$, and $\Delta K_s\la0.05$ mag (corresponding to $u\!\simeq\!24.4$, $i\!\simeq\!23.4$, $K_s\!\simeq\!22.0$ AB mag in our dataset) will have the potential to select very clean samples of extragalactic GCs.~We will discuss and quantify the exact contamination fractions of such samples in future papers.

\begin{figure}[!t]
\centering
\includegraphics[width=8.9cm, bb=20 0 590 415]{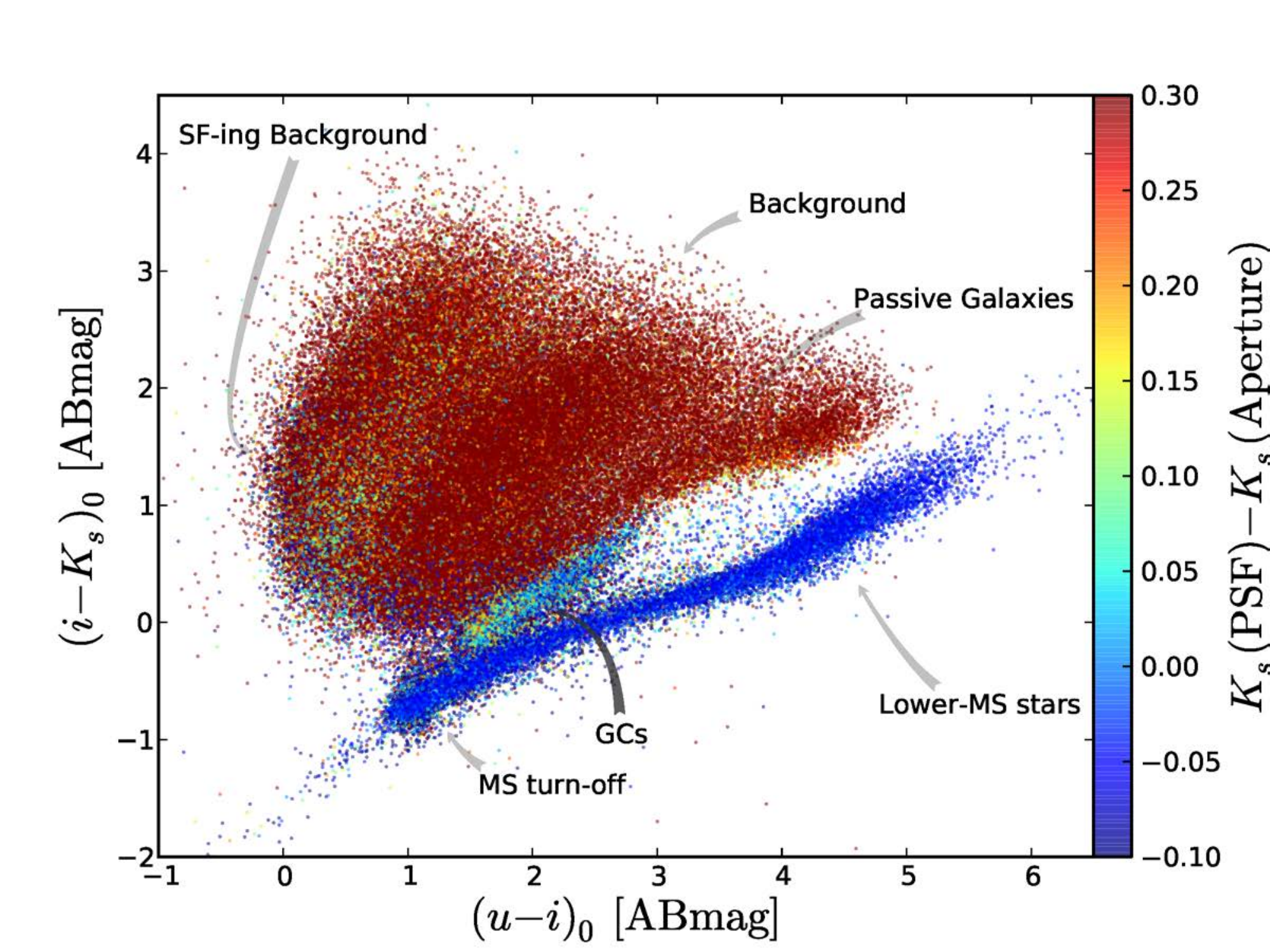}
\caption{Illustration of the $K_s$-band morphology variations across the $uiK_s$ diagram. As a measurement of compactness, this figure uses the difference between the $K_s$-band PSF magnitude (resulting from the combined used of {\sc PSFEx} and {\sc SExtractor}) and a magnitude obtained within an aperture with 3\arcsec\ diameter.}
\label{fig:uik_morph}
\end{figure}

With the superior spatial resolution of our $K_s$-band data (median seeing $0.54\arcsec$, see Section~\ref{txt:stacks} and Figure~\ref{fig:seeing}) we search for additional diagnostic tools related to the morphology of objects in our survey field.~For this, we devise a simple magnitude difference between a 3\arcsec\ diameter aperture magnitude and a PSF magnitude.~The PSF magnitude produces integrated colors for stars very similar to the total integrated colors for GCs, and a color roughly representative of the central colors for galaxies.~The results for all our objects with $K_s$-band photometry are shown in Figure~\ref{fig:uik_morph} as a color parametrization of the $uiK_s$ plane according to the morphological measure of their radially symmetric  compactness.~Intriguing correlations between this simple morphology parameter and $uiK_s$ colors appear, that will be studied in subsequent papers of this series.~We note here, that the figure clearly shows that GCs are marginally resolved in the WIRCam $K_s$ images: their observed shapes are slightly less concentrated than those of stars.~The accuracy of measuring relative GC sizes is sufficient to even make out the classic size difference between blue and red GCs \citep[e.g.][and references therein]{webb12}. Attempts to identify GCs solely on the basis of their morphology, however, produce samples whose colors reveal a high level of contamination by stars and remote galaxies. Our final classification algorithm (in preparation) will therefore be based on both colors and morphology.

\section{Summary}
\label{sec:sum}
In this paper, we have described the NGVS-IR survey, its motivation, observational strategy, and science goals. We discuss in detail the data reduction procedures and present $K_s$-band imaging data of the central 4\;deg$^2$ of the Virgo galaxy cluster observed with WIRCam mounted on the Canada-France-Hawaii Telescope (CFHT). These NGVS-IR images have a median seeing of $0.54\arcsec$ and a 90\% completeness magnitude of better than $K_s\!=\!23.3$ AB mag, thus, superseding in all aspects the 2MASS and UKIDSS mosaics of the same region.

Using the full near-UV to near-IR SED coverage of our NGVS+NGVS-IR data we investigate the $uiK_s$ color-color plane as a diagnostic tool to identify and select characteristic object classes.~We also study the $gzK_s$ color-color plane as the closest equivalent to the $BzK$ plot from which star-forming galaxies at redshift $z\!>\!1.4$ are typically selected.~With the described $uiK_s$ diagnostics we can identify several distinct and isolated groups of objects in the $uiK_s$ diagram that correspond to {\it i)} star-forming galaxies out to redshifts $\sim\!1.5$ and higher, {\it ii)} a mixed distribution of galaxies with a variety of (non-negligible) levels of on-going star formation, {\it iii)} a redshift sequence of passively evolving old galaxies, {\it iv)} a relatively smooth sequence of Virgo GCs and UCDs, and {\it v)} a very distinct sequence of foreground Milky Way stars. 

According to the photometric error distributions of our NGVS+NGVS-IR data in the $uiK_s$ color-color diagram (see right panels of Figure~\ref{fig:giKuik_errdistrib}), we can isolate the Virgo GC population, virtually free of contamination from foreground stars and background galaxies, if we restrict our selection to objects with photometric errors $\Delta u\!\la\!0.05$, $\Delta i\!\la\!0.05$ and $\Delta K_s\!\la\!0.05$ mag.~Independent of the photometric quality of the dataset such a selection would be fundamentally impossible from the $gzK_s$ diagram.~The $uiK_s$  diagram is to our knowledge the most efficient and least biased extragalactic GC selection technique that is relying purely on photometric colors and is likely to prove extremely efficient in the study of distant star cluster systems with the combined near-UV/optical/near-IR instrumentation that will be available on the upcoming generation of survey telescopes, such as the Large Synoptic Survey Telescope (LSST) and the Euclid spacecraft, as well as future facilities, like for instance E-ELT, JWST, etc.

\acknowledgments
{\it Acknowledgments} -- This research was supported by CONICYT through Gemini-CONICYT Project No.~32100022, FONDECYT Regular Project No.~1121005, FONDECYT Postdoctoral Fellowship Project No.~3130750, FONDAP Center for Astrophysics (15010003), and BASAL Center for Astrophysics and Associated Technologies (PFB-06), the French Agence Nationale de la Recherche (ANR) Grant Programme Blanc VIRAGE (ANR10-BANC-0506-01).~R.P.~Mu\~noz and A.~Lan\c{c}on acknowledge support from the Scientific Council of Strasbourg University.~A.~Lan\c{c}on is grateful for the hospitality and financial support during her stay at the Instituto de Astrof\'isica of Pontificia Universidad Cat\'olica de Chile which was funded by the UMI International Academic Exchange Fund.~E.~W.~Peng acknowledges support from the National Natural Science Foundation of China (grant 11173003), and from the Laboratoire International Associ\'e ``ORIGINS''.~C.~Liu acknowledges support from the National Natural Science Foundation of China (grant 11203017).

Based on observations obtained with WIRCam, a joint project of CFHT, Taiwan, Korea, Canada, France, at the Canada-France-Hawaii Telescope (CFHT) which is operated by the National Research Council (NRC) of Canada, the Institute National des Sciences de l'Univers of the Centre National de la Recherche Scientifique of France, and the University of Hawaii.~This publication makes use of data products from the Two Micron All Sky Survey, which is a joint project of the University of Massachusetts and the Infrared Processing and Analysis Center/California Institute of Technology, funded by the National Aeronautics and Space Administration and the National Science Foundation.~This work is based in part on data obtained as part of the UKIRT Infrared Deep Sky Survey (found on the UKIRT site).~This research has made use of the VizieR catalogue access tool and the Aladin plot tool at CDS, Strasbourg, France.~We thank Sebastien Foucaud for providing a version of Swarp software that includes the sigma clipping rejection method.~The authors acknowledge useful discussions with Simon Angel, Sibilla Perina, Ryan Quadri, Alvio Renzini, Mirko Simunovic and Jin-cheng Yu.

\vspace{0.25cm}
\noindent {\it Facilities:} \facility{CFHT (WIRCam, MegaCam)}.

\clearpage

%==========================
\appendix

%==========================
\section{WIRCam $K_s$ photometry - additional information}
%-----------------------------------------------------
\subsection{AB magnitudes versus Vega-based magnitudes}
%...........
\label{app:ABvsVega}

The constant difference between Vega-based magnitudes and AB magnitudes 
for the WIRCam Ks observations can be obtained from synthetic photometry.
The CFHT WIRCam web pages\footnote{
{\tt http://www.cfht.hawaii.edu/Instruments/WIRCam/ dietWIRCam.html} 
} warn that various authors have used conversion constants that differ 
by up to 0.2 magnitudes. How the various values have been obtained,
however, is not documented systematically and is rarely indicated
in journal publications.  We detail our own derivation below.  

\begin{figure*}
\centering
\includegraphics[width=7cm]{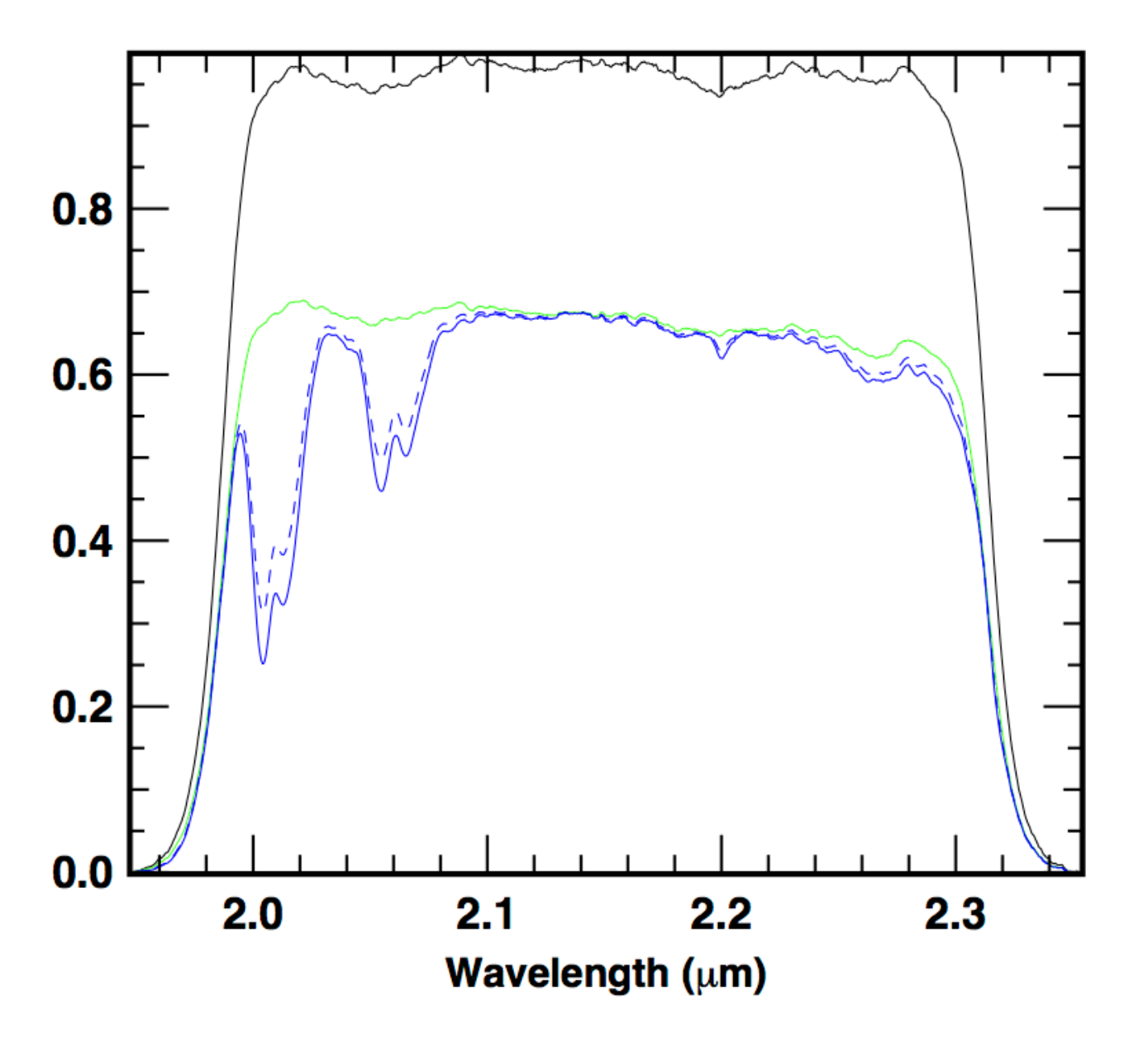}
\caption[WIRCam Ks Transmission]{Components of the WIRCam Ks transmission used
to compute synthetic photometry. The uppermost curve shows the filter
transmission. The efficiency of the optics reduces the global efficiency
to values between 60 and 70 percent (green curve). The additional 
effect of telluric features is shown for airmass of 1 (dotted) and
1.5 (solid).}
\label{fig:KsTransmission}
\end{figure*}

Figure\,\ref{fig:KsTransmission} shows the components of the WIRCam
Ks transmission curve. The data for the instrument components
were obtained from the CFHT WIRCam Throughput web pages 
(files {\tt cfh8302.dat} for the filter, and file 
{\tt WIRCamOpticsResponseCurve.xls} for the optics). The optics 
transmission curve includes all optics except the telescope mirror 
and tip-tilt plate. We thus implicitely assume these two components
have a flat response. The atmospheric transmission for Mauna Kea 
was obtained from the website of the Gemini Observatory. 
It is based on the ATRAN code (Lord, s.D. 1992, NASA Technical Memorandum 103957)
and assumes a water vapor column of 1mm.

\begin{table*}
\begin{center}
\caption{Synthetic AB magnitude of Vega, for various transmission curves and Vega models. An airmass of 1.5 is assumed in the last line of this table. The recommended value is highlighted.}
\label{tab:Vega2AB}
\begin{tabular}{lcc} 
\hline\hline
                             &  alpha.lyr.stis.003 & VegaLCB \\ \hline
filter only               &  1.821  &  1.828 \\
filter+optics           & 1.819   & 1.826 \\
filter+optics+atm.  & \fbox{1.827}  &  1.834 \\ 
\hline
\end{tabular}
\end{center}
\end{table*}

Table\,\ref{tab:Vega2AB} provides the conversion constants obtained for WIRCam Ks
under various assumptions. Three models for Vega were used.
{\tt alpha.lyr.stis.003} and {\tt alpha.lyr.stis.005} are two
versions of the standard Vega spectrum distributed by the 
Space Telescope Science Institute (Bohlin 2007 ASP Conf Ser. 364, 315,
and {\tt calspec} webpage on {\tt www.stsci.edu}). 
They differ only shortward of
0.53\,$\mu$m, which affects the NGVS $u^*$ and $g$ bands but not WIRCam Ks. 
{\tt VegaLCB} is the reference spectrum  distributed 
with the population synthesis code {\sc P\'egase} (Fioc \& Rocca-Volmerange
1997). The differences with the
HST spectra are (1) a globally lower flux level (0.7\,\%), (2) a 
lower spectral resolution (which leads to small differences of the 
integrals over hydrogen features), and (3) local differences in the shape
of the spectrum, such as between 2.3 and 2.5\,$\mu$m or around 1.5\,$\mu$m.
The conversion constant we recommend is
$$ K_s(\mathrm{AB})-K_s(\mathrm{Vega}) = 1.827\,\mathrm{mag} $$
This value  is essentially identical to
the value obtained by St\'ephane Arnouts at CFHT (1.824\,mag).

Users of {\sc P\'egase} should be reminded that code provides 
Vega-type magnitudes and colors under the assumption that 
Vega has a magnitude of 0.03 (and not 0.0) in all filters. 
Therefore, the WIRCam Ks magnitudes produced for synthetic populations 
by {\sc P\'egase} in the AB-system and in the Vega-based system, 
using {\tt VegaLCB} as a reference, differ by 1.804 instead of 1.834.
Rather than using Vega, our customized version of {\sc P\'egase} 
computes AB synthetic magnitudes directly.

\section{Extinction coefficients}
\label{app:extinction}

We have derived extinction coefficients by comparing the synthetic 
colors of stellar model spectra and of reddened versions thereof, 
using the extinction law of Cardelli et al. (1989) with R$_{V}$=3.1.
The stellar models used were representations of Vega, of the Sun
and of a red star (in practice, the solar model seen through 3 magnitudes
of $V$-band extinction). Table~\ref{tab:ExtinctionCoefs} lists the 
results.

\begin{table*}
\begin{center}
\caption[]{Extinction coefficients for NGVS+NGVS-IR filters.}
\begin{tabular}{llcccccc}
\hline\hline
 & & $u^*$ & $g$ & $r$ & $i$ & $z$ & $K_s$ \\ 
\hline
Vega & A($\lambda$)/A(V) & 1.490 & 1.190 & 0.874 & 0.674 & 0.498 & 0.118 \\
     & Effective $\lambda$ ($\mu$m) &  0.3895 & 0.4803 & 0.6212 & 0.7493 & 0.8849 & 2.144 \vspace{0.1cm} \\
Sun &A($\lambda$)/A(V) & 1.492 & 1.160 & 0.868 & 0.668 & 0.498 & 0.120 \\
     & Effective $\lambda$ ($\mu$m) & 0.3876 & 0.4906 & 0.6252 & 0.7532 & 0.885 & 2.144 \vspace{0.1cm} \\
Red star & A($\lambda$)/A(V) & 1.462 & 1.120 & 0.858 & 0.654 & 0.492 & 0.120 \\
     & Effective $\lambda$ ($\mu$m) &  0.3967 & 0.5038 & 0.6309 & 0.7620 & 0.8904 & 2.146 \\
\hline
\end{tabular}
\label{tab:ExtinctionCoefs}
\end{center}
\end{table*}

One may compare the values obtained here for CFHT/MegaCam
with those provided for a G2V star on the stellar isochrone
web site of Padova Observatory 
({\tt http://stev.oapd.inaf.it/cgi-bin/cmd}), 
where the same extinction law is assumed.
Their values of A($\lambda$)/A(V) for Megacam $u^*, g, r, i, z$ are
1.466, 1.167, 0.860, 0.656, 0.500. No comparison value is available
through that web interface for WIRCam/Ks.

Note that the foreground extinction towards Virgo is smaller
A(V)=0.1. Uncertainties associated with neglecting color terms or 
using one or the other reference for extinction coefficients will therefore
be smaller than $\sim$0.5\,\%.


\begin{thebibliography}{}
\bibitem[Ames(1930)]{ames30} Ames, A.\ 1930, Annals of Harvard College Observatory, 88, 1 
\bibitem[Arimoto(1996)]{arimoto96} Arimoto, N.\ 1996, From Stars to Galaxies: the Impact of Stellar Physics on Galaxy Evolution, 98, 287
\bibitem[Bekki et al.(2003)]{bekki03} Bekki, K., Couch, W.~J., Drinkwater, M.~J., \& Shioya, Y.\ 2003, \mnras, 344, 399
\bibitem[Bertin et al.(2002)]{bertin02} Bertin, E., Mellier, Y., Radovich, M., et al.\ 2002, Astronomical Data Analysis Software and Systems XI, 281, 228
\bibitem[Bertin(2006)]{bertin06} Bertin, E.\ 2006, Astronomical Data Analysis Software and Systems XV, 351, 112 
\bibitem[Bertin \& Arnouts(1996)]{bertin96} Bertin, E., \& Arnouts, S.\ 1996, A\&AS, 117, 393
\bibitem[Bertin(2011)]{bertin11} Bertin, E.\ 2011, Astronomical Data Analysis Software and Systems XX, 442, 435 
\bibitem[Binggeli et al.(1985)]{binggeli85} Binggeli, B., Sandage, A., \& Tammann, G.~A.\ 1985, \aj, 90, 1681
\bibitem[Bielby et al.(2012)]{bielby12} Bielby, R., Hudelot, P., McCracken, H.~J., et al.\ 2012, \aap, 545, A23
\bibitem[Blakeslee et al.(2009)]{blakeslee09} Blakeslee, J.~P., Jord{\'a}n, A., Mei, S., et al.\ 2009, \apj, 694, 556
\bibitem[Blakeslee(2012)]{blakeslee12} Blakeslee, J.~P.\ 2012, Ap\&SS, 341, 179
\bibitem[Cohen et al.(1998)]{cohen98} Cohen, J.~G., Blakeslee, J.~P., \& Ryzhov, A.\ 1998, \apj, 496, 808
\bibitem[C{\^o}t{\'e} et al.(2004)]{cote04} C{\^o}t{\'e}, P., Blakeslee, J.~P., Ferrarese, L., et al.\ 2004, \apjs, 153, 223
\bibitem[Cutri et al.(2003)]{cutri03} Cutri, R.~M., Skrutskie, M.~F., van Dyk, S., et al.\ 2003, VizieR Online Data Catalog, 2246, 0
\bibitem[Dahlen et al.(2013)]{dahlen13} Dahlen, T., Mobasher, B., Faber, S.~M., et al.\ 2013, \apj, 775, 93
\bibitem[Daddi et al.(2004)]{daddi04} Daddi, E., Cimatti, A., Renzini, A., et al.\ 2004, \apj, 617, 746
\bibitem[Dalton et al.(2006)]{dalton06} Dalton, G.~B., Caldwell, M., Ward, A.~K., et al.\ 2006, \procspie, 6269
\bibitem[Faber et al.(2007)]{faber07} Faber, S.~M., Willmer, C.~N.~A., Wolf, C., et al.\ 2007, \apj, 665, 265
\bibitem[Fellhauer \& Kroupa(2002)]{fellhauer02} Fellhauer, M., \& Kroupa, P.\ 2002, \mnras, 330, 642
\bibitem[Ferrarese et al.(2012)]{ferrarese12} Ferrarese, L., C{\^o}t{\'e}, P., Cuillandre, J.-C., et al.\ 2012, \apj, 200, 4
\bibitem[Fioc \& Rocca-Volmerange(1997)]{fioc97} Fioc, M., \& Rocca-Volmerange, B.\ 1997, \aap, 326, 950
\bibitem[Forte et al.(2013)]{forte13} Forte, J.~C., Faifer, F.~R., Vega, E.~I., et al.\ 2013, \mnras, 431, 1405
\bibitem[Gavazzi et al.(2003)]{gavazzi03} Gavazzi, G., Boselli, A., Donati, A., Franzetti, P., \& Scodeggio, M.\ 2003, \aap, 400, 451 
\bibitem[Hanes \& Brodie(1986)]{hanes86} Hanes, D.~A., \& Brodie, J.~P.\ 1986, \apj, 300, 279
\bibitem[Hanes et al.(2001)]{hanes01} Hanes, D.~A., C{\^o}t{\'e}, P., Bridges, T.~J., et al.\ 2001, \apj, 559, 812 
\bibitem[Harris et al.(2006)]{harris06} Harris, W.~E., Whitmore, B.~C., Karakla, D., et al.\ 2006, \apj, 636, 90
\bibitem[Ha{\c s}egan et al.(2005)]{hasegan05} Ha{\c s}egan, M., Jord{\'a}n, A., C{\^o}t{\'e}, P., et al.\ 2005, \apj, 627, 203
\bibitem[Hempel \& Kissler-Patig(2004)]{hempel04} Hempel, M., \& Kissler-Patig, M.\ 2004, \aap, 428, 459
\bibitem[Hilker(2011)]{hilker11} Hilker, M.\ 2011, EAS Publications Series, 48, 219
\bibitem[Jensen et al.(2003)]{jensen03} Jensen, J.~B., Tonry, J.~L., Barris, B.~J., et al.\ 2003, \apj, 583, 712
\bibitem[Jones et al.(2006)]{jones06} Jones, J.~B., Drinkwater, M.~J., Jurek, R., et al.\ 2006, \aj, 131, 312 
\bibitem[Jord{\'a}n et al.(2002)]{jordan02} Jord{\'a}n, A., C{\^o}t{\'e}, P., West, M.~J., \& Marzke, R.~O.\ 2002, \apjl, 576, L113
\bibitem[Jord{\'a}n et al.(2009)]{jordan09} Jord{\'a}n, A., Peng, E.~W., Blakeslee, J.~P., et al.\ 2009, \apjs, 180, 54
\bibitem[Kissler-Patig et al.(2002)]{kisslerpatig02} Kissler-Patig, M., Brodie, J.~P., \& Minniti, D.\ 2002, \aap, 391, 441
\bibitem[Kotulla et al.(2009)]{kotulla09} Kotulla, R., Fritze, U., Weilbacher, P., \& Anders, P.\ 2009, \mnras, 396, 462
\bibitem[Lan{\c c}on \& Mouhcine(2000)]{lancon00} Lan{\c c}on, A., \& Mouhcine, M.\ 2000, Massive Stellar Clusters, 211, 34
\bibitem[Lane et al.(2007)]{lane07} Lane, K.~P., Almaini, O., Foucaud, S., et al.\ 2007, \mnras, 379, L25
\bibitem[Lawrence et al.(2007)]{lawrence07} Lawrence, A., Warren, S.~J., Almaini, O., et al.\ 2007, \mnras, 379, 1599
\bibitem[Lawrence et al.(2012)]{lawrence12} Lawrence, A., Warren, S.~J., Almaini, O., et al.\ 2012, VizieR Online Data Catalog, 2314, 0
\bibitem[Lin et al.(2012)]{lin12} Lin, L., Dickinson, M., Jian, H.-Y., et al.\ 2012, \apj, 756, 71
\bibitem[Liu et al.(2005)]{liu05} Liu, Y., Zhou, X., Ma, J., et al.\ 2005, \aj, 129, 2628
\bibitem[Ly et al.(2012)]{ly12} Ly, C., Malkan, M.~A., Kashikawa, N., et al.\ 2012, \apj, 757, 63
\bibitem[McCracken et al.(2010)]{mccracken10} McCracken, H.~J., Capak, P., Salvato, M., et al.\ 2010, \apj, 708, 202
\bibitem[McCracken et al.(2012)]{mccracken12} McCracken, H.~J., Milvang-Jensen, B., Dunlop, J., et al.\ 2012, \aap, 544, A156
\bibitem[Merson et al.(2013)]{merson13} Merson, A.~I., Baugh, C.~M., Helly, J.~C., et al.\ 2013, \mnras, 429, 556
\bibitem[Mouhcine et al.(2005)]{mouhcine05} Mouhcine, M., Gonz{\'a}lez, R.~A., \& Liu, M.~C.\ 2005, \mnras, 362, 1208
\bibitem[Kim et al.(2013)]{kim13} Kim, H.-S., Yoon, S.-J., Sohn, S.~T., et al.\ 2013, \apj, 763, 40
\bibitem[King(1978)]{king78} King, I.~R.\ 1978, \apj, 222, 1
\bibitem[Kurk et al.(2013)]{kurk13} Kurk, J., Cimatti, A., Daddi, E., et al.\ 2013, \aap, 549, A63
\bibitem[McDonald et al.(2011)]{mcdonald11} McDonald, M., Courteau, S., Tully, R.~B., \& Roediger, J.\ 2011, \mnras, 414, 2055
\bibitem[Mei et al.(2007)]{mei07} Mei, S., Blakeslee, J.~P., C{\^o}t{\'e}, P., et al.\ 2007, \apj, 655, 144
\bibitem[Misgeld \& Hilker(2011)]{misgeld11} Misgeld, I., \& Hilker, M.\ 2011, \mnras, 414, 3699
\bibitem[Park \& Choi(2005)]{park05} Park, C., \& Choi, Y.-Y.\ 2005, \apjl, 635, L29
\bibitem[Park et al.(2012)]{park12} Park, H.~S., Lee, M.~G., \& Hwang, H.~S.\ 2012, \apj, 757, 184
\bibitem[Paudel et al.(2013)]{paudel13} Paudel, S., Duc, P.-A., C{\^o}t{\'e}, P., et al.\ 2013, \apj, 767, 133
\bibitem[Peng et al.(2008)]{peng08} Peng, E.~W., Jord{\'a}n, A., C{\^o}t{\'e}, P., et al.\ 2008, \apj, 681, 197
\bibitem[Peng et al.(2009)]{peng09} Peng, E.~W., Jord{\'a}n, A., Blakeslee, J.~P., et al.\ 2009, \apj, 703, 42
\bibitem[Pessev et al.(2008)]{pessev08} Pessev, P.~M., Goudfrooij, P., Puzia, T.~H., \& Chandar, R.\ 2008, \mnras, 385, 1535
\bibitem[Pota et al.(2013)]{pota13} Pota, V., Forbes, D.~A.,Romanowsky, A.~J., et al.\ 2013, \mnras, 428, 389
\bibitem[Puget et al.(2004)]{pug04} Puget, P., Stadler, E., Doyon, R., et al.\ 2004, \procspie, 5492, 978
\bibitem[Puzia et al.(2002)]{puzia02} Puzia, T.~H., Zepf, S.~E., Kissler-Patig, M., et al.\ 2002, \aap, 391, 453
\bibitem[Puzia et al.(2007)]{puzia07} Puzia, T.~H., Mobasher, B., \& Goudfrooij, P.\ 2007, \aj, 134, 1337
\bibitem[Raimondo et al.(2010)]{raimondo10} Raimondo, G., Cantiello, M., Brocato, E., \& Biscardi, I.\ 2010, Memorie della Societa Astronomica Italiana Supplementi, 14, 63
\bibitem[Rangel et al.(2013)]{rangel13} Rangel, C., Nandra, K., Laird, E.~S., \& Orange, P.\ 2013, \mnras, 428, 3089
\bibitem[Reaves(1956)]{reaves56} Reaves, G.\ 1956, \aj, 61, 69
\bibitem[Reaves(1983)]{reaves83} Reaves, G.\ 1983, \apjs, 53, 375
\bibitem[Riffel et al.(2011)]{riffel11} Riffel, R., Ruschel-Dutra, D., Pastoriza, M.~G., et al.\ 2011, \mnras, 410, 2714 
\bibitem[Romanowsky et al.(2012)]{romanowsky12} Romanowsky, A.~J., Strader, J., Brodie, J.~P., et al.\ 2012, \apj, 748, 29
\bibitem[Schlafly \& Finkbeiner(2011)]{schlafly11} Schlafly, E.~F., \& Finkbeiner, D.~P.\ 2011, \apj, 737, 103
\bibitem[Schuberth et al.(2012)]{schuberth12} Schuberth, Y., Richtler, T., Hilker, M., et al.\ 2012, \aap, 544, A115
\bibitem[Sick et al.(2013)]{sick13} Sick, J., Courteau, S., Cuillandre, J.-C., et al.\ 2013, arXiv:1303.6290
\bibitem[Skrutskie et al.(2006)]{skrutskie06} Skrutskie, M.~F., Cutri, R.~M., Stiening, R., et al.\ 2006, \aj, 131, 1163
\bibitem[Strader et al.(2011)]{strader11} Strader, J., Romanowsky, A.~J., Brodie, J.~P., et al.\ 2011, \apjs, 197, 33
\bibitem[Tamura et al.(2006a)]{tamura06a} Tamura, N., Sharples, R.~M., Arimoto, N., et al.\ 2006a, \mnras, 373, 588 
\bibitem[Tamura et al.(2006b)]{tamura06b} Tamura, N., Sharples, R.~M., Arimoto, N., et al.\ 2006b, \mnras, 373, 601
\bibitem[Taylor et al.(2011)]{taylor11} Taylor, E.~N., Hopkins, A.~M., Baldry, I.~K., et al.\ 2011, \mnras, 418, 1587
\bibitem[Tonry et al.(1990)]{tonry90} Tonry, J.~L., Ajhar, E.~A., \& Luppino, G.~A.\ 1990, \aj, 100, 1416
\bibitem[de Vaucouleurs et al.(1991)]{RC3} de Vaucouleurs, G., de Vaucouleurs, A., Corwin, H.~G., Jr., et al.\ 1991, Third Reference Catalogue of Bright Galaxies.~Volume I: Explanations and references.~Volume II: Data for galaxies between 0$^{h}$ and 12$^{h}$.~Volume III: Data for galaxies between 12$^{h}$ and 24$^{h}$., by de Vaucouleurs, G.; de Vaucouleurs, A.; Corwin, H.~G., Jr.; Buta, R.~J.; Paturel, G.; Fouqu{\'e}, P..~Springer, New York, NY (USA), 1991, 2091 p., ISBN 0-387-97552-7, Price US\$ 198.00.~ISBN 3-540-97552-7, Price DM 448.00.~ISBN 0-387-97549-7 (Vol.~I), ISBN 0-387-97550-0 (Vol.~II), ISBN 0-387-97551-9 (Vol.~III)
\bibitem[van Dokkum(2001)]{dokkum01} van Dokkum, P.~G.\ 2001, \pasp, 113, 1420
\bibitem[Vogel(1979)]{vogel79} Vogel, H.\ 1979, Mathematical Biosciences, 44, 179-189
\bibitem[Webb et al.(2012)]{webb12} Webb, J.~J., Harris, W.~E., \& Sills, A.\ 2012, \apjl, 759, L39
\bibitem[Worthey(1994)]{worthey94} Worthey, G.\ 1994, \apjs, 95, 107
\bibitem[Yoon et al.(2011)]{yoon11} Yoon, S.-J., Sohn, S.~T., Lee, S.-Y., et al.\ 2011, \apj, 743, 149
\bibitem[Yuma et al.(2011)]{yuma11} Yuma, S., Ohta, K., Yabe, K., Kajisawa, M., \& Ichikawa, T.\ 2011, \apj, 736, 92
\bibitem[Yuma et al.(2012)]{yuma12} Yuma, S., Ohta, K., \& Yabe, K.\ 2012, \apj, 761, 19
\end{thebibliography}
\end{document}